\documentclass[aps,prx,superscriptaddress,reprint,twocolumn,longbibliography]{revtex4-1}

\usepackage{graphicx}
\usepackage{amsmath}
\usepackage{amssymb}
\usepackage{dsfont}
\usepackage{here}

\newcommand{\mD}[0]{\mathcal{D}}
\newcommand{\mP}[0]{\mathcal{P}}
\newcommand{\mT}[0]{\mathcal{T}}
\newcommand{\mI}[0]{\mathcal{I}}
\newcommand{\mM}[0]{\mathcal{M}}
\newcommand{\tr}[0]{{\rm Tr}}
\newcommand{\eff}[0]{^{\rm eff}}
\newcommand{\E}[0]{{\rm E}}

\begin{document}

\title{Fluctuation theorems in feedback-controlled open quantum systems:\\quantum coherence and absolute irreversibility}

\author{Y\^uto Murashita}
\email{murashita@cat.phys.s.u-tokyo.ac.jp}
\author{Zongping Gong}
\author{Yuto Ashida}
\affiliation{Department of Physics, University of Tokyo, 7-3-1 Hongo, Bunkyo-ku, Tokyo, 113-0033, Japan}
\author{Masahito Ueda}
\affiliation{Department of Physics, University of Tokyo, 7-3-1 Hongo, Bunkyo-ku, Tokyo, 113-0033, Japan}
\affiliation{RIKEN Center for Emergent Matter Science (CEMS), Wako, Saitama 351-0198, Japan}
\date{\today}

\begin{abstract}
Thermodynamics of quantum coherence has attracted growing attention recently, where the thermodynamic advantage of quantum superposition is characterized in terms of quantum thermodynamics.
We investigate thermodynamic effects of quantum coherent driving in the context of the fluctuation theorem.
We adopt a quantum-trajectory approach to investigate open quantum systems under feedback control.
In these systems, the measurement backaction in the forward process plays a key role, and therefore the corresponding time-reversed quantum measurement and post-selection must be considered in the backward process in sharp contrast to the classical case.
The state reduction associated with quantum measurement, in general, creates a zero-probability region in the space of quantum trajectories of the forward process, which causes singularly strong irreversibility with divergent entropy production (i.e., absolute irreversibility) and hence makes the ordinary fluctuation theorem break down.
In the classical case, the error-free measurement ordinarily leads to absolute irreversibility because the measurement restricts classical paths to the region compatible with the measurement outcome.
In contrast, in open quantum systems, absolute irreversibility is suppressed even in the presence of the projective measurement due to those quantum rare events that go through the classically forbidden region with the aid of quantum coherent driving.
This suppression of absolute irreversibility exemplifies the thermodynamic advantage of quantum coherent driving.
Absolute irreversibility is shown to emerge in the absence of coherent driving after the measurement, especially in systems under time-delayed feedback control.
We show that absolute irreversibility is mitigated by increasing  the duration of quantum coherent driving or decreasing the delay time of feedback control.
\end{abstract}

\maketitle

\section{Introduction}
Nanotechnology has commanded the attention of researchers to thermodynamics in micro- and nano-scale devices.
The interplay between thermodynamics and quantum coherence has emerged as one of the central issues in physics.
The pioneering work by Scully et al. \cite{Scully_2003} demonstrated that quantum coherence in the thermal environment changes its effective temperature, enabling one to extract positive work from an isothermal environment.
Later on, it was shown that the efficiency of photon quantum heat engines can be enhanced by quantum coherence between degenerate energy levels of atoms since quantum coherence breaks the detailed balance \cite{DillenschneiderLutz2009,Scully_2011}.
Furthermore, several researches have found evidence that quantum coherence is utilized in biological processes at room temperature \cite{Lloyd2011}, and, in particular, that the photosynthetic transport of excitons is boosted by the interplay between quantum coherence and environmental noise \cite{EngelFleming2007,Collini_Scholes2010,Lloyd_2008,PlenioHuelga2008}.

Meanwhile, quantum information theory has reignited interest in the quantitative characterization of quantum coherence.
Quantum resource theory, which originates from  theory of entanglement, was applied for quantitative measures of quantum coherence \cite{Plenio_2014,Lostaglio_2015,WinterYang2016,Plenio_2016}.
Although the free energy corresponding to quantum coherence cannot be converted into work under thermal operations  which is known as work locking \cite{HorodeckiOppenheim2013,Skrzypczyk2014,Lostaglio_2015nc}, we can extract work from it by catalystically using a coherent resource \cite{Aberg2014,Korzekwa2016}.
Thus, quantum coherence can be harnessed as a thermodynamic resource with the help of quantum information theory.

While thermal operations in resource theory are defined by their mathematical properties, it is unclear how to connect work defined in it with work measured experimentally in the context of nonequilibrium statistical mechanics \cite{AZUe15}.
The quantum fluctuation theorem \cite{Tasaki2000,Mukamel2003,DeRoeckMaes2004,EspositoMukamel2006,TalknerLutzHanggi2007,TalknerHanggi2007,Crooks2008,CampisiTalknerHanggi2009,EspositoHarbolaMukamel2009,CampisiHanggiTalkner2011} is expected to deepen our understanding of thermodynamics of quantum coherence, yet quantum coherence does not manifest itself in the quantum fluctuation theorem.
This is because entropy production can be defined via the diagonal decomposition in parallel with the classical case and consequently the quantum fluctuation theorem assumes the same form as that of the classical one \cite{Jarzynski1997,Crooks1999,Seifert2005,Jarzynski2011,Seifert2012,EspositoVandenBroeck2015}.

The fluctuation theorem describes thermodynamics of small systems and is intimately related to information processing \cite{ParrondoHorowitzSagawa2015}.
In manipulating small systems, measurement-based feedback control, which is the modern incarnation of Maxwell's demon \cite{MaruyamaNoriVedral2009,Sagawa2012}, plays a central role, where information-theoretic quantities are treated on an equal footing with thermodynamic quantities \cite{SagawaUeda2008, SagawaUeda2009,Sagawa2012}.
In fact, the integral fluctuation theorems in the feedback-controlled process \cite{SagawaUeda2010,HorowitzVaikuntanathan2010,SagawaUeda2012PRE,AbreuSeifert2012,LahiriRanaJayannavar2012,FunoWatanabeUeda2013,Rastegin2013} and in the erasure process \cite{GooldPaternostroModi2015} have been derived.

In this paper, we investigate the fluctuation theorem in open quantum systems under feedback control.
We show that quantum coherence does change the form of the quantum fluctuation theorem under projective-measurement-based feedback control.
In fact, quantum coherent driving affects the degree of absolute irreversibility of the fluctuation theorem \cite{Sung2005,LuaGrosberg2005,Gro05Aug,Jar05,MurashitaFunoUeda2014,AshidaFunoMurashitaUeda2014,FunoMurashitaUeda2015}, which corresponds to singularly irreversible events with divergent entropy production and inevitable information loss of imperfect feedback operations.
By quantum coherent driving, we mean that the driving Hamiltonian has nonzero off-diagonal elements with respect to the eigenspace of the projective measurement.
Since absolute irreversibility is suppressed by quantum coherent driving, the second-law-like inequality associated with the fluctuation theorem indicates that quantum coherent driving is thermodynamically advantageous for work extraction over classical driving whose Hamiltonian has only block diagonal elements.
Our result gives yet another insight into thermodynamics of quantum coherence complementary to the previous quests to thermodynamically evaluate quantum coherence from the quantum-information-theoretic viewpoints.

The technical novelty of our work is that we explicitly construct the backward process of the feedback-controlled process by invoking backward measurement and post-selection, which are not required in classical systems with no quantum backaction of the measurement.
Consequently, we can derive the fluctuation theorem with absolute irreversibility in a concise and physically transparent manner.

From a physical viewpoint, we find that quantum coherent driving manifests its thermodynamic effect as the change of absolute irreversibility.
While classical driving after the projective measurement causes absolute irreversibility, quantum coherent driving suppresses it due to quantum rare events, which go through classically-prohibited paths with the assistance of quantum coherence.
Moreover, by inserting a delay time before quantum coherent driving, we can change the degree of absolute irreversibility.
As the delay time decreases, the degree of absolute irreversibility is alleviated, implying thermodynamic advantage of quantum coherent driving.

This article is organized as follows.
In Sec. II, we formulate our problem of an open quantum system.
The time evolution of the system is governed by the Lindblad equation.
We unravel the Lindblad equation into a stochastic Schr\"odinger equation and discuss the stochastic energetics of an individual quantum trajectory.
Moreover, we introduce the time-reversed dynamics of this system to derive fluctuation theorems.
In Sec. III, we introduce the measurement-based feedback protocol and show that both measurement and post-selection are needed in the backward protocol.
By doing so, we realize the time-reversed operation of the forward quantum measurement and cancel out its backaction in the transition probability.
Then, we derive the detailed fluctuation theorem under feedback control by utilizing the backward process.
In Sec. IV, we derive the fluctuation theorems in the presence of absolute irreversibility with the help of the time-reversed protocol.
The associated inequalities give a tighter restriction on the work extraction under feedback control than the second law of information thermodynamics.
In Sec. V, we argue that quantum rare events suppress absolute irreversibility even for a projective measurement-based feedback control.
Then, we show that absolute irreversibility emerges from incoherent feedback control, especially due to the time delay of a feedback operation.
We numerically study a two-level system under a time-delayed feedback control and demonstrate the validity of the fluctuation theorems with absolute irreversibility derived in Sec. IV.
In Sec. VI, we discuss a possible experimental implementation to vindicate the fluctuation theorems.
In Sec. VII, we conclude this paper.

\section{Quantum stochastic thermodynamics}

\subsection{Lindblad equation}
Recently, due to experimental advances \cite{Barreiro_Blatt2011,Petersson_Petta2012,DevoretSchoelkopf2013}, thermodynamics \cite{DeffnerLutz2011,Deffner2013,Lorenzo_Palma2015,Binder_Goold2015,KutvonenSagawaAlaNissila2016,TalknerHanggi2016,Pigeon_Paternostro2016,AlonsLutzRomito2016,Goold_Skrzypczyk2016} and fluctuation theorems \cite{Horowitz2012,LeggioNapoliMessinaBreuer2013,HekkingPekola2013,HorowitzParrondo2013,GongAshidaUeda2016} of open quantum systems have intensively been studied. 
We consider open quantum systems coupled to an ideal environment with inverse temperature $\beta$, and assume that the energy levels of the system are non-degenerate.
For an appropriate separation of timescales \cite{GongAshidaUeda2016}, the time evolution of the system can  be described by the Lindblad equation \cite{Lindblad1976}:
\begin{eqnarray}
\label{Lindbladeq}
	\dot \rho_t
	&=&
	-\frac{i}{\hbar}[H(\lambda_t)+h_t,\rho_t]
	+\sum_{k,l} \mD[L_{kl}(\lambda_t)]\rho_t,
\end{eqnarray}
where $\lambda_t$ is the control parameter and $\mD$ is a super-operator defined by 
$
	\mD[c]\rho
	=
	c\rho c^\dag
	- \{c^\dag c,\rho\}/2
$.
We divide the time-dependent driving into the inclusive part $H(\lambda_t)$ and the exclusive part $h_t$,
so that only the former is included in the energy of the system $E_t = \tr [\rho_t H(\lambda_t)]$.
The operator $L_{kl}(\lambda)$ describes a quantum jump of the system from the $k$th energy eigenstate to the $l$th one due to the interaction with the environments.
Then, the jump operator $L_{kl}(\lambda)$ satisfies
$
	[L_{kl}(\lambda),H(\lambda)]
	= 
	\Delta_{kl}(\lambda) L_{kl}(\lambda),
$
where $\Delta_{kl}(\lambda) = E_k(\lambda)-E_l(\lambda)$ and $E_k(\lambda)$ is the $k$th eigenenergy of $H(\lambda)$.
When the total Hamiltonian for the composite of the system and environments is time-reversal symmetric, the jump operator satisfies the detailed balance condition
\begin{equation}
\label{DBC}
	L^\dag_{lk}(\lambda)
	=
	L_{kl}(\lambda) e^{-\beta \Delta_{kl}(\lambda)/2}.
\end{equation}

\subsection{Energetics at the trajectory level}
Here, we assume that the initial state of the system starts from an energy eigenstate $|a(\lambda_0)\rangle$.
During $0<t<T$, the system evolves under the dissipation due to the environment.
At the final time $t=T$, we perform the energy projective measurement and observe the energy eigenstate $|b(\lambda_T)\rangle$.

By the interaction with the environment, the state of the system  immediately changes into a mixed state at the ensemble level.
However, by monitoring the environment (see Sec.~VI for detail), we can keep the state of the system pure during the time evolution.
In this case, the evolution can be described by the following stochastic Schr\"odinger equation
\begin{eqnarray}
	\label{SSE}
	d|\psi_t\rangle
	&=&
	\left[
		-\frac{i}{\hbar} H\eff_t
		+ \frac{1}{2}\sum_{k,l} ||L_{kl}(\lambda_t)|\psi_t\rangle||^2
	\right] |\psi_t\rangle dt
	\nonumber\\
	&&
	+\sum_{k,l} \left[
		\frac{L_{kl}(\lambda_t)}{||L_{kl}(\lambda_t)|\psi_t\rangle||}-I
	\right] |\psi_t\rangle dN^{kl}_t,
\end{eqnarray}
where the non-Hermitian effective Hamiltonian is given by
\begin{equation}
	H\eff_t
	=
	H(\lambda_t) + h_t
	-\frac{i\hbar}{2}\sum_{k,l} L^\dag_{kl}(\lambda_t)L_{kl}(\lambda_t),
\end{equation}
and $dN^{kl}_t$'s are statistically independent Poisson increments.
The expectation value of $dN^{kl}_t$, $\E[dN^{kl}_t] = ||L_{kl}(\lambda_t)|\psi_t\rangle||^2 dt$, gives the probability of a quantum jump at time $t$.
The second term on the right-hand side of Eq.~(\ref{SSE}) represents non-unitary evolution due to the measurement backaction under the condition that no jump occurs.
The third term describes the state reduction due to a quantum jump.
Then, a single quantum trajectory $\psi:=\{|\psi_t\rangle\}_{t=0}^T$ in the Hilbert space can be completely specified by the initial and final quantum numbers and a history of quantum jumps as
\begin{equation}
	\psi
	\Leftrightarrow
	(a, (k_1,l_1,t_1),(k_2,l_2,t_2),\cdots,(k_N,l_N,t_N),b),
\end{equation}
where the subset $(k_n,l_n,t_n)$ indicates that the $n$th jump occurs at time $t=t_n$ from the $k_n$th energy eigenstate to the $l_n$th one.

Now we assign each trajectory $\psi$ with stochastic thermodynamic quantities.
The energy changes of the environment originate from the energy transfer between the system and the reservoir.
Therefore, by summing these changes, we can evaluate the flow of total energy from the system to the reservoir, which we identify as heat.
Therefore, the heat flow from time $t=0$ to $T$ should be defined by
\begin{equation}
	Q[\psi]
	=
	\sum_{k,l} \int_0^T dN^{kl}_t \Delta_{kl}(\lambda_t).
\end{equation}
By the first law of thermodynamics, the work performed on the system from $t=0$ to $T$ should be identified as
\begin{equation}
	W[\psi]
	=
	E_b(\lambda_T)-E_a(\lambda_0)
	+
	\sum_j \int_0^T dN^{kl}_t \Delta_{kl}(\lambda_t),
\end{equation}
where we note that $\Delta U = E_b(\lambda_T)-E_a(\lambda_0)$ is the energy difference of the system.
Thus, we can formulate the stochastic energetics for the individual quantum trajectory $\psi$, when the system starts from and ends at energy eigenstates.

\subsection{Time-reversed dynamics}
To derive the fluctuation theorem, it is convenient to construct the time-reversed dynamics \cite{Crooks1999,Crooks2000}.
Given a forward protocol $\lambda_t$ and $h_t$, we drive the system in the time-reversed manner according to the time-reversed protocol $\bar\lambda_t = \lambda_{\bar t}$ and $\bar h_t=\Theta h_{\bar t}\Theta^\dag$, where $\bar t = T-t$ and $\Theta$ is the antiunitary time-reversal operator.
Moreover, we perform the time-reversal operation on the Hamiltonian and the jump operators as $\bar H(\lambda)=\Theta H(\lambda)\Theta^\dag$ and $\bar L_{kl} (\lambda)= \Theta L_{kl}(\lambda) \Theta^\dag$, respectively.
Consequently, we obtain the following time-reversed evolution:
\begin{equation}
	\label{TRLBE}
	\dot \rho_t^{\rm rev}
	=
	-\frac{i}{\hbar} [\bar H(\bar\lambda_t)+\bar h_t,\rho_t^{\rm rev}]
	+\sum_{k,l} \mD[\bar L_{kl}(\bar\lambda_t)]\rho_t^{\rm rev}.
\end{equation}
By unravelling Eq.~(\ref{TRLBE}), we obtain the time-reversed stochastic Schr\"odinger equation:
\begin{eqnarray}
	\label{TRSSE}
	d|\psi_t\rangle^{\rm rev}
	&=&
	\left[
		-\frac{i}{\hbar} \bar H_t\eff
		+\frac{1}{2}\sum_{k,l} ||\bar L_{kl}(\bar \lambda_t)|\psi_t\rangle^{\rm rev}||^2
	\right]
	|\psi_t\rangle^{\rm rev} dt
	\nonumber\\
	&&
	+ \sum_{k,l}\left[
		\frac{\bar L_{kl}(\bar\lambda_t)}{||\bar L_{kl}(\bar \lambda_t)|\psi_t\rangle^{\rm rev}||} -I
	\right]
	|\psi_t\rangle^{\rm rev} d\bar N^{kl}_t,
\end{eqnarray}
where $\bar H_t\eff=\Theta H_{\bar t}^{\rm eff\dag} \Theta^\dag$ and $\E[d\bar N^{kl}_t]=||\bar L_{kl}(\bar\lambda_t)|\psi_t\rangle^{\rm rev}||^2 dt$.

\subsection{Fluctuation theorem}
For later reference, we illustrate the essence of the derivation of the detailed fluctuation theorem \cite{Horowitz2012,HorowitzParrondo2013}.
The backward trajectory $\bar \psi$ should start from $|\bar b(\bar\lambda_0)\rangle=\Theta|b(\lambda_T)\rangle$, undergo quantum jumps $(\bar k_n, \bar l_n, \bar t_n)=(l_{N+1-n},k_{N+1-n},T-t_{N+1-n})$ and end at $|\bar a(\bar\lambda_T)\rangle=\Theta|a(\lambda_0)\rangle$.
The transition amplitude of the forward trajectory $\psi$ involves the following factor
\begin{equation}
\label{UL}
	U\eff(t_{n+1},t_n)L_{k_nl_n}(\lambda_{t_n})
	,
\end{equation}
where
$
	U\eff(s,t)
	=
	\mT \exp[
	-\frac{i}{\hbar}
		\int_{t}^{s} H\eff_u du
	]
$
and $\mT$ is the time-ordering operator.
In contrast, the backward transition amplitude for $\bar\psi$ involves the factor
\begin{eqnarray}
	&&\bar L_{\bar k_{N+1-n}\bar l_{N+1-n}}(\bar\lambda_{\bar t_{N+1-n}})
	\bar U\eff(\bar t_{N+1-n},\bar t_{N-n})
	\nonumber\\
	&&=
	\Theta L_{l_nk_n}(\lambda_{t_n})
	\Theta^\dag
	\mT \exp\left[
	-\frac{i}{\hbar}
		\int_{\bar t_{N-n}}^{\bar t_{N+1-n}}
		\Theta H^{\rm eff\dag}_{\bar t} \Theta^\dag d\bar t
	\right]
	\nonumber\\
	&&=
	\Theta L_{l_nk_n}(\lambda_{t_n}) \bar \mT \exp\left[
		\frac{i}{\hbar}
		\int_{t_n}^{t_{n+1}} H^{\rm eff\dag}_t dt
	\right]
	\Theta^\dag,
	\label{TRLU}
\end{eqnarray}
where $\bar U\eff(s,t)= \mT\exp[-\frac{i}{\hbar}\int_t^s\bar H\eff_u du] $ and $\bar\mT$ is the anti-time-ordering operator.
By taking the Hermitian conjugate of Eq.~(\ref{TRLU}), we obtain
$
	\Theta 
		U\eff(t_{n+1},t_n)L^\dag_{l_nk_n}(\lambda_{t_n})
	\Theta^\dag,
$
which can be written from the detail balance condition~(\ref{DBC}) as
\begin{equation}
\label{LU}
	\Theta 
		U\eff(t_{n+1},t_n)L_{k_nl_n}(\lambda_{t_n})
	\Theta^\dag
	e^{-\beta \Delta_{k_nl_n}(\lambda_{t_n})/2}.
\end{equation}
By multiplying these factors, the time-reversal operators in the adjacent factors are canceled out.
Thus, comparing Eqs.~(\ref{UL}) and (\ref{LU}), we conclude that the ratio of the Hermitian-conjugate backward transition amplitude to the forward transition amplitude is given by $\exp[-\beta\sum_n \Delta_{k_nl_n}(\lambda_{t_n})/2]=\exp[-\beta Q[\psi]/2]$.
Therefore, the ratio of the transition probabilities is
\begin{equation}
	\frac{\bar\mP[\bar\psi|\bar b]}{\mP[\psi|a]}
	=
	e^{-\beta Q[\psi]}.
\end{equation}
If we set the initial states of the forward and backward processes to thermal equilibrium states, we obtain the detailed fluctuation theorem
\begin{equation}
\label{DFT}
	\frac{\mP[\bar\psi]}{\mP[\psi]}
	=
	e^{-\beta(W[\psi]-\Delta F)},
\end{equation}
where $\Delta F=F(\lambda_T)-F(\lambda_0)$ is the free-energy difference.

In this way, the detailed fluctuation theorem can be derived in the absence of feedback control.
However, in the presence of feedback control, we have to manage the measurement backaction.
In the next section, we discuss how to do it.

\section{Construction of backward protocol under feedback control}

\subsection{Protocol of feedback control}
\begin{figure}
	\includegraphics[bb=0 0 708 386, width=\columnwidth]{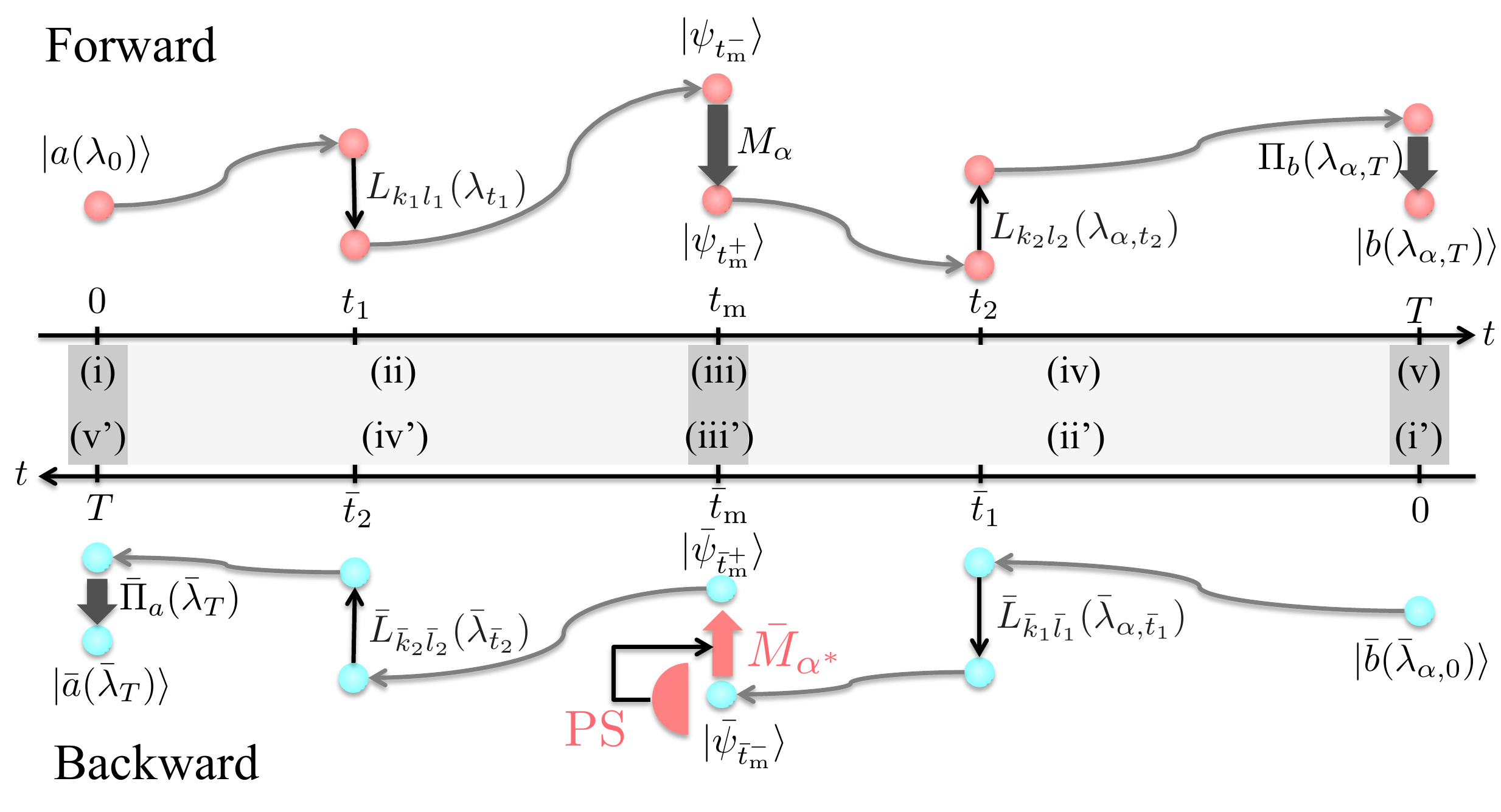}
	\caption{\label{fig:SI}
		Schematic illustration of the forward and backward quantum trajectories.
		In the forward process, we perform measurements at $t=0$, $t_{\rm m}$, and $T$. In the backward process, we perform measurements at $t=0$, $\bar{t}_{\rm m}$, and $T$ and post-selection (PS) to cancel out the effect of the measurement backaction in the forward protocol on the transition probability.
	}
\end{figure}

We introduce the protocol of discrete feedback control as schematically illustrated in Fig.~\ref{fig:SI} \cite{SagawaUeda2008,GongAshidaUeda2016}, which is a quantum analog of the protocol of the classical discrete feedback control in Ref.~\cite{SagawaUeda2010}.
At the level of a single quantum trajectory, the entire process can be described by the following five steps:
{\bf (i)} At the initial time $t=0$, we prepare a thermal equilibrium state
$
	\rho_0 = \exp[{-\beta (H(\lambda_0)-F(\lambda_0))}]
$
with
$
	F(\lambda)
	=
	-\beta^{-1}\ln\tr [e^{-\beta H(\lambda)}]
$
being the free energy of the system.
Then, we perform a projective energy measurement $\Pi(\lambda_0)$ to determine the initial energy $E_a(\lambda_0)$ and the initial state
$
	|\psi_{0^+}\rangle = |a(\lambda_0)\rangle
$
of the system.
{\bf (ii)} During $0<t<t_{\rm m}$, we drive the system under a prescribed protocol $(\lambda_t, h_t)$.
Thus, the system evolves according to Eq.~(\ref{SSE}) with the initial condition $|\psi_{0^+}\rangle=|a(\lambda_0)\rangle$.
{\bf (iii)} At $t=t_{\rm m}$, we perform a measurement to extract information of the system.
We denote a set of measurement operators by
$
	{\rm M}_A=\{M_\alpha|\alpha\in A\},
$
which satisfy
$
	\sum_\alpha M^\dag_\alpha M_\alpha = I.
$
Then, the state of the system $|\psi_{t_{\rm m}^-}\rangle$ just before the measurement reduces to a post-measurement state $|\psi_{t_{\rm m}^+}\rangle = M_\alpha|\psi_{t_{\rm m}^-}\rangle/||M_\alpha|\psi_{t_{\rm m}^-}\rangle||$ with the probability $||M_\alpha|\psi_{t_{\rm m}^-}\rangle||^2$.
{\bf (iv)} During $t_{\rm m}<t<T$, we drive the system according to a protocol $(\lambda_{\alpha,t}, h_{\alpha,t})$, which depends on the measurement outcome $\alpha$.
{\bf (v)} Finally, at time $t=T$, we perform a projective energy measurement $\Pi(\lambda_{\alpha,T})$ to determine the final energy $E_b(\lambda_{\alpha,T})$ and the final state $|\psi_T\rangle=|b(\lambda_{\alpha,T})\rangle$ of the system.

Under the protocol of feedback control, we obtain the joint probability distribution $\mP[\psi,\alpha]$ of the quantum trajectory $\psi$ and the measurement outcome $\alpha$.
From $\mP[\psi,\alpha]$, we can calculate the marginal probability distribution $p(\alpha)$ and the conditional probability distribution $\mP[\psi|\alpha]$, both of which play crucial roles in the construction of the time-reversed protocol and the derivation of the fluctuation theorem.
It is noteworthy that the set $(\psi,\alpha)$ has a one-to-one correspondence via Eq.~(\ref{SSE}) with the following sequence
\begin{eqnarray}
	\label{CT}
	&&(a,(k_1,l_1,t_1),\cdots,(k_{N_{\rm m}},l_{N_{\rm m}},t_{N_{\rm m}}),\alpha,
	\nonumber\\
	&&\ \ \ \ \ 
	(k_{N_{\rm m}+1},l_{N_{\rm m}+1},t_{N_{\rm m}+1}),\cdots,(k_N,l_N,t_N),b),
\end{eqnarray}
where $t_n$'s satisfy  $0<t_1<\cdots<t_{N_{\rm m}}<t_{\rm m}<t_{N_{\rm m}+1}<\cdots<t_N<T$.
Thus, a quantum trajectory of the system can be translated into a history of quantum jumps and outcomes of the measurements at $t=0,\ t_{\rm m}$, and $T$.
We note that, under this description, the heat can be rewritten as
\begin{equation}
	Q[\psi,\alpha]
	=
	\sum_n \Delta_{k_nl_n}(\lambda_{\alpha, t_n}),
\end{equation}
where for simplicity we define $\lambda_{\alpha,t}:=\lambda_t\ (^\forall\alpha\in A)$ for $t<t_{\rm m}$.
The work is given by
\begin{equation}
	W[\psi,\alpha]
	=
	E_b(\lambda_{\alpha,T}) - E_a(\lambda_0)
	+\sum_n \Delta_{k_nl_n}(\lambda_{\alpha, t_n}).
\end{equation}

\subsection{Backward protocol with post-selection}
First of all, we have to fix the measurement outcome $\alpha$ to construct the time-reversed process, since in the time-reversed process the protocol $\bar \lambda_{\alpha,t}=\lambda_{\alpha,\bar t}$ and $\bar h_{\alpha,t}=h_{\alpha, \bar t}$ before the measurement time ($0<t<\bar t_{\rm m}$) should depend on the forward measurement outcome $\alpha$.
Therefore, for the time-reversed process, we choose $\alpha$ according to the probability distribution $p(\alpha)$ in the forward protocol.
We note that this is the standard procedure when we derive the classical fluctuation theorem under feedback control \cite{SagawaUeda2010,HorowitzVaikuntanathan2010}.

Next, we discuss the quantum nature of our setup, namely, the measurement backaction.
In the presence of feedback control, the transition amplitude for $(\psi,\alpha)$ involves the term
\begin{equation}
	U\eff_\alpha(t_{N_{\rm m}+1},t_{\rm m})
	M_\alpha
	U\eff(t_{\rm m},t_{N_{\rm m}}).
\end{equation}
By taking the Hermitian conjugate, we obtain
\begin{eqnarray}
	&&\bar \mT \exp\left[
		\frac{i}{\hbar}
		\int_{t_{N_{\rm m}}}^{t_{\rm m}} H^{\rm eff\dag}_t dt
	\right]
	M_\alpha^\dag
	\bar \mT \exp\left[
		\frac{i}{\hbar}
		\int_{t_{\rm m}}^{t_{N_{\rm m}+1}} H^{\rm eff\dag}_{\alpha,t} dt
	\right]
	\nonumber\\
	&&=
	\Theta^\dag \bar U\eff(\bar t_{N-N_{\rm m}+1},\bar t_{\rm m})
	\Theta M^\dag_\alpha \Theta^\dag
	\bar U\eff_\alpha(\bar t_{\rm m}, \bar t_{N-N_{\rm m}}) \Theta.
\end{eqnarray}
Therefore, we should perform the operation $\bar M_{\alpha^*} := \Theta M^\dag_\alpha \Theta^\dag$ at time $t=\bar t_{\rm m}$ in the backward process to cancel out the effect of measurement backaction in forming the ratio of the forward and backward transition probabilities.
Unfortunately, the operation $\rho\mapsto\bar M_{\alpha^*}\rho \bar M_{\alpha^*}^\dag$ is not a completely positive trace-preserving (CPTP) map, and therefore we cannot realize it directly.
Nevertheless, we can realize $\bar M_{\alpha^*}$ as one operator in the measurement set ${\rm M}_{\bar A_\alpha}:=\{\bar M_{\bar \alpha}|\bar \alpha\in \bar A_\alpha\}$  (see Ref.~\cite{GongAshidaUeda2016}).
In other words, we can construct ${\rm M}_{\bar A_\alpha}$ so that one of the measurement operator in ${\rm M}_{\bar A_\alpha}$ coincides with $\bar M_{\alpha^*}$.
Thus, we should perform measurement in the backward protocol to deal with the measurement backaction.

However, another problem arises when we perform the measurement in the backward protocol.
The operation $\bar M_{\alpha^*}$ does not preserve the norm.
Moreover, operations $\bar M_{\bar \alpha}$ with $\bar \alpha \neq \alpha^*$ are not required from a physical point of view,
since we can arbitrarily choose them as long as they satisfy the condition $\bar M_{\alpha^*}^\dag \bar M_{\alpha^*}+\sum_{\bar \alpha \neq \alpha^*} \bar M_{\bar \alpha}^\dag \bar M_{\bar \alpha} = I$.
To overcome these problems, we conduct post-selection and discard the events with outcomes $\bar \alpha \neq \alpha^*$.
This selective evolution $|\bar\psi_{\bar t_{\rm m}^-}\rangle \to |\bar\psi_{\bar t_{\rm m}^+}\rangle =  \bar M_{\alpha^*} |\bar\psi_{\bar t_{\rm m}^-}\rangle/||\bar M_{\alpha^*} |\bar\psi_{\bar t_{\rm m}^-}\rangle||$ ensures the normalization of probability without unphysical events.

Thus, the backward protocol conditioned on the forward measurement outcome $\alpha$ is as follows.
({\bf i'}) At the initial time $t=0$, we prepare a thermal equilibrium state $\bar \rho_{\alpha,0} = \exp[-\beta(\bar H(\bar \lambda_{\alpha,0})-F(\bar \lambda_{\alpha,0}))]$.
Then, we perform a projective energy measurement $\Pi(\bar \lambda_{\alpha,0})$ to determine the initial energy $E_b(\bar \lambda_{\alpha, 0})$ and the initial state $|\bar\psi_{0^+}\rangle = |\bar b(\bar\lambda_{\alpha,0})\rangle$.
{\bf (ii')} During $0<t<\bar t_{\rm m}$, we drive the system according to Eq.~(\ref{TRSSE}) with the time-reversed protocol $\bar\lambda_{\alpha,t}$ and $\bar h_{\alpha,t}$.
{\bf (iii')} At $t=\bar t_{\rm m}$, we perform the measurement ${\rm M}_{\bar A_\alpha}$ on the state $|\bar\psi_{\bar t_{\rm m}^-}\rangle$ and obtain the measurement outcome $\bar \alpha$.
Then, we post-select events with the measurement outcome $\bar\alpha=\alpha^*$.
Therefore, the state after this procedure is given by $|\bar\psi_{\bar t_{\rm m}^+}\rangle=\bar M_{\alpha^*}|\bar\psi_{\bar t_{\rm m}^-}\rangle/||\bar M_{\alpha^*}|\bar\psi_{\bar t_{\rm m}^-}\rangle||$.
We emphasize that the post-selection is crucial to treat the backaction of the quantum measurement.
We note that we cannot conduct post-selection when $||\bar M_{\alpha^*}|\bar\psi_{\bar t_{\rm m}^-}\rangle||=0$.
In this case, we discard the measurement outcome $\bar \alpha$ instead of post-selection.
{\bf (iv')} During $\bar t_{\rm m}<t<T$, we drive the system according to $\bar\lambda_t$ and $\bar h_t$.
{\bf (v')} At time $t=T$, we perform a projective energy measurement $\Pi(\bar\lambda_T)$ and determine the final energy $E_a(\bar\lambda_T)$ and the final state $|\bar a(\bar\lambda_T)\rangle$.

The joint probability distribution of $(\bar\psi,\bar\alpha)$ conditioned by $\alpha$ is written as $\bar\mP[\bar\psi,\bar\alpha|\alpha]$.
By the post-selection, the conditional probability of a quantum trajectory $\bar\psi$ being obtained reduces to
\begin{equation}\label{TRCP}
	\bar\mP[\bar\psi|\alpha]
	=
	\left\{
	\begin{array}{ll}
	\cfrac{\bar\mP[\bar\psi,\bar\alpha=\alpha^*|\alpha]}
	{||\bar M_{\alpha^*}|\bar\psi_{\bar t_{\rm m}^-}\rangle||^2},
	& (||\bar M_{\alpha^*}|\bar\psi_{\bar t_{\rm m}^-}\rangle||\neq 0);\\[15pt]
	\displaystyle\sum_{\bar \alpha} \bar\mP[\bar\psi,\bar\alpha|\alpha].
	& (||\bar M_{\alpha^*}|\bar\psi_{\bar t_{\rm m}^-}\rangle||=0).
	\end{array}
	\right.
\end{equation}
We note that Eq.~(\ref{TRCP}) is normalized with respect to $\bar\psi$, which is why we introduce the post-selection in the time-reversed process.

\subsection{Detailed fluctuation theorem}
Here, we utilize the time-reversed protocol to derive the fluctuation theorem.
By the construction of the backward probability, we obtain
\begin{eqnarray}
	\frac{\bar\mP[\bar\psi,\bar\alpha=\alpha^*|\alpha]}
	{\mP[\psi,\alpha]}
	=
	e^{-\beta (W[\psi,\alpha]-\Delta F_\alpha)}.
\end{eqnarray}
Therefore, when $||\bar M_{\alpha^*}|\bar\psi_{\bar t_{\rm m}^-}\rangle||\neq0$, we obtain the detailed fluctuation theorem
\begin{eqnarray}
	\label{HVDFT}
	\frac{\bar\mP[\bar\psi|\alpha]}{\mP[\psi|\alpha]}
	=
	e^{-\beta(W[\psi,\alpha]-\Delta F_\alpha)-I[\psi,\alpha]},
\end{eqnarray}
where we define the unaveraged relevant information gain \cite{GongAshidaUeda2016} by
\begin{eqnarray}
	\label{RIG}
	I[\psi,\alpha]
	=
	-\ln p(\alpha)
	+ \ln||\bar M_{\alpha^*}|\bar\psi_{\bar t_{\rm m}^-}\rangle||^2.
\end{eqnarray}
We note that when $||\bar M_{\alpha^*}|\bar\psi_{\bar t_{\rm m}^-}\rangle||=0$, the denominator of the left-hand side of the detailed fluctuation theorem~(\ref{HVDFT}) vanishes and therefore the fraction is ill-defined.
The averaged relevant information is defined by $\mI(\rho:{\rm M}_X):= H(p_\rho^{{\rm M}_X}||p_{\rho_{\rm u}}^{{\rm M}_X})$ \cite{Wehrl1977,BalianVeneroniBalazs1986,SlomczynskiSzymusiak2016}, where $H(\cdot|\cdot)$ is the classical relative entropy, $p_\rho^{{\rm M}_X}$ represents the probability distribution of the outcomes of the measurement ${\rm M}_X$ on the state $\rho$, and $\rho_{\rm u}$ is the maximally mixed state.
Thus, the relevant information $\mI(\rho:{\rm M}_X)$ provides a measure of how distinguishable the state $\rho$ is from the maximally mixed state $\rho_{\rm u}$ with respect to the measurement ${\rm M}_X$.
The average of the relevant information gain (\ref{RIG}) gives change in the relevant information upon the measurement ${\rm M}_A$, where the relevant information is evaluated with respect to the subsequent protocol during $t_{\rm m}<t\le T$, which is regarded as an effective continuous measurement \cite{GongAshidaUeda2016}.
We note that Eq.~(\ref{HVDFT}) has already been derived in Ref.~\cite{GongAshidaUeda2016}.
The difference is that the backward process is explicitly derived in this paper, while it is not in Ref.~\cite{GongAshidaUeda2016}.
The construction of the backward process is essential for later physical discussions.

We emphasize that the relevant information gain is different from the QC-mutual information gain used in the previous researches on feedback-controlled quantum systems under the unitary evolution (see, e.g., Refs.~\cite{SagawaUeda2008,FunoWatanabeUeda2013,FunoMurashitaUeda2015}).
To operationally achieve the QC-mutual information gain, we have to invoke two rank-1 projective measurements just before and after the measurement whose outcome is used for feedback control.
These additional projective measurements destroy quantum coherence at the time of measurement.
As a consequence, we can describe the quantum system on the basis of classical probability distributions.
In contrast, to operationally achieve the relevant information gain, we do not have to invoke such additional measurements.
As a result, quantum coherence can be preserved upon the measurement.
Consequently, we can observe genuinely quantum phenomena as in Sec.~V, which cannot be captured by the classical probability theory.
We note that the quantitative difference between the relevant information and the QC-mutual information is discussed in Ref.~\cite{GongAshidaUeda2016}.

\section{Fluctuation theorems with absolute irreversibility}

\subsection{Absolute irreversibility}

In ordinary situations, the detailed fluctuation theorem~(\ref{HVDFT}) leads to the integral fluctuation theorem
$
	\langle e^{-\beta(W-\Delta F)-I} \rangle
	=
	1
$ \cite{GongAshidaUeda2016}
as in the classical case \cite{HorowitzVaikuntanathan2010}.
However, under measurement-based feedback control, $\mP[\psi|\alpha]$ sometimes vanishes because quantum trajectories incompatible with the measurement outcome are prohibited. 
These trajectories come under a special class of irreversibility which we refer to as absolute irreversibility, because the corresponding thermodynamic irreversibility as represented by the left-hand side of Eq.~(\ref{HVDFT}) diverges.
Here, we explain the general notion of absolute irreversibility \cite{MurashitaFunoUeda2014}.

In the fluctuation theorem, we compare the forward and backward probability measures and especially focus on their ratio.
In measure theory in mathematics, it is known that we can discuss the ratio only when a condition called absolute continuity (see below) is satisfied \cite{Hal74,Bar95}.
Let $\mM[\cdot|\alpha]$ and $\bar\mM[\cdot|\alpha]$ denote the forward and backward measures conditioned by $\alpha$.
In ordinary situations (i.e., when both of them are absolutely continuous with respect to the Lebesgue measure $\mD\psi$), we can relate them with the probability density as $\mM[\mD\psi|\alpha]=\mP[\psi|\alpha]\mD\psi$ and $\bar\mM[\mD\bar\psi|\alpha]=\bar\mP[\bar\psi|\alpha]\mD\bar\psi$.
The measure $\bar\mM[\cdot|\alpha]$ is said to be absolutely continuous with respect to $\mM[\cdot|\alpha]$ if for any set of paths $\Psi$
\begin{equation}
	\mM[\Psi|\alpha]=0\ \Rightarrow\ \bar\mM[\bar\Psi|\alpha]=0.
\end{equation}
Under this condition of absolute continuity, we can transform the measure as
\begin{equation}
	\bar\mM[\mD\bar\psi|\alpha]
	=
	\frac{\bar\mP[\bar\psi|\alpha]}{\mP[\psi|\alpha]}
	\mM[\mD\psi|\alpha]
\end{equation}
by using the probability ratio, which can be rewritten in terms of the entropy production and the information gain.

However, absolute continuity does not always hold especially under measurement-based feedback control.
When absolute continuity is not satisfied, we have to decompose the backward probability measure into two parts:
\begin{equation}
	\bar\mM[\cdot|\alpha]
	=
	\bar\mM_{\rm AC}[\cdot|\alpha]
	+
	\bar\mM_{\rm S}[\cdot|\alpha],
\end{equation}
where $\bar\mM_{\rm AC}[\cdot|\alpha]$ and $\bar\mM_{\rm S}[\cdot|\alpha]$ are absolutely continuous and singular with respect to $\mM[\cdot|\alpha]$, respectvely.
This decomposition is called the Lebesgue decomposition and the uniqueness of the decomposition is mathematically guaranteed by the Lebesgue decomposition theorem \cite{Hal74,Bar95}.
For the absolutely continuous part, we can invoke the Radon-Nikod\'ym theorem \cite{Hal74,Bar95} to transform the measure as
\begin{equation}\label{ACRN}
	\bar\mM_{\rm AC}[\mD\bar\psi|\alpha]
	=
	\frac{\bar\mP[\bar\psi|\alpha]}{\mP[\psi|\alpha]}
	\mM[\mD\psi|\alpha].
\end{equation}
In general, the singular part $\bar\mM_{\rm S}[\cdot|\alpha]$ may involve such singular measures as the delta-function-like localization and the Cantor measure.
However, we restrict ourselves to the simplest case of the singularity since it is enough for the aim of this paper.
The simplest singularity emerges when the probability ratio on the right-hand side of Eq.~(\ref{ACRN}) diverges because the denominator vanishes and the numerator does not, i.e.,
\begin{equation}
	\label{AI}
	\mP[\psi|\alpha]=0\ \&\ \bar\mP[\bar\psi|\alpha]\neq0.
\end{equation}	
These trajectories are called absolutely irreversible since they are not even stochastically reversible and consequently the left-hand side of Eq.~(\ref{HVDFT}) diverges \cite{MurashitaFunoUeda2014,FunoMurashitaUeda2015}.
It is noteworthy that absolute irreversibility encompasses such thermodynamically fundamental processes as free expansion \cite{MurashitaFunoUeda2014, FunoMurashitaUeda2015}, gas mixing \cite{MurashitaUeda2017}, and spontaneous symmetry breaking in second-order phase transitions \cite{Hoang_2016}.
Moreover, under measurement-based feedback control, absolute irreversibility naturally emerges because the measurement precludes the possibility of those forward trajectories that are incompatible with the measurement outcome \cite{MurashitaFunoUeda2014}.
The correspondence between the classification of irreversibility and that of probability measure is summarized in Table~\ref{tab:irrev}.

\begin{table}[b]
\caption{\label{tab:irrev}
Correspondence between the classification of irreversibility and that of probability measure.
Here, irreversibility is classified into ordinary irreversibility and absolute irreversibility.
Ordinary irreversibility is characterized by the absolute continuity of the probability measures and therefore has a finite probability ratio~(\ref{HVDFT}).
Consequently, the quantity $\beta(W-\Delta F)+I$, which characterizes thermodynamic irreversibility under feedback control, is finite and well-defined.
In contrast, absolute irreversibility is characterized by the singularity of the probability measures and therefore the probability ratio diverges.
As a result, $\beta(W-\Delta F)+I$ is negatively divergent.
}
{\renewcommand\arraystretch{1.1}
\begin{center}\begin{tabular}{|c|c|c|}
\hline
Class of irreversibility & Ordinary & Absolute\\
\hline
Class of measure & absolutely continuous & singular\\
\hline
\rule[-12pt]{0pt}{30pt}
Probability ratio&
$\displaystyle \frac{\bar\mP[\bar\psi|\alpha]}{\mP[\psi|\alpha]}=$finite &
$\displaystyle \frac{\bar\mP[\bar\psi|\alpha]}{0}=\infty$\\
\hline
Entropy-production-&$\beta(W-\Delta F)+I$&$\beta(W-\Delta F)+I$\\
like quantity&={\rm finite}&$=-\infty$\\
\hline
\end{tabular}
\end{center}}
\end{table}

\subsection{Fluctuation theorems}
Here, we derive the fluctuation theorems in the presence of absolute irreversibility.
For the ordinary irreversible part, i.e. the absolutely continuous part, we use the Radon-Nikod\'ym theorem~(\ref{ACRN}) to derive the integral fluctuation theorem:
\begin{eqnarray}
	\int\bar\mM_{\rm AC}[\mD\bar\psi|\alpha]
	&=&
	\int\frac{\bar\mP[\bar\psi|\alpha]}{\mP[\psi|\alpha]}
	\mM[\mD\psi|\alpha]
	\nonumber\\
	&=&
	\int e^{-\beta(W-\Delta F)-I}
	\mM[\mD\psi|\alpha]
	\nonumber\\
	&=&
	\langle e^{-\beta(W-\Delta F)-I} |\alpha \rangle,
\end{eqnarray}
where
$
	\langle\cdot|\alpha\rangle
$
denotes the ensemble average conditioned by the measurement outcome $\alpha$.
On the other hand, from the normalization of probability measure, we have
\begin{equation}
	1=
	\int\bar\mM_{\rm AC}[\mD\bar\psi|\alpha]
	+
	\int\bar\mM_{\rm S}[\mD\bar\psi|\alpha].
\end{equation}
As a result, we obtain the conditional integral fluctuation theorem
\begin{eqnarray}
	\label{CIFT}
	\langle e^{-\beta(W-\Delta F)-I} |\alpha \rangle
	=
	1-\lambda(\alpha),
\end{eqnarray}
where $\lambda(\alpha)$ is the absolutely irreversible probability defined by
\begin{eqnarray}
	\label{DefL}
	\lambda(\alpha)
	&=&
	\int\bar\mM_{\rm S}[\mD\bar\psi|\alpha]
	\nonumber\\
	&=&
	\int_{\mP[\psi|\alpha]= 0} \mD\psi \bar\mP[\bar\psi|\alpha].
\end{eqnarray}
Thus, the absolutely irreversible probability is the sum of the probabilities of the time-reversed trajectories that have vanishing forward trajectories.
We note that while $\mathcal{\bar P}[\bar\psi|\alpha]$ is not uniquely defined by Eq.~(\ref{TRCP}) for the case of $||\bar M_{\alpha^*}|\bar\psi_{\bar t_{\rm m}}\rangle||=0$, $\lambda(\alpha)$  as the integral of $\mathcal{\bar P}[\bar\psi|\alpha]$ can be uniquely determined.
By averaging Eq.~(\ref{CIFT}) over $\alpha$, we obtain
\begin{equation}
	\label{IFT}
	\langle e^{-\beta(W-\Delta F)-I} \rangle
	=
	1-\bar\lambda,
\end{equation}
where
\begin{equation}
	\bar\lambda=\sum_\alpha p(\alpha)\lambda(\alpha)
\end{equation}
is the averaged absolutely irreversible probability.
Moreover, if we define the unavailable information by $I_{\rm u}(\alpha)=-\ln(1-\lambda(\alpha))$ \cite{AshidaFunoMurashitaUeda2014}, which is the inevitable loss of information due to absolute irreversibility, we obtain
\begin{eqnarray}
	\label{UIFT}
	\langle e^{-\beta(W-\Delta F)-(I-I_{\rm u})} \rangle = 1.
\end{eqnarray}
Thus, the fluctuation theorems have been derived in the presence of absolute irreversibility.

From Eq. (\ref{IFT}) and Jensen's inequality, we obtain
\begin{eqnarray}
	\label{2LAI}
	-\beta\langle W\rangle
	\le
	-\beta\Delta F + \langle I \rangle + \ln (1-\bar \lambda).
\end{eqnarray}
Since $\ln(1-\bar{\lambda})\leq 0$, this inequality gives a tighter bound for work extraction than the conventional second law of information thermodynamics:
\begin{equation}
	\label{2L}
	-\beta\langle W\rangle
	\le
	-\beta\Delta F + \langle I \rangle.
\end{equation}
Thus, absolute irreversibility quantifies how much work extraction capability is reduced.
Applying Jensen's inequality to Eq. (\ref{UIFT}), we obtain
\begin{eqnarray}
	\label{2LUI}
	-\beta\langle W\rangle
	\le
	-\beta\Delta F + \langle I-I_{\rm u} \rangle.
\end{eqnarray}
This inequality gives an even tighter bound on work extraction than Eq.~(\ref{2LAI}) because the concavity of the logarithmic function guarantees $\langle\ln(1-\lambda(\alpha))\rangle\le\ln(1-\bar\lambda)$.
We note that the relevant information is expected to give a tighter bound than the QC-mutual information.
This is because, unlike the QC-mutual information, the relevant information is protocol-dependent and even takes a negative value if we choose a bad feedback protocol \cite{GongAshidaUeda2016}.
However, there is no general magnitude relation between the relevant information and the QC-mutual information \cite{GongAshidaUeda2016}.

\section{Interplay between quantum coherence and absolute irreversibility}

\subsection{Suppression of absolute irreversibility by quantum coherence}
In previous studies \cite{MurashitaFunoUeda2014,FunoMurashitaUeda2015}, a projective (error-free) measurement, in general, causes absolute irreversibility.
However, absolute irreversibility can be suppressed in quantum stochastic thermodynamics even under a projective measurement.
Here, we show that this difference is a consequence of quantum superposition.

In quantum stochastic thermodynamics, quantum superposition, in general, suppresses absolute irreversibility even in the presence of a projective measurement.
Suppose that every off-diagonal element of a driving Hamiltonian with respect to the eigenspace of the projective measurement is nonzero.
Then, the post-measurement state immediately spreads over the entire Hilbert space due to the time evolution of this Hamiltonian.
Therefore, all types of quantum jumps are allowed after the projective measurement.
Thus, we see that there are no trajectories with vanishing probability in the forward process.
We note that the condition that every off-diagonal element is nonzero is a sufficient condition for the suppression of absolute irreversibility.

To be specific, suppose that we perform a projective energy measurement on a two-level system with states $e$ and $g$, and then coherently drive the system according to the outcome.
Classically, the jump $e\to g$ is prohibited just after we observe the state $g$.
However, under quantum coherent driving with nonzero off-diagonal elements with respect to the energy eigenbasis, the jump $e\to g$ can occur immediately after the observation of $g$, because the coherent driving instantaneously brings the state into a superposition of $e$ and $g$.
Similarly, the quantum jump $g\to e$ can occur immediately after the observation of $e$.
We call these phenomena quantum rare events.
Thus, even immediately after the projective measurement, both jumps are allowed thanks to the quantum rare events.
This is why the fluctuation theorem~(\ref{IFT}) with $\bar\lambda=0$ is valid even in the presence of the projective measurement as numerically demonstrated in Ref.~\cite{GongAshidaUeda2016}.
We note that $I[\psi,\alpha]$ takes on very large negative values for quantum rare events \cite{GongAshidaUeda2016}.

\subsection{Emergence of absolute irreversibility in the absence of quantum coherence}

We have seen that coherent driving plays a crucial role in suppressing absolute irreversibility.
However, incoherent driving immediately after the measurement, in general, gives rise to absolute irreversibility.

As a typical example, we consider a case in which the feedback process is delayed for time $t_{\rm de}$.
At time $t=t_{\rm m}$, we perform a projective energy measurement and obtain an outcome $\alpha$.
During $t_{\rm m}<t<t_{\rm m}+t_{\rm de}$, the system evolves under its bare Hamiltonian.
At $t=t_{\rm m}+t_{\rm de}$, we start a coherent driving based on the measurement outcome.
In this forward protocol, the first jump during $t_{\rm m}<t<t_{\rm m}+t_{\rm de}$ must be a jump from the state $\alpha$, and other jumps are prohibited.
In contrast, in the time-reversed protocol, there are no such restrictions.
As a consequence, there exists quantum trajectories satisfying the singularity condition~(\ref{AI}),
and hence we have the fluctuation theorem~(\ref{IFT}) with nonvanishing $\bar\lambda$.
Thus, the suppression of quantum coherence due to time delay gives rise to absolute irreversibility.

We will revisit the two-level system with the projective energy measurement discussed in Sec.~V~A.
When we observe the state $g$ at $t=t_{\rm m}$, the first jump during $t_{\rm m}<t<t_{\rm m}+t_{\rm de}$ cannot be $e\to g$.
This is because the free Hamiltonian does not mix different energy eigenstates and therefore the observed state stays the same until the first jump occurs.
As a result, quantum trajectories $\psi$ with $\alpha=g,k_{N_{\rm m}+1}=e, l_{N_{\rm m}+1}=g$ and $t_{\rm m}<t_{N_{\rm m}+1}<t_{\rm m}+t_{\rm de}$ have vanishing probability.
In contrast, in the backward protocol, quantum trajectories $\bar\psi$ with $\alpha^*=g$, $\bar k_{N-N_{\rm m}}=g, \bar l_{N-N_{\rm m}}=e$ and $\bar t_{\rm m}-t_{\rm de}<\bar t_{N-N_{\rm m}}<\bar t_{\rm m}$ is possible.
Therefore, these trajectories $\psi$ are absolutely irreversible.
Thus, under the projective measurement and feedback control with time delay, absolute irreversibility arises, which we numerically demonstrate in the next section.

We can interpret the emergence of absolute irreversibility in terms of the information gain.
Under coherent driving, quantum rare events arise, although their probability may be extremely small.
As a consequence, the information gain takes on a very large negative value.
However, when we do not have quantum coherence, the probability of such events is exactly zero and therefore the information gain negatively diverges $I\to -\infty$.
This singularity renders the fluctuation theorem~(\ref{IFT}) with $\bar\lambda=0$ inapplicable and leads to non-vanishing $\bar\lambda$.

\begin{figure}
	\begin{tabular}{rl}
		&(a)\\
		&\includegraphics[width=0.8\columnwidth, bb=0 0 360 205]{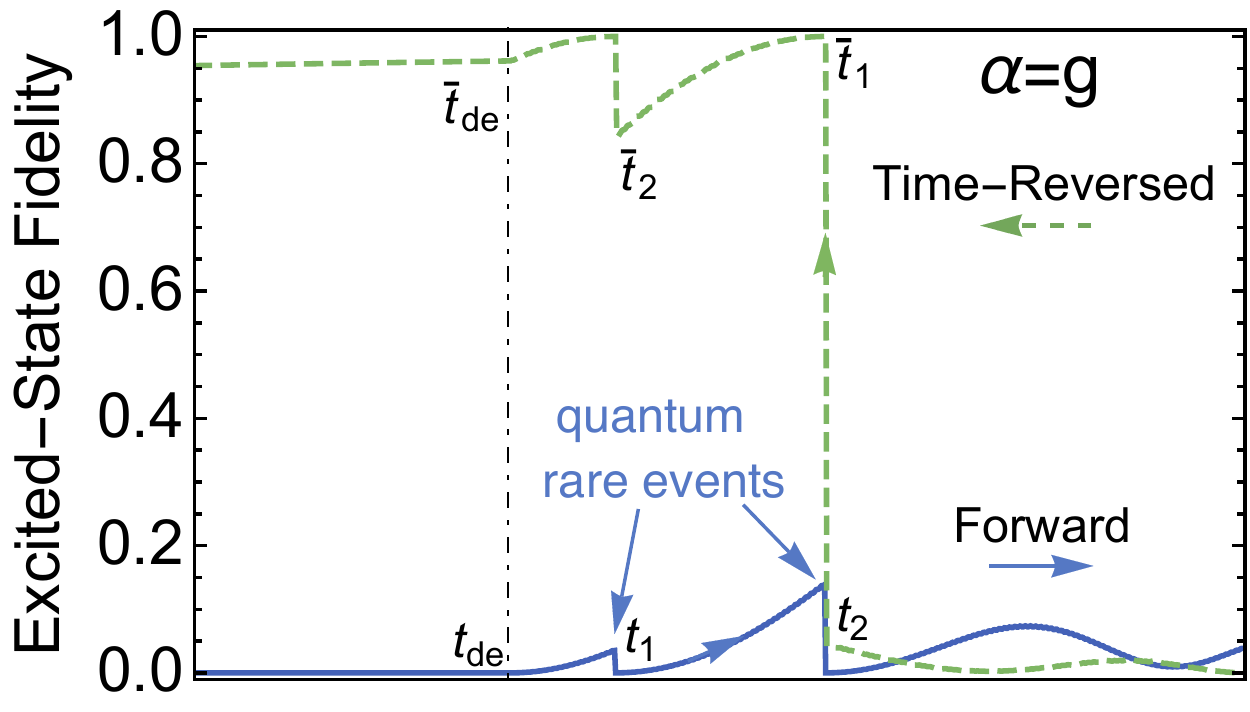}\\
		&(b)\\
		&\includegraphics[width=0.8\columnwidth, bb=0 0 360 205]{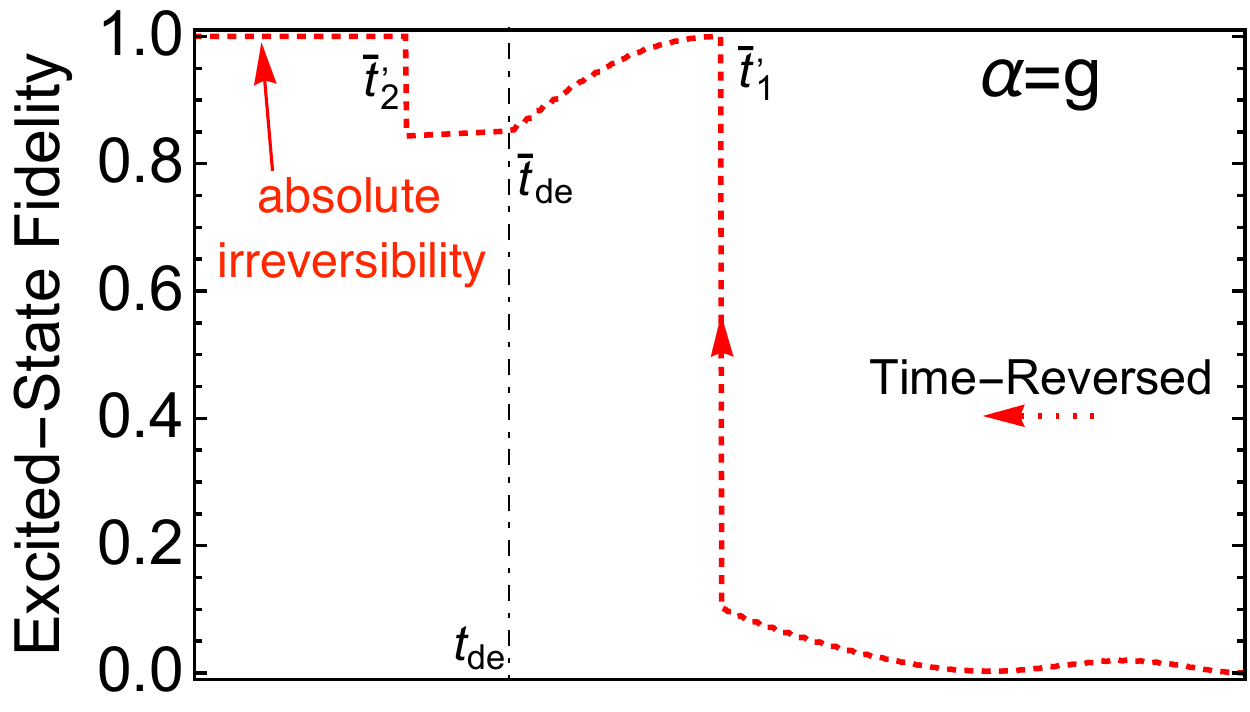}
	\end{tabular}
	\caption{\label{fig:QR}
		Examples of quantum trajectories.
		The abscissa represents time and the ordinate shows the exited-state fidelity.
		The forward measurement outcome is fixed as $\alpha=g$.
		(a) An example of quantum rare events.
		In the forward process, the system jumps downward at $t=t_1$ and $t_2$.
		This is counterintuitive because for downward jumps to occur the system must be in the excited state, but the probability of the system remaining in the excited state is rather small.
		By these jumps, the forward and backward trajectories show very different behavior in the excited-state fidelity $|\langle e|\psi_t\rangle|^2$.
		As a consequence of these quantum rare events, the probability that the final state of the backward process is found to be in the ground state becomes very small,
		and the relevant information gain~(\ref{RIG}) takes a large negative value.
		(b) An example of absolutely irreversible events.
		In this backward process, the last jump of the system at $t=\bar t_2'$ is upward.
		However, in the forward process, the system cannot experience the reverse (i.e., downward) jump because it starts from the ground state and coherent driving is absent.
		This backward event causes absolute irreversibility because it has no counterpart in the forward process.
	}
\end{figure}

\subsection{Numerical vindication}
We demonstrate the emergence of absolute irreversibility in a two-level system by numerical simulations.
The time-evolution equation of the two-level system is given by
\begin{eqnarray}\label{mod}
	\dot \rho_t
	=
	&-&\frac{i}{2}
	[\omega_t\sigma_z+\epsilon\theta(t-t_{\rm de})\sigma_x\cos\omega_{\rm d} t, \rho_t]
	\nonumber\\
	&+& \sum_{j=\pm} \gamma_j(\omega_t)\mD[\sigma_j]\rho_t,
\end{eqnarray}
where we introduce a delay time $t_{\rm de}$ in the Heaviside step function $\theta(\cdot)$.
We drive the energy gap $\omega_t$ linearly as $\omega_t=\omega_0+\Delta\omega t/T$.
The heat bath is assumed to have an Ohmic spectrum, and therefore the jump rates are given by $\gamma_\pm(\omega)=\kappa\omega[\coth(\beta\hbar\omega/2)\mp1]/2$ which satisfy the detailed balance condition~(\ref{DBC}).

\begin{figure}
	\begin{tabular}{rl}
		\includegraphics[width=0.5\columnwidth, bb=0 0 360 250]{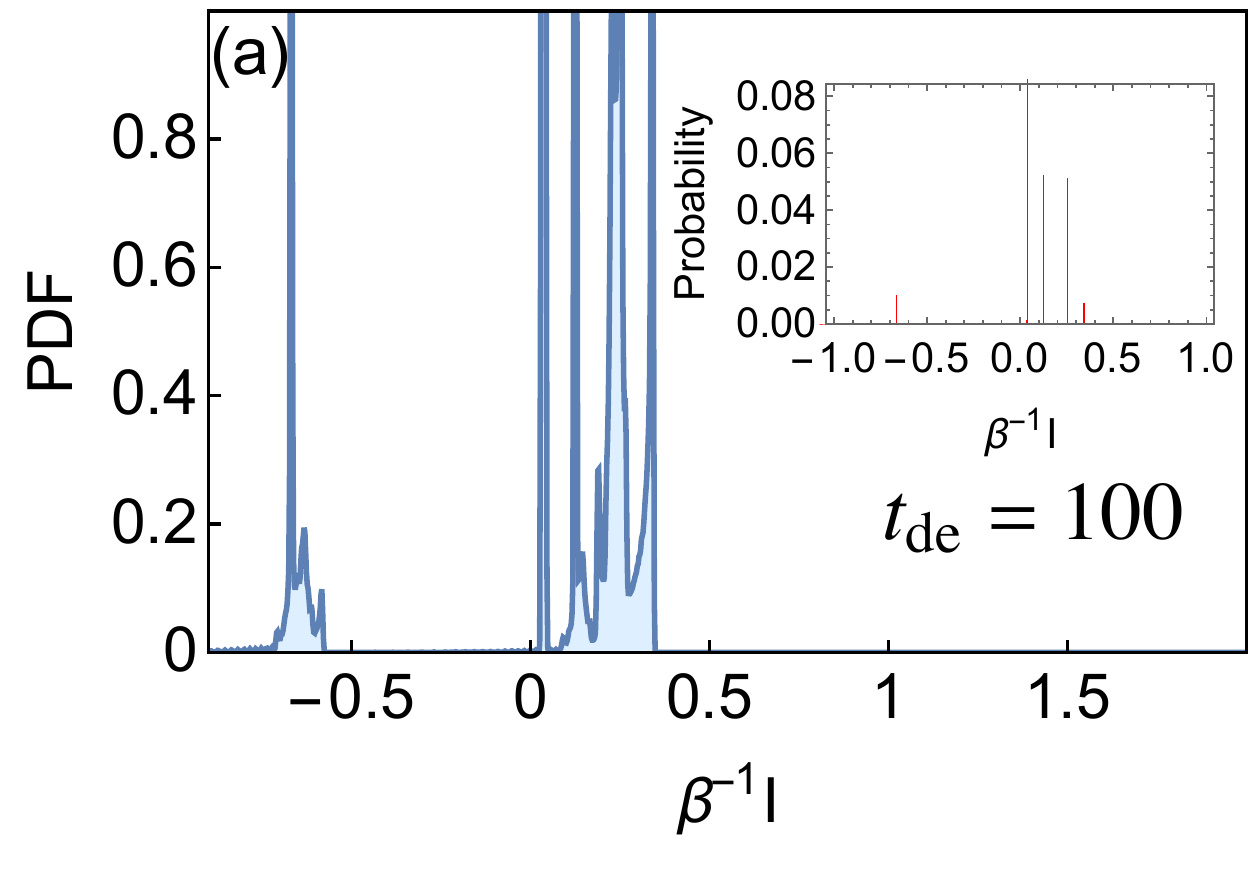}
		&\includegraphics[width=0.5\columnwidth, bb=0 0 360 250]{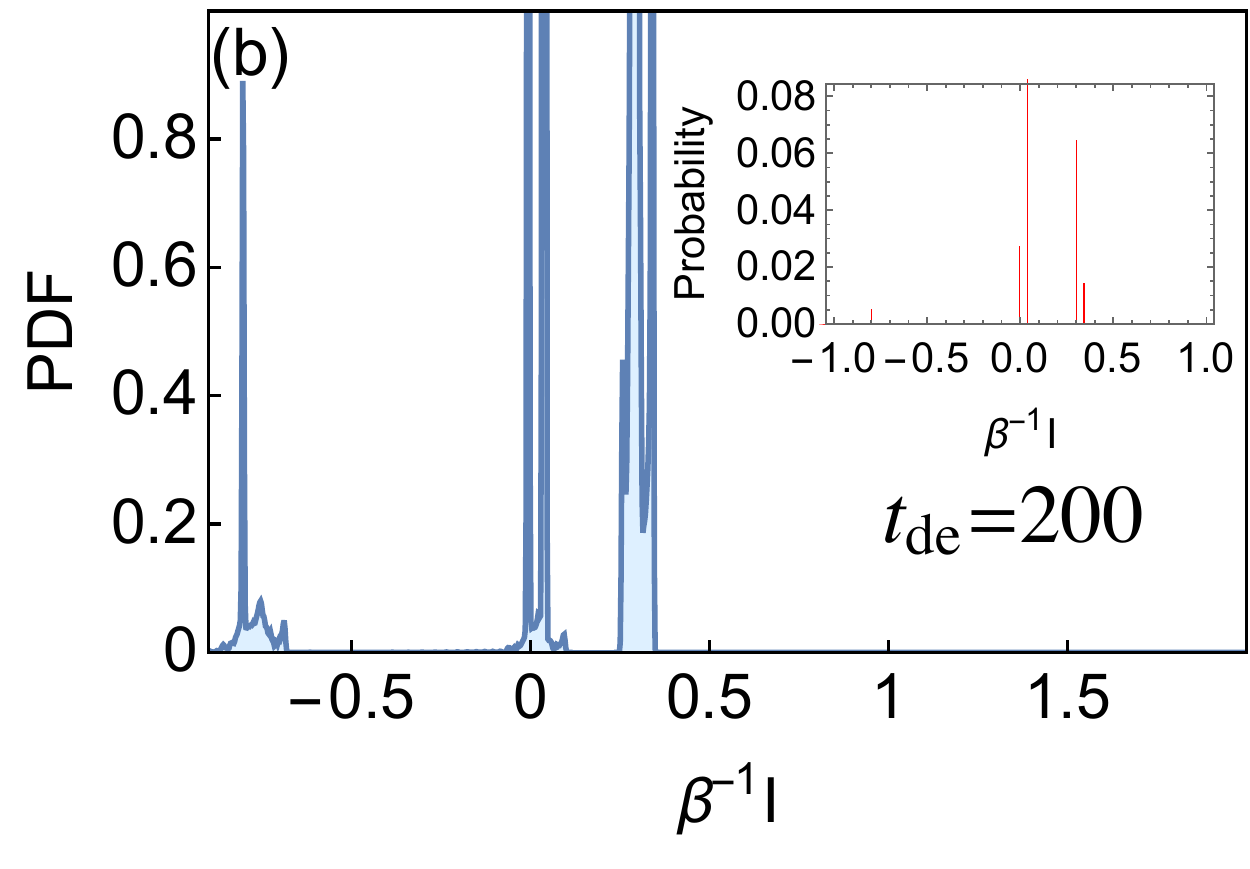}\\
		\includegraphics[width=0.5\columnwidth, bb=0 0 360 250]{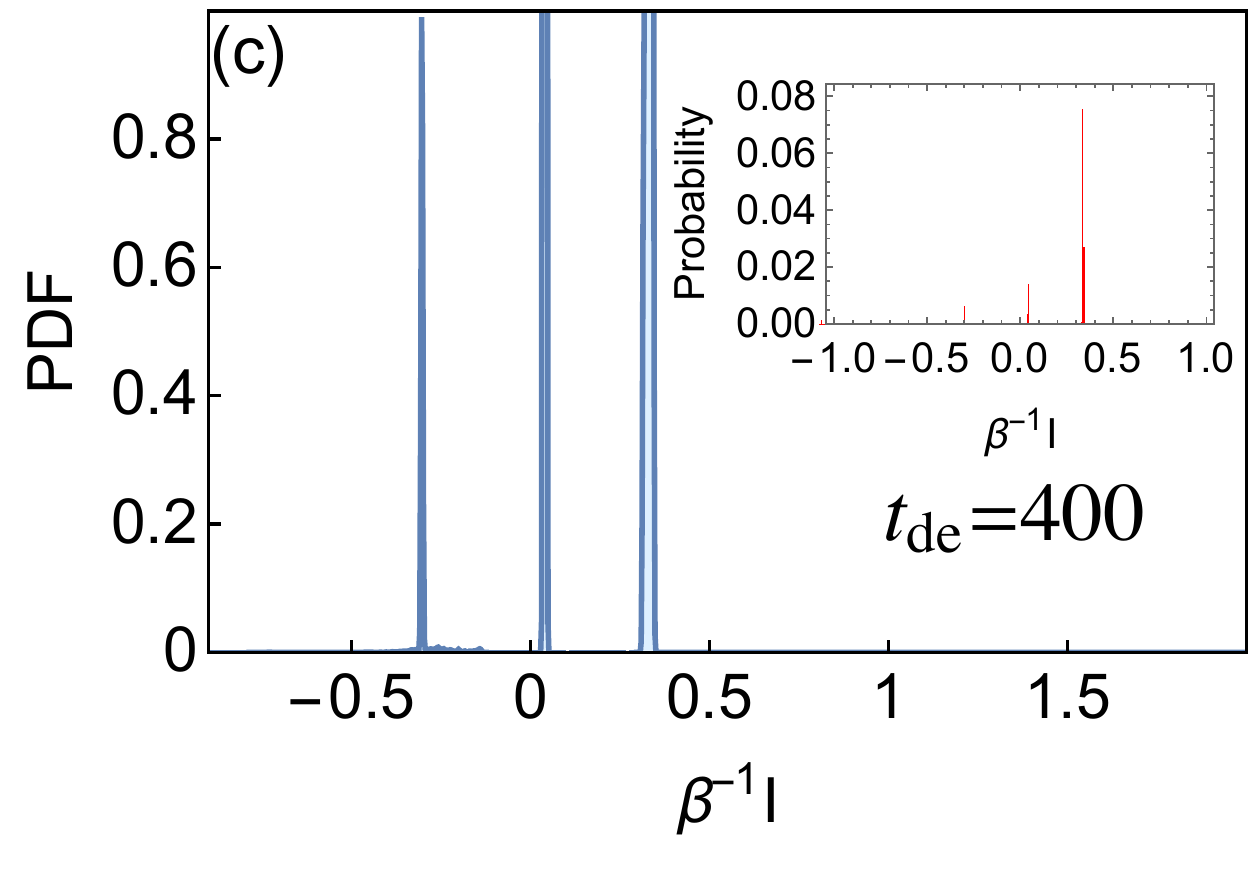}
		&\includegraphics[width=0.5\columnwidth, bb=0 0 360 250]{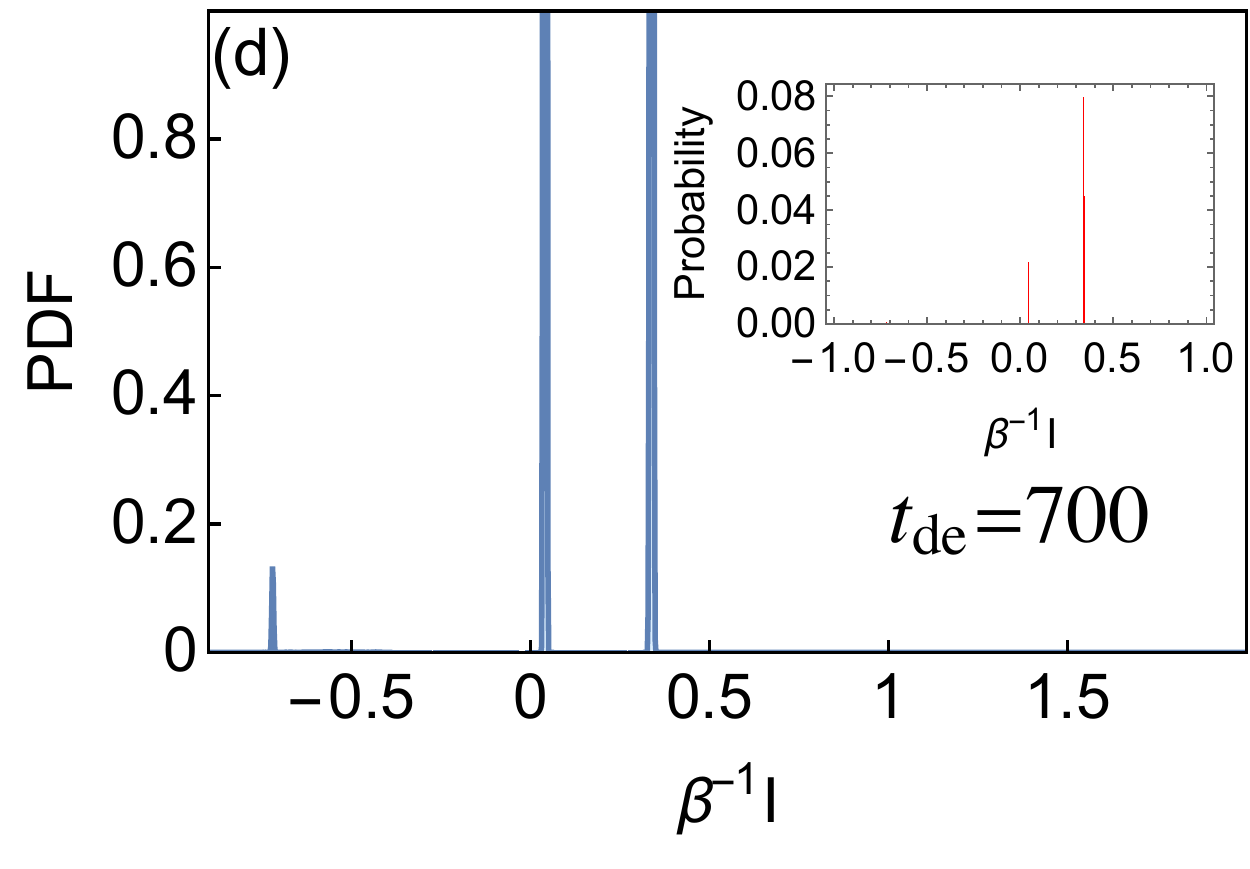}
	\end{tabular}
	\caption{\label{fig:Hist}
		Probability density function (PDF) of the relevant information gain for the varying delay time $t_{\rm de}$.
		With larger $t_{\rm de}$, the probability distribution becomes sharper because quantum coherent effects are suppressed.
		(Inset) Probability distribution of the relevant information, which is constituted of a set of $\delta-$functions in terms of the probability density.
	}
\end{figure}

\begin{figure}
	\begin{center}
	\begin{tabular}{l}
		(a)\\
		\includegraphics[width=0.8\columnwidth, bb=0 0 360 223]{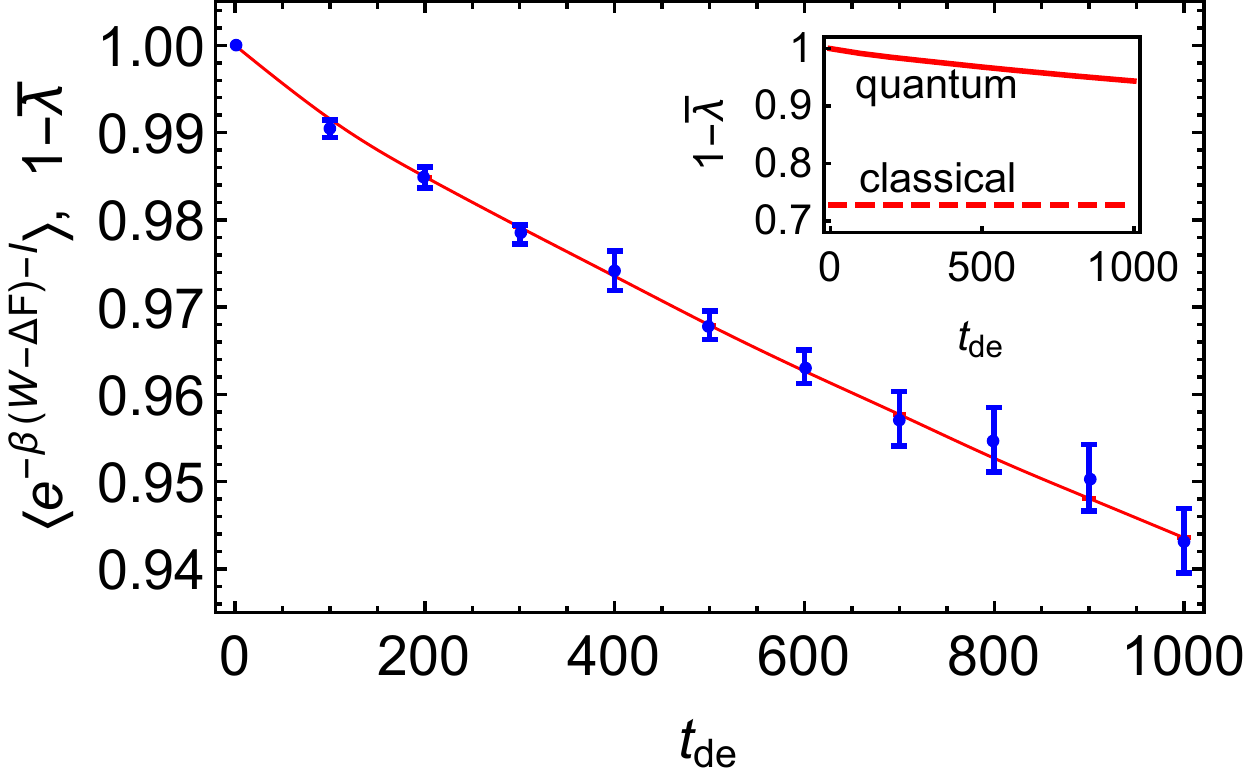}\\
		(b)\\
		\includegraphics[width=0.8\columnwidth, bb=0 0 360 223]{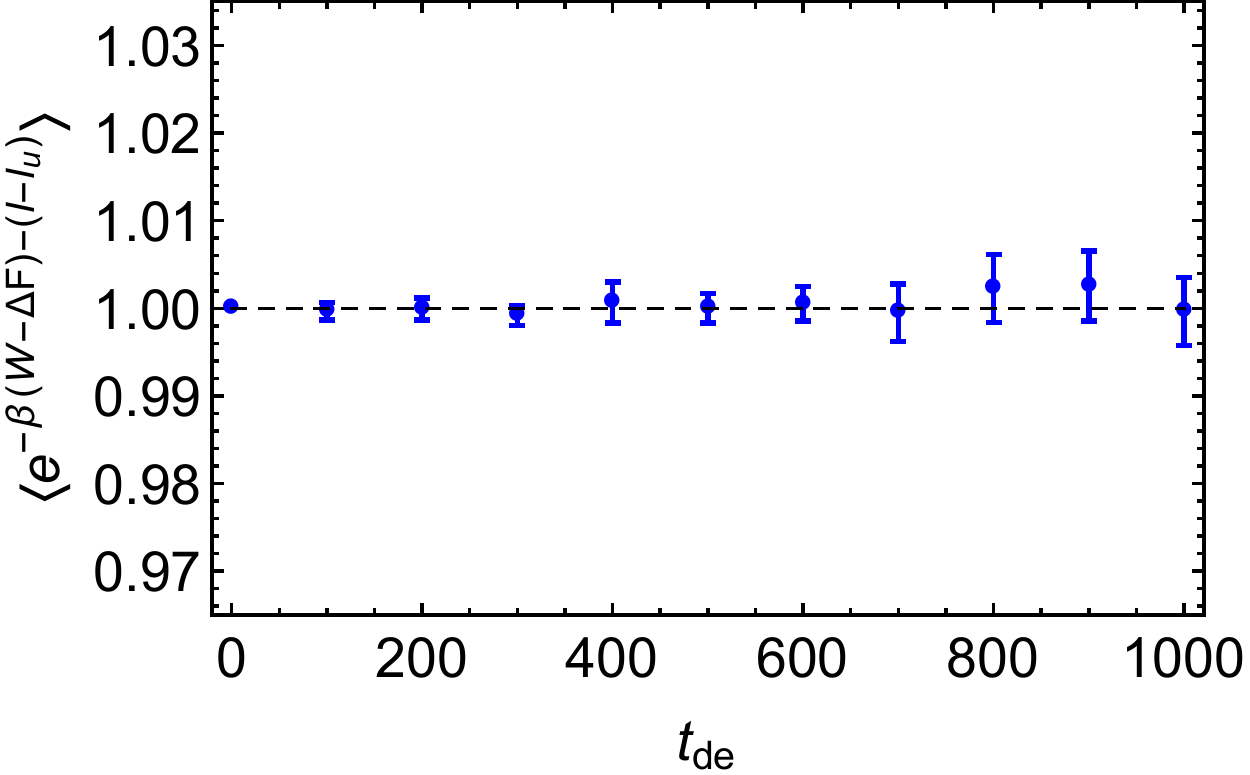}
	\end{tabular}
	\end{center}
	\caption{\label{fig:VFT}
		Numerical test of the fluctuation theorems with absolute irreversibility against variations of the delay time $t_{\rm d}$. 
		(a) Verification of the fluctuation theorem~(\ref{IFT}).
		The blue dots show the values of $\langle e^{-\beta(W-\Delta F)-I}\rangle$ which are numerically calculated by the stochastic wave-function method.
		The red curve is the value of $1-\bar\lambda$ evaluated from the path integral~(\ref{DefL}).
		(Inset) Comparison between the quantum and classical cases.
		The red solid (dashed) curve is the value of $1-\bar\lambda$ in the presence (absence) of quantum coherent driving.
		(b) Verification of the fluctuation theorem with unavailable information~(\ref{UIFT}).
		For each $t_{\rm de}$, we sample $2\times10^7$ trajectories.
		The duration of the numerical simulations is set to be $T=2000$ and $t_{\rm de}$ is varied over the range of $[0,1000]$.
		The parameters of the feedback protocol are set to be $\omega_{\rm d}=0.1\pi$, $\omega_0=0.3$, $\Delta \omega =0.1$, $\epsilon_g=0.002$ and $\epsilon_e=0.008$.
		The parameters of the heat bath are chosen to be $\beta=5$ and $\kappa=0.001$.
	}
\end{figure}

To realize a feedback protocol, we prepare a thermal equilibrium state with the inverse temperature $\beta$ under the inclusive Hamiltonian $H(\omega)=\hbar\omega\sigma_z/2$.
Then, we immediately perform the projective energy measurement to obtain an outcome $\alpha=g$ or $e$.
Thus, we set $t_{\rm m}=0$ in this protocol.
According to the outcome, we change the strength $\epsilon_\alpha$ of the exclusive driving $h_{\alpha,t}=\hbar\epsilon_\alpha\theta(t-t_{\rm de})\sigma_x\cos\omega_{\rm d}t/2$.
To extract more work, when we obtain the outcome $g$, we should decrease $\epsilon_g$ to suppress the upward jump, which contributes to a negative extracted work.
In contrast, when we obtain the outcome $e$, we should increase $\epsilon_e$.
Therefore, to perform an efficient feedback control, we should set $\epsilon_g<\epsilon_e$.

\begin{figure}
	\includegraphics[bb=0 0 360 233,width=0.8\columnwidth]{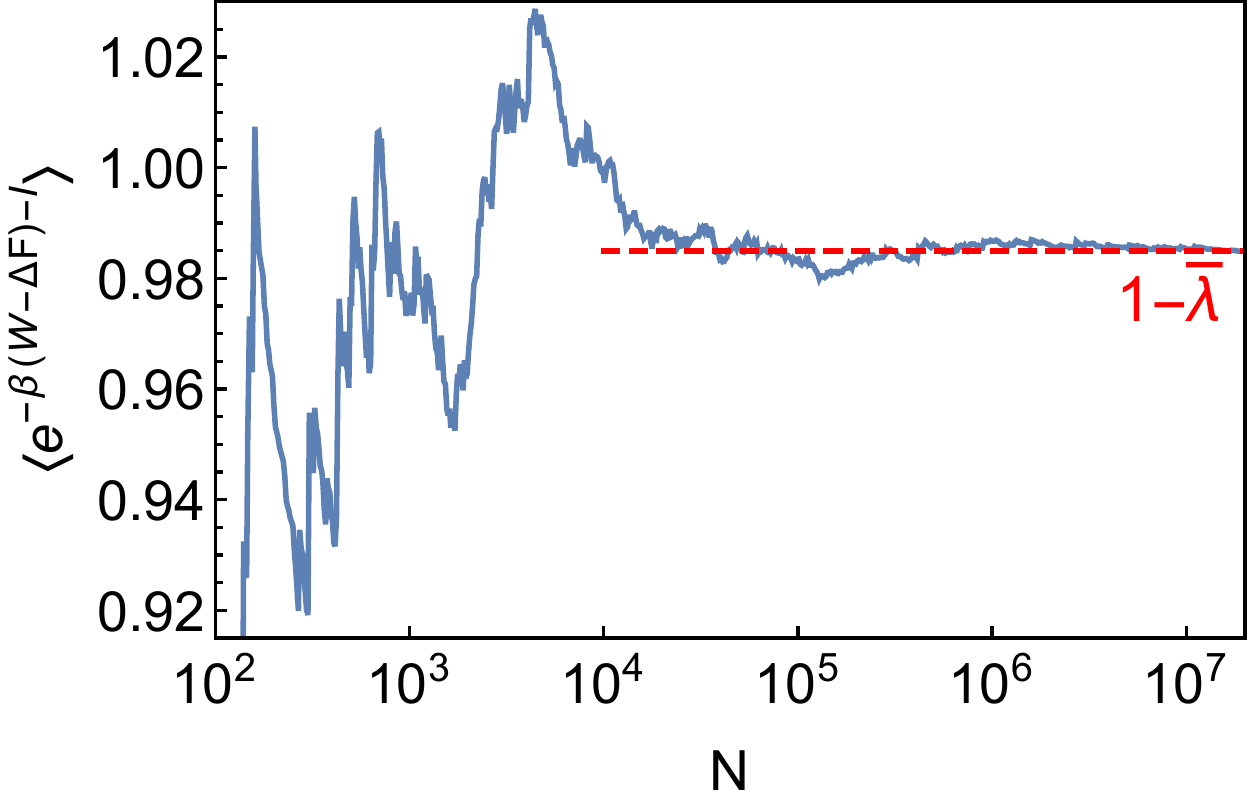}
	\caption{\label{fig:conv}
		A typical convergent behavior of the left-hand side of the fluctuation theorem~(\ref{IFT}) for $t_{\rm de}=200$.
		The abscissa represents the accumulated number of samples and the ordinate shows the numerically calculated average of the exponential function in the angle brackets in Eq.~(\ref{IFT}).
		We see that the average undergoes sudden increases.
		This behavior implies that sampling quantum rare events with large negative $I$ and therefore with large $e^{-\beta(W-\Delta F)-I}$ is crucial for the convergence.
	}
\end{figure}

We conduct Monte Carlo simulations based on the stochastic wave function approach \cite{HekkingPekola2013,GongAshidaUeda2016} to numerically verify the fluctuation theorems with absolute irreversibility.
We generate quantum trajectories and record their thermodynamic quantities and information gain.

Figure~\ref{fig:QR}~(a) illustrates quantum rare events when the system is initially found to be in the ground state $\alpha=g$.
Then, the system undergoes two downward jumps, although the excited-state fidelity is small.
We note that in the absence of quantum coherent driving after $t=t_{\rm m}=0$, the first jump must be upward.
Thus, these events are quantum rare events caused by quantum coherence.
We note that the forward and backward events look very different with respect to the excited-state fidelity.
As a result, the probability $||\bar M_{g^*}|\bar\psi_{\bar \bar t_{\rm m}^-}\rangle||^2$ that the system is found to be in the ground state upon the backward measurement is small.
Consequently, the relevant information gain~(\ref{RIG}) takes a large negative value and therefore quantum rare events significantly contribute to the average in the fluctuation theorem.

Figure~\ref{fig:QR}~(b) shows an absolutely irreversible event in the backward process.
This event ends up with the upward jump during $0<t<t_{\rm de}$.
However, in the forward process, the downward jump is prohibited as the first jump of the system during $0<t<t_{\rm de}$ after the system is found to be in the ground state.
Thus, the event does not have the corresponding event in the forward process, which means that it is absolutely irreversible.
To calculate the absolutely irreversible probability $\lambda(\alpha)$, we accumulate the probability of these events in the backward process.

By the Monte Carlo simulations, we can obtain the probability distribution function of work and information gain.
Figure~\ref{fig:Hist} shows the probability density functions of the relevant information gain.
With a smaller delay time, the probability distribution is broader due to quantum coherence.
With a larger delay time, the distribution is sharper because quantum coherence is suppressed.

Using the probability density functions of $W+\beta^{-1}I$, we can calculate the average in the fluctuation theorems.
In Fig.~\ref{fig:VFT}~(a), we verify the fluctuation theorem~(\ref{IFT}) by varying the delay time $t_{\rm de}$.
The blue and purple dots are obtained by calculation of the left-hand sides of the fluctuation theorem.
The red curve shows the value of $1-\bar\lambda$ calculated from the definition~(\ref{DefL}).
The deviation of this quantity from unity indicates the emergence of absolute irreversibility in the absence of quantum coherent driving.
We find that they agree within the error bars which increase as $t_{\rm de}$ increases.

The inset of Fig.~\ref{fig:VFT}~(a) compares the value of $1-\bar\lambda$ (red solid curve) with that in the classical case (red dashed line).
Here, by the classical case, we mean the situation where the off-diagonal Hamiltonian in Eq.~(\ref{mod}) proportional to $\sigma_x$ vanishes.
This dynamics is equivalent to the one where we couple the system with an environment with infinitely large phase decoherence while keeping the original non-diagonal Hamiltonian.
The difference between the curve and the dashed line indicates the thermodynamic benefit of quantum coherent driving after $t=t_{\rm de}$.

Moreover, we verify the validity of the generalized fluctuation theorem~ (\ref{UIFT}) as shown in Fig~\ref{fig:VFT}~(b).
We again observe an increase of errors as the delay time increases.

To understand the increase of the errors, we note that the system tends to behave more classically for larger $t_{\rm de}$.
In the classical limit, the information gain $I$ negatively diverges for quantum rare events.
Therefore, with large $t_{\rm de}$, $I$ takes on a large negative value.
This large negative value contributes to the left-hand side of Eq.~(\ref{CIFT}) significantly, although its probability is extremely small.
In Fig.~\ref{fig:conv}, the numerically obtained average $\langle e^{-\beta(W-\Delta F)-I}\rangle$ with $N$ sample trajectories is shown.
We observe that the average suddenly increases several times before it converges.
This behavior implies that quantum rare events with large negative $I$ and large positive $e^{-\beta(W-\Delta F)-I}$ play a vital role for the convergence of the average.
Therefore, the convergence tends to be slower and slower as the delay time increases, implying larger statistical error (see Fig.~\ref{fig:VFT}).

\begin{figure}
	\centering
	\includegraphics[width=0.9\columnwidth]{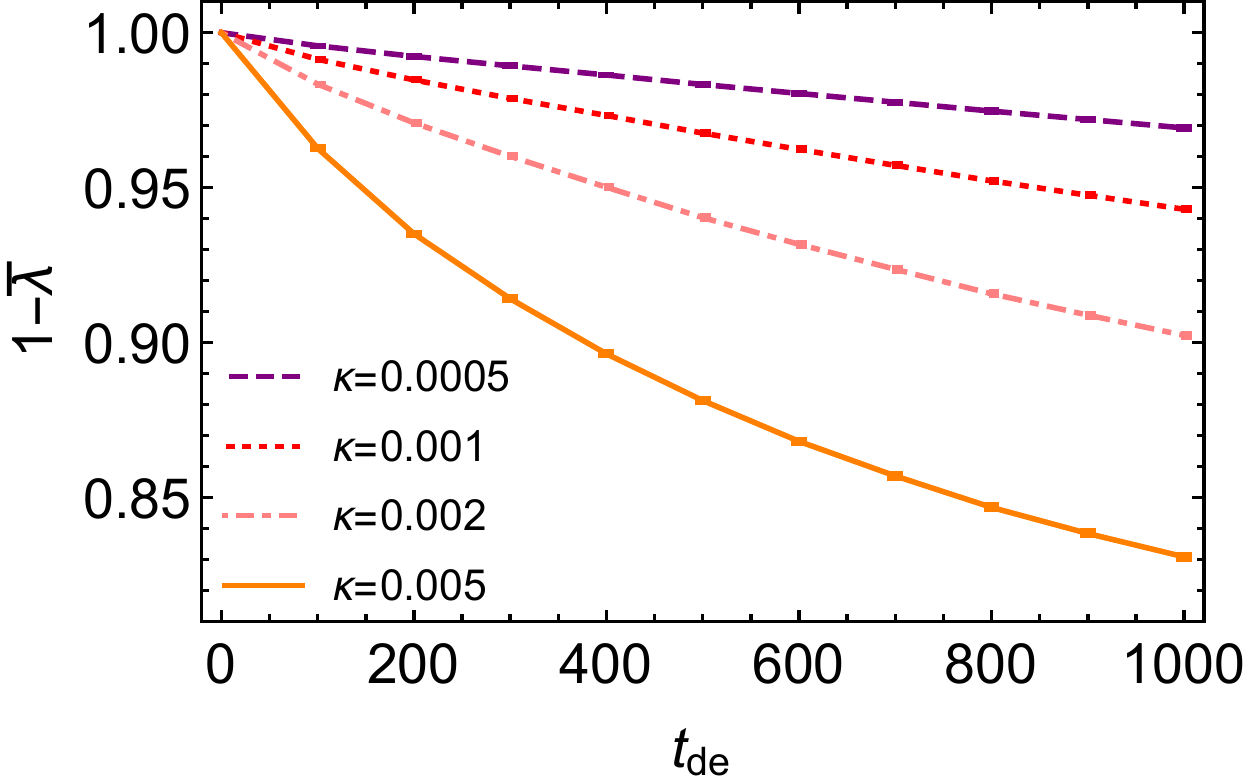}
	\caption{\label{fig:decoherence}
		Effects of decoherence due to the heat bath.
		The curves represent the numerically calculated values of $1-\bar\lambda$ against the delay time $t_{\rm de}$ for different values of $\kappa$, which quantifies the strength of the system-bath interaction.
		The parameters are set as in the caption of Fig.~\ref{fig:VFT} except $\kappa$.
	}
\end{figure}

Finally, we discuss how decoherence due to the heat bath affects the degree of absolute irreversibility.
In Fig.~\ref{fig:decoherence}, we change the strength of the interaction between the system and the heat bath denoted by $\kappa$ and plot $1-\bar\lambda$ against the delay time $t_{\rm de}$.
We see that, as $\kappa$ increases, the degree of absolute irreversibility also increases for a given $t_{\rm de}$.
Thus, a larger decoherence due to the heat bath causes a greater thermodynamic disadvantage as the increasing degree of absolute irreversibility in the fluctuation theorem.

\section{Towards Experimental Realization}

\subsection{Sequence of small systems as an environment}
To implement quantum stochastic thermodynamics, we should record quantum jumps by monitoring the environment.
However, this monitoring is difficult when the environment is infinitely large.
This difficulty can be overcome if we consider a situation in which an environment can be modeled as a collection of small systems interacting sequentially with a system of our interest; we call each of the small systems an environment \cite{HorowitzParrondo2013}.
Specifically, we prepare each environment in a canonical distribution with inverse temperature $\beta$, and perform the projective energy measurement on it to find the initial energy of the environment.
Then, we let the environment interact with the system during a time interval $\tau$.
After the interaction is switched off, we again perform the projective energy measurement on the environment to find the final energy of the environment.
We repeat this procedure for a sequence of identically prepared environments. 
Based on this procedure, we can evaluate the heat flow from the system to the environment by accumulating the energy increment of each environment. 
By constructing the environment in this way instead of invoking an infinite environment, we can circumvent the conceptual and practical difficulty in evaluating heat on the basis of the two-point energy measurement on the infinite reservoir \cite{EspositoHarbolaMukamel2009,CampisiHanggiTalkner2011}.

\begin{figure}
	\includegraphics[width=\columnwidth, bb=0 0 635 267]{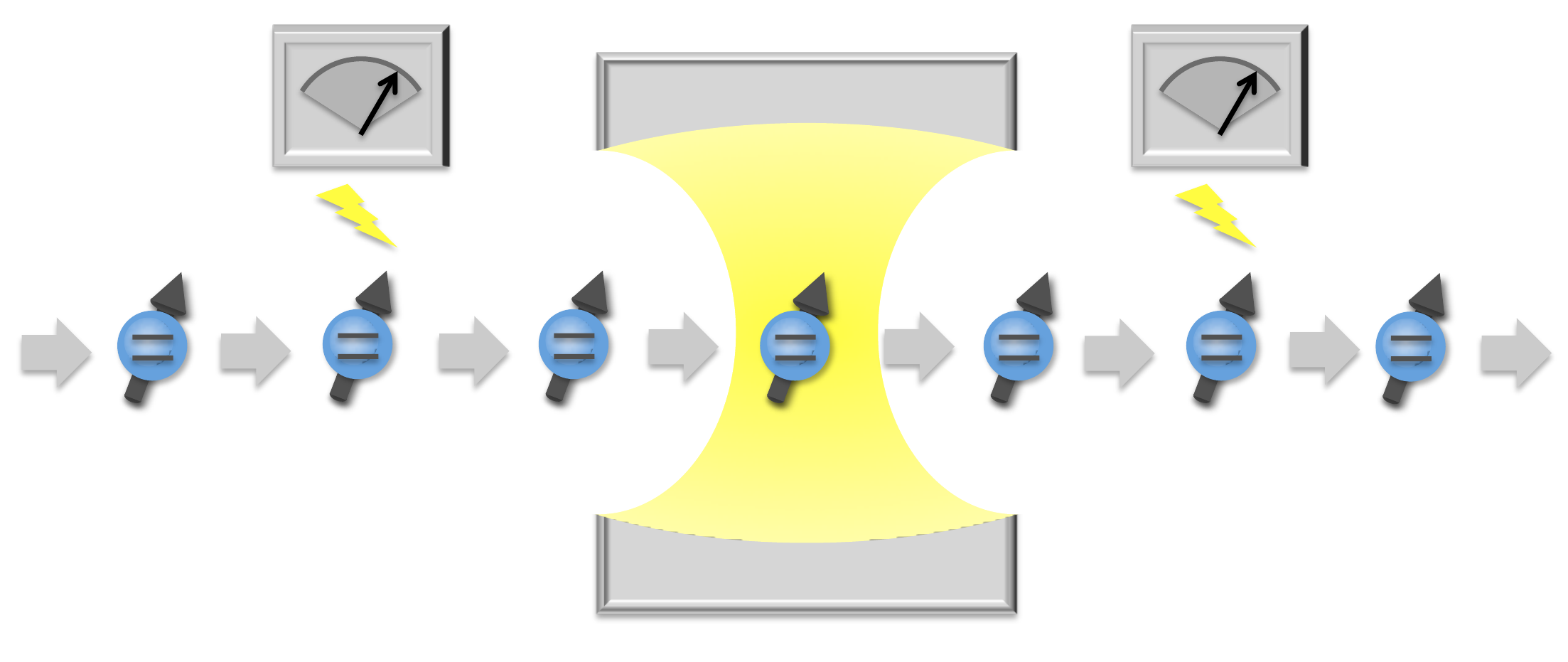}
	\caption{\label{fig:sketch}
		Schematic illustration of a cavity QED system as a model for quantum stochastic thermodynamics.
		The photon field in the cavity is the system and a sequence of two-level atoms, which are injected through the cavity, play the role of environments for the system.
		To determine the energy increment of each individual environment, we perform energy projective measurements on each atom before and after injection.
	}
\end{figure}

A prototypical example is a cavity QED system as schematically illustrated in Fig.~\ref{fig:sketch}.
Here, two-level atoms, which are regarded as environments, are sequentially injected through the cavity and interact with the photon field during a time interval $\tau$.
This example has been theoretically studied in Refs.~\cite{UedaImotoNagaokaOgawa1992,KistOrszagBrunDavidovich1999} and experimentally realized in Refs.~\cite{Guerlin...Haroche2007,Deleglise...Haroche2008}. 
In this setup, by taking the limit of continuous injection $\tau\to 0$ and assuming an appropriate condition for the coupling strength between the system and an environment, we can confirm that the time evolution of the system is governed by the Lindblad equation~(\ref{Lindbladeq}) as described in detail in Ref.~\cite{HorowitzParrondo2013}.

\subsection{Implementation in a circuit QED system}
\begin{figure}
	\includegraphics[width=1.0\columnwidth, bb=0 0 522 308]{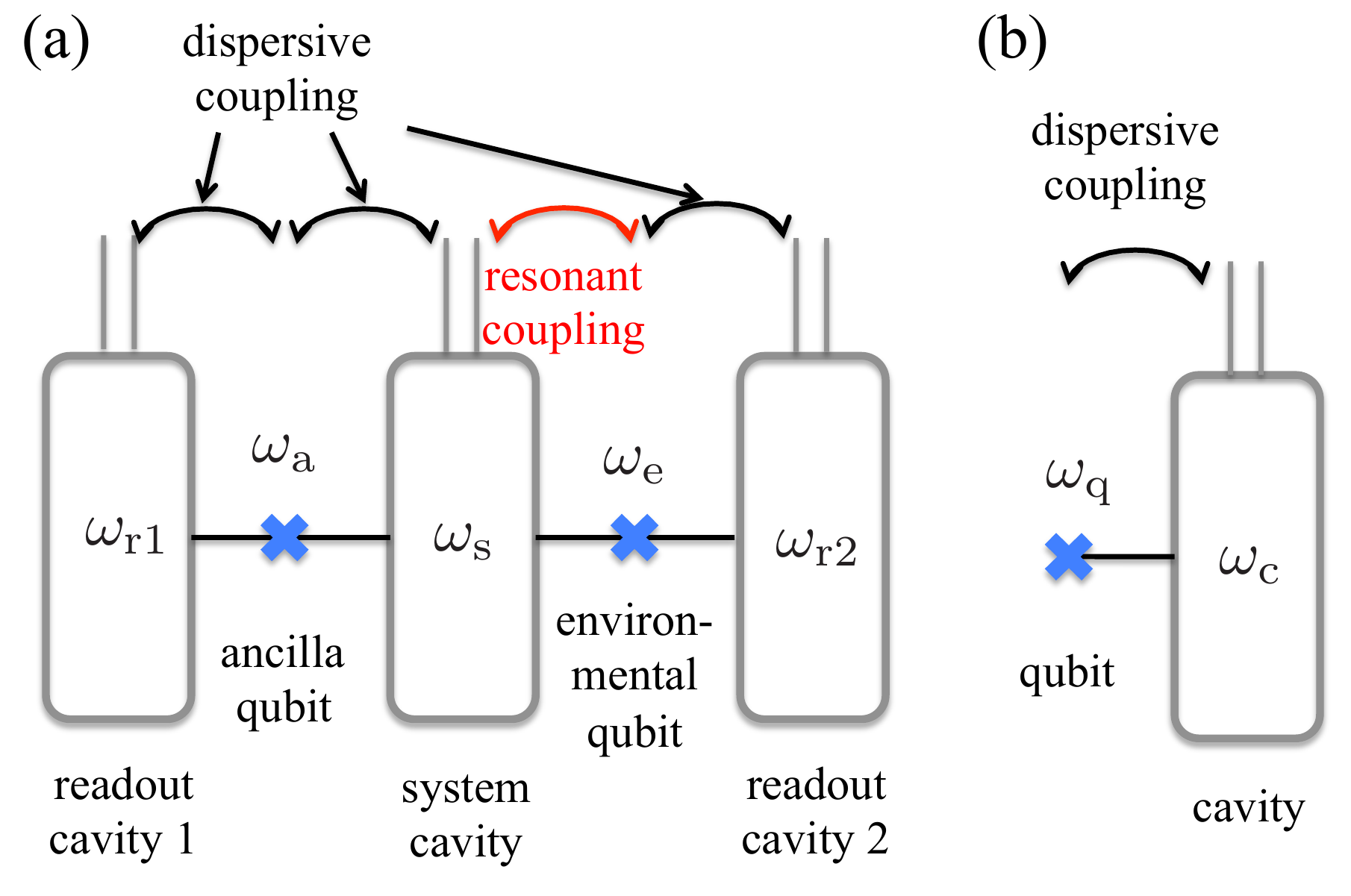}
	\caption{\label{fig:SCQA}
		Schematic illustrations of experimental setups.
		(a) Proposed experimental setup consisting of three cavities and two superconducting qubits.
		The middle cavity is the system of our interest.
		The environment is engineered by the environmental qubit, which is resonantly coupled to the system cavity.
		The ancilla qubit, which is dispersively coupled to the system cavity, is utilized for the energy projective measurement on the system cavity.
		The cavities at both ends are for the energy projective measurement on the adjacent qubits.
		(b) A pair of a dispersively coupled qubit and cavity.
		By the dispersive shift in Eq.~(\ref{JCeq}), quantum non-demolition measurement on the qubit (cavity) can be realized by the projective measurement on the cavity (qubit).
	}
\end{figure}

The second example is a circuit QED system \cite{HarocheRaimond}.
We consider a setup consisting of three cavities and two superconducting qubits as illustrated in Fig.~\ref{fig:SCQA}~(a).
The middle cavity with a resonant frequency $\omega_{\rm s}$, which we shall henceforth refer to as the ``system,'' has a long lifetime $\tau_{\rm s}$.
We engineer the ``environment'' by the right qubit with a transition frequency $\omega_{\rm e}$ and the coupling strength $g_{\rm se}$.
We assume that the environment is nearly resonant to the system; $|\omega_{\rm e} - \omega_{\rm s}|\ll g_{\rm se}$ and therefore energy exchanges occur between them.
We note that this environmental qubit corresponds to the two-level atoms in Fig.~\ref{fig:sketch}.
The left qubit with a resonant frequency $\omega_{\rm a}$ is an ``ancilla'' to realize the non-demolition measurement of the photon-number state of the system \cite{Schuster_2007}.
The coupling strength with the system $g_{\rm sa}$ is assumed to be in the dispersive regime $|\omega_{\rm s}-\omega_{\rm a}|\gg g_{\rm sa}$.
The leftmost (rightmost) cavity with resonant frequency $\omega_{\rm r1(2)}$ couples to the adjacent ancilla (environment) qubit dispersively as $|\omega_{\rm a}-\omega_{\rm r1}|\gg g_{\rm ar1}$ ($|\omega_{\rm e}-\omega_{\rm r2}|\gg g_{\rm er2}$),
and is utilized for fast readout of the ancilla (environment) qubit states \cite{Wallraff_2005}.

To clarify the role of the dispersive regime, we here restrict our attention to the qubit with a resonant frequency $\omega_{\rm q}$ and the cavity with a resonant frequency $\omega_{\rm c}$ as depicted in Fig.~\ref{fig:SCQA}~(b).
The dispersive regime refers to the situation where the qubit-cavity detuning $\Delta$ is much larger than the coupling strength $g$, i.e., $|\Delta|:=|\omega_{\rm q}-\omega_{\rm c}|\gg g$.
Therefore, we may assume that there are no real energy exchanges between the qubit and the cavity.
Up to the second order in $g/\Delta\ (\ll 1)$, the effective Hamiltonian reads
\begin{equation}\label{JCeq}
	H
	=
	\hbar \omega_{\rm c} (a^\dag a + 1/2) + \hbar \omega_{\rm q} \sigma_z/2
	+\hbar \chi (a^\dag a+1/2)\sigma_z,
\end{equation}
where $a$ is the mode of the cavity and $\sigma_z$ is the pseudo-spin operator of the qubit; we define the dispersive shift $\chi=g^2/\Delta$.
The third term on the right-hand side of Eq. ~(\ref{JCeq}) shifts the cavity frequency depending on the state of an atom, and also shifts the frequency of an atomic frequency depending on the photon number in the cavity.
The interaction term commutes with the bare Hamiltonians of both the atom and the cavity, and therefore the quantum non-demolition measurement can be achieved by this interaction.

To verify the fluctuation theorems in this scheme, we need (1) engineering of an environment by the environmental qubit and (2) the projective energy measurement on the system.
Furthermore, the component (1) can be divided into (1-a) the projective measurement on the environment and (1-b) fast preparation of a thermal state of the environment.
In the following, we discuss how to realize each part.

(1-a) Due to the dispersive coupling between the environmental qubit and the rightmost readout cavity, the resonant frequency of the readout cavity shifts by $\chi\sigma_z$.
Therefore, by measuring a phase shift of a microwave pulse reflected by the cavity, we can perform the energy projective measurement on the environmental qubit.

(1-b) We inject a pulse at frequency $\omega_{\rm e}$ with a fixed width via the rightmost reservoir and coherently rotate the state of the environmental qubit.
Then, we remove the off-diagonal elements of the density operator of the environment by performing the projective measurement of the energy of the environmental qubit.
The fraction of the environmental qubit being excited determines the temperature of the engineered environment.

(2) First, we measure a phase shift of a microwave pulse reflected by the leftmost readout cavity.
Then, we repeatedly inject $\pi$-pulses at frequency $\omega_{\rm a}+\chi_{\rm sa}(2n+1)\ (n=0,1,\cdots)$ and measure the phase shift.
When it is different from the initial phase shift, $n$ is the photon number of the system.

With these technologies, we can conduct the feedback protocol introduced in Sec.~III~A.
(i) First, we prepare the system in a thermal equilibrium.
Then, we perform the initial projective energy measurement on the system.
(ii) We prepare the environment in the thermal equilibrium at the same temperature.
We then perform the projective measurement on the environment.
We let the system and the environment interact with each other during a short time interval $\tau$.
After the interaction, we perform the energy projective measurement on the environment.
When the outcomes of the measurements before and after the interaction differ, we record that a quantum jump occurs at this time.
The prescribed driving of the system can be injection of microwaves and/or change of the energy gap of the qubit \cite{Paauw_2009}.
(iii) The projective energy measurement is performed on the system.
We conduct different feedback controls on the system depending on the measurement outcomes.
Absolute irreversibility is expected to emerge when we apply a coherent driving after a nonvanishing delay time.
(iv) We resume the engineering of the environment.
How to drive the system depends on the measurement outcome.
(v) The final projective energy measurement is performed on the system.
With all these procedures, we can experimentally verify the fluctuation theorems.

In practice, the fact that the dimension of the system's Hilbert space is infinite can be problematic because it drastically increases the number of possible quantum trajectories and prevents convergence of the ensemble average.
This is the reason why we use the two-level qubit, rather than the harmonic oscillator, for numerical simulations.
Therefore, it is preferable that we use the qubit as a system.
This can be achieved in our proposed setup; we can block the second excited state and confine the system's Hilbert space to the two levels.
We can do so by coupling an additional dispersive qubit (not shown in the figure) to the system cavity and dynamically shifting the energy of the second excited state by the Rabi coupling \cite{Bretheau_2016}.

\section{Conclusions}
We have shown that quantum coherence can manifest its thermodynamic effects in the fluctuation theorem.
While quantum information theory has revealed intriguing thermodynamic properties of quantum coherence, the statistical-mechanical formulation of quantum thermodynamics by the fluctuation theorem has previously provided little insight on quantum coherence.
In this sense, our study gives a complementary view of thermodynamic properties of quantum coherence.
Moreover, in contrast to the resource-theoretic approach, ours gives the definition of work with clear operational meanings and therefore has immediate relevance to state-of-the-art experiments.
Furthermore, our formulation gives a trajectory-level description of thermodynamics of open quantum systems and is applicable even under measurement-based feedback control.

Specifically, we develop theory of stochastic thermodynamics in open quantum systems under feedback control.
The open quantum system obeys the Lindblad equation, which can be unraveled into a stochastic Schr\"odinger equation.
We define stochastic thermodynamic quantities along individual quantum trajectories.
We introduce the time-reversed process since it is convenient for the derivation of the fluctuation theorem.

In terms of quantum trajectories, we introduce the measurement-based feedback protocol.
The quantum path under this protocol is completely characterized by the initial and final quantum numbers, the measurement outcome and the history of quantum jumps.
We introduce the time-reversed dynamics to derive the fluctuation theorem.
To cancel out the backaction of the measurement in the original process, we should perform a non-CPTP operation in the time-reversed process.
To circumvent this problem, both measurement and post-selection are needed in the time-reversed protocol.
This construction makes stark contrast to the classical case, where neither of them is needed in the time-reversed protocol.
Based on the explicit construction of the time-reversed protocol, we derive the detailed fluctuation theorem in the presence of absolute irreversibility.

To derive the integral fluctuation theorems, we should take account of absolute irreversibility.
Absolute irreversibility refers to singularly irreversible situations with divergent entropy-production-like quantity, and typically emerges when the forward paths are restricted by the consistency with the measurement outcome.
The derived integral fluctuation theorems with absolute irreversibility give tighter inequalities than the conventional second law of information thermodynamics.

Based on the derived fluctuation theorems, we consider the interplay between quantum coherence and absolute irreversibility.
Absolute irreversibility is suppressed by quantum coherence, which spreads the state onto the entire Hilbert space even just after the projective measurement.
Conversely, absolute irreversibility emerges in the absence of quantum coherence.
To demonstrate these effects, we consider a practical situation with direct relevance to experiments, i.e., an open quantum system under time-delayed feedback control.
We numerically investigate the qubit system under time-delayed feedback control and verify the fluctuation theorem with absolute irreversibility.
We show how absolute irreversibility due to an experimental delay time characterizes thermodynamic operations in a practical situation.
We also find that the quantum rare events prevent the convergence of the statistical average in the fluctuation theorem.

Finally, we have considered possible experimental implementations of quantum stochastic thermodynamics.
Experimentally, it is convenient to realize the heat bath as a sequence of small systems such as two-level systems.
We propose a setup in a circuit QED system, where we repeatedly initialize a superconducting qubit to model the heat bath.

\emph{Note added}.--- 
Recently, two papers \cite{Santos_2017,FrancicaGooldPlastina2017} appeared and discussed the role of coherence in dissipated work and entropy production in isolated and open quantum systems.
It is found that the averaged irreversible work or entropy production can be decomposed into two parts relevant to the generation of coherence and the change of population in energy eigenstates.
Although those two papers are irrelevant to feedback control or absolute irreversibility, the notion of coherence therein is also related to nonzero off-diagonal components in the basis of projective measurements.
It is an intriguing issue to generalize the formalism in Refs.~\cite{Santos_2017,FrancicaGooldPlastina2017} to feedback-controlled processes and to study whether and, if yes, how absolute irreversibility is related to the coherence-relevant entropy production.

\begin{acknowledgements}
Authors acknowledge fruitful comments by Yasunobu Nakamura and Yuta Masuyama on the experimental aspects of our work.
This work was supported by
KAKENHI Grant No. JP26287088 from the Japan Society for the Promotion of Science, 
a Grant-in-Aid for Scientific Research on Innovative Areas ``Topological Materials Science'' (KAKENHI Grant No. JP15H05855), 
the Photon Frontier Network Program from MEXT of Japan,
and the Mitsubishi Foundation.
YM was supported by the Japan Society for the Promotion of Science through the Program for Leading Graduate Schools (MERIT) and JSPS Fellowship (Grant No. JP15J00410).
ZG was supported by a MEXT scholarship.
YA acknowledges support from JSPS (Grant No. JP16J03613).
\end{acknowledgements}

\bibliography{reference}

\begin{thebibliography}{91}%
\makeatletter
\providecommand \@ifxundefined [1]{%
 \@ifx{#1\undefined}
}%
\providecommand \@ifnum [1]{%
 \ifnum #1\expandafter \@firstoftwo
 \else \expandafter \@secondoftwo
 \fi
}%
\providecommand \@ifx [1]{%
 \ifx #1\expandafter \@firstoftwo
 \else \expandafter \@secondoftwo
 \fi
}%
\providecommand \natexlab [1]{#1}%
\providecommand \enquote  [1]{``#1''}%
\providecommand \bibnamefont  [1]{#1}%
\providecommand \bibfnamefont [1]{#1}%
\providecommand \citenamefont [1]{#1}%
\providecommand \href@noop [0]{\@secondoftwo}%
\providecommand \href [0]{\begingroup \@sanitize@url \@href}%
\providecommand \@href[1]{\@@startlink{#1}\@@href}%
\providecommand \@@href[1]{\endgroup#1\@@endlink}%
\providecommand \@sanitize@url [0]{\catcode `\\12\catcode `\$12\catcode
  `\&12\catcode `\#12\catcode `\^12\catcode `\_12\catcode `\%12\relax}%
\providecommand \@@startlink[1]{}%
\providecommand \@@endlink[0]{}%
\providecommand \url  [0]{\begingroup\@sanitize@url \@url }%
\providecommand \@url [1]{\endgroup\@href {#1}{\urlprefix }}%
\providecommand \urlprefix  [0]{URL }%
\providecommand \Eprint [0]{\href }%
\providecommand \doibase [0]{http://dx.doi.org/}%
\providecommand \selectlanguage [0]{\@gobble}%
\providecommand \bibinfo  [0]{\@secondoftwo}%
\providecommand \bibfield  [0]{\@secondoftwo}%
\providecommand \translation [1]{[#1]}%
\providecommand \BibitemOpen [0]{}%
\providecommand \bibitemStop [0]{}%
\providecommand \bibitemNoStop [0]{.\EOS\space}%
\providecommand \EOS [0]{\spacefactor3000\relax}%
\providecommand \BibitemShut  [1]{\csname bibitem#1\endcsname}%
\let\auto@bib@innerbib\@empty
\bibitem [{\citenamefont {Scully}\ \emph {et~al.}(2003)\citenamefont {Scully},
  \citenamefont {Zubairy}, \citenamefont {Agarwal},\ and\ \citenamefont
  {Walther}}]{Scully_2003}%
  \BibitemOpen
  \bibfield  {author} {\bibinfo {author} {\bibfnamefont {Marlan~O.}\
  \bibnamefont {Scully}}, \bibinfo {author} {\bibfnamefont {M.~Suhail}\
  \bibnamefont {Zubairy}}, \bibinfo {author} {\bibfnamefont {Girishi~S.}\
  \bibnamefont {Agarwal}}, \ and\ \bibinfo {author} {\bibfnamefont {Herbert}\
  \bibnamefont {Walther}},\ }\bibfield  {title} {\enquote {\bibinfo {title}
  {Extracting work from a single heat bath via vanishing quantum coherence},}\
  }\href@noop {} {\bibfield  {journal} {\bibinfo  {journal} {Science}\ }\textbf
  {\bibinfo {volume} {299}},\ \bibinfo {pages} {862--864} (\bibinfo {year}
  {2003})}\BibitemShut {NoStop}%
\bibitem [{\citenamefont {Dillenschneider}\ and\ \citenamefont
  {Lutz}(2009)}]{DillenschneiderLutz2009}%
  \BibitemOpen
  \bibfield  {author} {\bibinfo {author} {\bibfnamefont {R.}~\bibnamefont
  {Dillenschneider}}\ and\ \bibinfo {author} {\bibfnamefont {E.}~\bibnamefont
  {Lutz}},\ }\bibfield  {title} {\enquote {\bibinfo {title} {Energetics of
  quantum correlations},}\ }\href@noop {} {\bibfield  {journal} {\bibinfo
  {journal} {Europhys. Lett.}\ }\textbf {\bibinfo {volume} {88}},\ \bibinfo
  {pages} {50003} (\bibinfo {year} {2009})}\BibitemShut {NoStop}%
\bibitem [{\citenamefont {Scully}\ \emph {et~al.}(2011)\citenamefont {Scully},
  \citenamefont {Chapin}, \citenamefont {Dorfman}, \citenamefont {Kim},\ and\
  \citenamefont {Svidzinsky}}]{Scully_2011}%
  \BibitemOpen
  \bibfield  {author} {\bibinfo {author} {\bibfnamefont {Marlan~O.}\
  \bibnamefont {Scully}}, \bibinfo {author} {\bibfnamefont {Kimberly~R.}\
  \bibnamefont {Chapin}}, \bibinfo {author} {\bibfnamefont {KOnstantin~E.}\
  \bibnamefont {Dorfman}}, \bibinfo {author} {\bibfnamefont {Moochan~Barnabas}\
  \bibnamefont {Kim}}, \ and\ \bibinfo {author} {\bibfnamefont {Anatoly}\
  \bibnamefont {Svidzinsky}},\ }\bibfield  {title} {\enquote {\bibinfo {title}
  {Quantum heat engine power can be increased by noise-induced coherence},}\
  }\href@noop {} {\bibfield  {journal} {\bibinfo  {journal} {Proc. Natl. Acad.
  Sci. USA}\ }\textbf {\bibinfo {volume} {108}},\ \bibinfo {pages} {15097}
  (\bibinfo {year} {2011})}\BibitemShut {NoStop}%
\bibitem [{\citenamefont {Lloyd}(2011)}]{Lloyd2011}%
  \BibitemOpen
  \bibfield  {author} {\bibinfo {author} {\bibfnamefont {Seth}\ \bibnamefont
  {Lloyd}},\ }\bibfield  {title} {\enquote {\bibinfo {title} {Quantum coherence
  in biological systems},}\ }\href@noop {} {\bibfield  {journal} {\bibinfo
  {journal} {J. Phys.: Conf. Ser.}\ }\textbf {\bibinfo {volume} {302}},\
  \bibinfo {pages} {012037} (\bibinfo {year} {2011})}\BibitemShut {NoStop}%
\bibitem [{\citenamefont {Engel}\ \emph {et~al.}(2007)\citenamefont {Engel},
  \citenamefont {Calhoun}, \citenamefont {Read}, \citenamefont {Ahn},
  \citenamefont {Mancal}, \citenamefont {Cheng}, \citenamefont {Blankenship},\
  and\ \citenamefont {Fleming}}]{EngelFleming2007}%
  \BibitemOpen
  \bibfield  {author} {\bibinfo {author} {\bibfnamefont {Gregory~S.}\
  \bibnamefont {Engel}}, \bibinfo {author} {\bibfnamefont {Tessa~R.}\
  \bibnamefont {Calhoun}}, \bibinfo {author} {\bibfnamefont {Elizabeth~L.}\
  \bibnamefont {Read}}, \bibinfo {author} {\bibfnamefont {Tae-Kyu}\
  \bibnamefont {Ahn}}, \bibinfo {author} {\bibfnamefont {Tom\'as}\ \bibnamefont
  {Mancal}}, \bibinfo {author} {\bibfnamefont {Yuan-Chung}\ \bibnamefont
  {Cheng}}, \bibinfo {author} {\bibfnamefont {Robert~E.}\ \bibnamefont
  {Blankenship}}, \ and\ \bibinfo {author} {\bibfnamefont {Graham~R.}\
  \bibnamefont {Fleming}},\ }\bibfield  {title} {\enquote {\bibinfo {title}
  {Evidence for wavelike energy transfer through quantum coherence in
  photosynthetic systems},}\ }\href@noop {} {\bibfield  {journal} {\bibinfo
  {journal} {Nature (London)}\ }\textbf {\bibinfo {volume} {446}},\ \bibinfo
  {pages} {782} (\bibinfo {year} {2007})}\BibitemShut {NoStop}%
\bibitem [{\citenamefont {Collini}\ \emph {et~al.}(2010)\citenamefont
  {Collini}, \citenamefont {Wong}, \citenamefont {Wilk}, \citenamefont {Curmi},
  \citenamefont {Brumer},\ and\ \citenamefont {Scholes}}]{Collini_Scholes2010}%
  \BibitemOpen
  \bibfield  {author} {\bibinfo {author} {\bibfnamefont {Elisabetta}\
  \bibnamefont {Collini}}, \bibinfo {author} {\bibfnamefont {Cathy~Y.}\
  \bibnamefont {Wong}}, \bibinfo {author} {\bibfnamefont {Krystyna~E.}\
  \bibnamefont {Wilk}}, \bibinfo {author} {\bibfnamefont {Paul M.~G.}\
  \bibnamefont {Curmi}}, \bibinfo {author} {\bibfnamefont {Paul}\ \bibnamefont
  {Brumer}}, \ and\ \bibinfo {author} {\bibfnamefont {Gregory~D.}\ \bibnamefont
  {Scholes}},\ }\bibfield  {title} {\enquote {\bibinfo {title} {Coherently
  wired light-harvesting in photosynthetic marine algae at ambient
  temperature},}\ }\href@noop {} {\bibfield  {journal} {\bibinfo  {journal}
  {Nature (London)}\ }\textbf {\bibinfo {volume} {463}},\ \bibinfo {pages}
  {644} (\bibinfo {year} {2010})}\BibitemShut {NoStop}%
\bibitem [{\citenamefont {Mohseni}\ \emph {et~al.}(2008)\citenamefont
  {Mohseni}, \citenamefont {Rebentrost}, \citenamefont {Lloyd},\ and\
  \citenamefont {Aspuru-Guzik'}}]{Lloyd_2008}%
  \BibitemOpen
  \bibfield  {author} {\bibinfo {author} {\bibfnamefont {Masoud}\ \bibnamefont
  {Mohseni}}, \bibinfo {author} {\bibfnamefont {Patrick}\ \bibnamefont
  {Rebentrost}}, \bibinfo {author} {\bibfnamefont {Seth}\ \bibnamefont
  {Lloyd}}, \ and\ \bibinfo {author} {\bibfnamefont {Al\'an}\ \bibnamefont
  {Aspuru-Guzik'}},\ }\bibfield  {title} {\enquote {\bibinfo {title}
  {Environment-assisted quantum walk in photosynthetic energy transfer},}\
  }\href@noop {} {\bibfield  {journal} {\bibinfo  {journal} {J. Chem. Phys.}\
  }\textbf {\bibinfo {volume} {129}},\ \bibinfo {pages} {174106} (\bibinfo
  {year} {2008})}\BibitemShut {NoStop}%
\bibitem [{\citenamefont {Plenio}\ and\ \citenamefont
  {Huelga}(2008)}]{PlenioHuelga2008}%
  \BibitemOpen
  \bibfield  {author} {\bibinfo {author} {\bibfnamefont {M.~B.}\ \bibnamefont
  {Plenio}}\ and\ \bibinfo {author} {\bibfnamefont {S.~F.}\ \bibnamefont
  {Huelga}},\ }\bibfield  {title} {\enquote {\bibinfo {title}
  {Dephasing-assisted transport: quantum networks and biomolecules},}\
  }\href@noop {} {\bibfield  {journal} {\bibinfo  {journal} {New J. Phys.}\
  }\textbf {\bibinfo {volume} {10}},\ \bibinfo {pages} {113019} (\bibinfo
  {year} {2008})}\BibitemShut {NoStop}%
\bibitem [{\citenamefont {Baumgratz}\ \emph {et~al.}(2014)\citenamefont
  {Baumgratz}, \citenamefont {Cramer},\ and\ \citenamefont
  {Plenio}}]{Plenio_2014}%
  \BibitemOpen
  \bibfield  {author} {\bibinfo {author} {\bibfnamefont {T.}~\bibnamefont
  {Baumgratz}}, \bibinfo {author} {\bibfnamefont {M.}~\bibnamefont {Cramer}}, \
  and\ \bibinfo {author} {\bibfnamefont {M.~B.}\ \bibnamefont {Plenio}},\
  }\bibfield  {title} {\enquote {\bibinfo {title} {Quantifying coherence},}\
  }\href@noop {} {\bibfield  {journal} {\bibinfo  {journal} {Phys. Rev. Lett.}\
  }\textbf {\bibinfo {volume} {113}},\ \bibinfo {pages} {140401} (\bibinfo
  {year} {2014})}\BibitemShut {NoStop}%
\bibitem [{\citenamefont {Lostaglio}\ \emph
  {et~al.}(2015{\natexlab{a}})\citenamefont {Lostaglio}, \citenamefont
  {Korzekwa}, \citenamefont {Jennings},\ and\ \citenamefont
  {Rudolph}}]{Lostaglio_2015}%
  \BibitemOpen
  \bibfield  {author} {\bibinfo {author} {\bibfnamefont {Matteo}\ \bibnamefont
  {Lostaglio}}, \bibinfo {author} {\bibfnamefont {Kamil}\ \bibnamefont
  {Korzekwa}}, \bibinfo {author} {\bibfnamefont {David}\ \bibnamefont
  {Jennings}}, \ and\ \bibinfo {author} {\bibfnamefont {Terry}\ \bibnamefont
  {Rudolph}},\ }\bibfield  {title} {\enquote {\bibinfo {title} {Quantum
  coherence, time-translation symmetry, and thermodynamics},}\ }\href@noop {}
  {\bibfield  {journal} {\bibinfo  {journal} {Phys. Rev. X}\ }\textbf {\bibinfo
  {volume} {5}},\ \bibinfo {pages} {021001} (\bibinfo {year}
  {2015}{\natexlab{a}})}\BibitemShut {NoStop}%
\bibitem [{\citenamefont {Winter}\ and\ \citenamefont
  {Yang}(2016)}]{WinterYang2016}%
  \BibitemOpen
  \bibfield  {author} {\bibinfo {author} {\bibfnamefont {Andreas}\ \bibnamefont
  {Winter}}\ and\ \bibinfo {author} {\bibfnamefont {Dong}\ \bibnamefont
  {Yang}},\ }\bibfield  {title} {\enquote {\bibinfo {title} {Operational
  resource theory of coherence},}\ }\href@noop {} {\bibfield  {journal}
  {\bibinfo  {journal} {Phys. Rev. Lett.}\ }\textbf {\bibinfo {volume} {116}},\
  \bibinfo {pages} {120404} (\bibinfo {year} {2016})}\BibitemShut {NoStop}%
\bibitem [{\citenamefont {Streltsov}\ \emph {et~al.}(2016)\citenamefont
  {Streltsov}, \citenamefont {Adesso},\ and\ \citenamefont
  {Plenio}}]{Plenio_2016}%
  \BibitemOpen
  \bibfield  {author} {\bibinfo {author} {\bibfnamefont {Alexander}\
  \bibnamefont {Streltsov}}, \bibinfo {author} {\bibfnamefont {Gerardo}\
  \bibnamefont {Adesso}}, \ and\ \bibinfo {author} {\bibfnamefont {Martin~B.}\
  \bibnamefont {Plenio}},\ }\bibfield  {title} {\enquote {\bibinfo {title}
  {Colloquium: Quantum coherence as a resource},}\ }\href@noop {} {\bibfield
  {journal} {\bibinfo  {journal} {arXiv}\ ,\ \bibinfo {pages} {1609.02439}}
  (\bibinfo {year} {2016})}\BibitemShut {NoStop}%
\bibitem [{\citenamefont {Horodecki}\ and\ \citenamefont
  {Oppenheim}(2013)}]{HorodeckiOppenheim2013}%
  \BibitemOpen
  \bibfield  {author} {\bibinfo {author} {\bibfnamefont {Michal}\ \bibnamefont
  {Horodecki}}\ and\ \bibinfo {author} {\bibfnamefont {Jonathan}\ \bibnamefont
  {Oppenheim}},\ }\bibfield  {title} {\enquote {\bibinfo {title} {Fundamental
  limitaions for quantum and nanoscale thermodynamics},}\ }\href@noop {}
  {\bibfield  {journal} {\bibinfo  {journal} {Nat. Comm.}\ }\textbf {\bibinfo
  {volume} {4}},\ \bibinfo {pages} {2059} (\bibinfo {year} {2013})}\BibitemShut
  {NoStop}%
\bibitem [{\citenamefont {Skrzypczyk}\ \emph {et~al.}(2014)\citenamefont
  {Skrzypczyk}, \citenamefont {Short},\ and\ \citenamefont
  {Popescu}}]{Skrzypczyk2014}%
  \BibitemOpen
  \bibfield  {author} {\bibinfo {author} {\bibfnamefont {Paul}\ \bibnamefont
  {Skrzypczyk}}, \bibinfo {author} {\bibfnamefont {Anthony~J.}\ \bibnamefont
  {Short}}, \ and\ \bibinfo {author} {\bibfnamefont {Sandu}\ \bibnamefont
  {Popescu}},\ }\bibfield  {title} {\enquote {\bibinfo {title} {Work extraction
  and thermodynamics for individual quantum systems},}\ }\href@noop {}
  {\bibfield  {journal} {\bibinfo  {journal} {Nat. Commun.}\ }\textbf {\bibinfo
  {volume} {5}},\ \bibinfo {pages} {4185} (\bibinfo {year} {2014})}\BibitemShut
  {NoStop}%
\bibitem [{\citenamefont {Lostaglio}\ \emph
  {et~al.}(2015{\natexlab{b}})\citenamefont {Lostaglio}, \citenamefont
  {Jennings},\ and\ \citenamefont {Rudolph}}]{Lostaglio_2015nc}%
  \BibitemOpen
  \bibfield  {author} {\bibinfo {author} {\bibfnamefont {Matteo}\ \bibnamefont
  {Lostaglio}}, \bibinfo {author} {\bibfnamefont {David}\ \bibnamefont
  {Jennings}}, \ and\ \bibinfo {author} {\bibfnamefont {Terry}\ \bibnamefont
  {Rudolph}},\ }\bibfield  {title} {\enquote {\bibinfo {title} {Description of
  quantum coherence in thermodynamic processes requires constraints beyond free
  energy},}\ }\href@noop {} {\bibfield  {journal} {\bibinfo  {journal} {Nat.
  Commun.}\ }\textbf {\bibinfo {volume} {6}},\ \bibinfo {pages} {6383}
  (\bibinfo {year} {2015}{\natexlab{b}})}\BibitemShut {NoStop}%
\bibitem [{\citenamefont {{\AA}berg}(2014)}]{Aberg2014}%
  \BibitemOpen
  \bibfield  {author} {\bibinfo {author} {\bibfnamefont {Johan}\ \bibnamefont
  {{\AA}berg}},\ }\bibfield  {title} {\enquote {\bibinfo {title} {Catalytic
  coherence},}\ }\href@noop {} {\bibfield  {journal} {\bibinfo  {journal}
  {Phys. Rev. Lett.}\ }\textbf {\bibinfo {volume} {113}},\ \bibinfo {pages}
  {150402} (\bibinfo {year} {2014})}\BibitemShut {NoStop}%
\bibitem [{\citenamefont {Korzekwa}\ \emph {et~al.}(2016)\citenamefont
  {Korzekwa}, \citenamefont {Lostaglio}, \citenamefont {Oppenheim},\ and\
  \citenamefont {Jennings}}]{Korzekwa2016}%
  \BibitemOpen
  \bibfield  {author} {\bibinfo {author} {\bibfnamefont {Kamil}\ \bibnamefont
  {Korzekwa}}, \bibinfo {author} {\bibfnamefont {Matteo}\ \bibnamefont
  {Lostaglio}}, \bibinfo {author} {\bibfnamefont {Jonathan}\ \bibnamefont
  {Oppenheim}}, \ and\ \bibinfo {author} {\bibfnamefont {David}\ \bibnamefont
  {Jennings}},\ }\bibfield  {title} {\enquote {\bibinfo {title} {The extraction
  of work from quantum coherence},}\ }\href@noop {} {\bibfield  {journal}
  {\bibinfo  {journal} {New J. Phys.}\ }\textbf {\bibinfo {volume} {18}},\
  \bibinfo {pages} {023045} (\bibinfo {year} {2016})}\BibitemShut {NoStop}%
\bibitem [{\citenamefont {An}\ \emph {et~al.}(2015)\citenamefont {An},
  \citenamefont {Zhang}, \citenamefont {Um}, \citenamefont {Lv}, \citenamefont
  {Lu}, \citenamefont {Zhang}, \citenamefont {Yin}, \citenamefont {Quan},\ and\
  \citenamefont {Kim}}]{AZUe15}%
  \BibitemOpen
  \bibfield  {author} {\bibinfo {author} {\bibfnamefont {Shuoming}\
  \bibnamefont {An}}, \bibinfo {author} {\bibfnamefont {Jing-Ning}\
  \bibnamefont {Zhang}}, \bibinfo {author} {\bibfnamefont {Mark}\ \bibnamefont
  {Um}}, \bibinfo {author} {\bibfnamefont {Dingshun}\ \bibnamefont {Lv}},
  \bibinfo {author} {\bibfnamefont {Yao}\ \bibnamefont {Lu}}, \bibinfo {author}
  {\bibfnamefont {Junhua}\ \bibnamefont {Zhang}}, \bibinfo {author}
  {\bibfnamefont {Zhang-Qi}\ \bibnamefont {Yin}}, \bibinfo {author}
  {\bibfnamefont {H.~T.}\ \bibnamefont {Quan}}, \ and\ \bibinfo {author}
  {\bibfnamefont {Kihwan}\ \bibnamefont {Kim}},\ }\bibfield  {title} {\enquote
  {\bibinfo {title} {Experimental test of the quantum jarzsynki equality with a
  trapped-ion system},}\ }\href@noop {} {\bibfield  {journal} {\bibinfo
  {journal} {Nat. Phys.}\ }\textbf {\bibinfo {volume} {11}},\ \bibinfo {pages}
  {193--199} (\bibinfo {year} {2015})}\BibitemShut {NoStop}%
\bibitem [{\citenamefont {Tasaki}(2000)}]{Tasaki2000}%
  \BibitemOpen
  \bibfield  {author} {\bibinfo {author} {\bibfnamefont {Hal}\ \bibnamefont
  {Tasaki}},\ }\bibfield  {title} {\enquote {\bibinfo {title} {Jarzynski
  realtions for quantum systems and some applications},}\ }\href@noop {}
  {\bibfield  {journal} {\bibinfo  {journal} {arXiv}\ ,\ \bibinfo {pages}
  {cond--mat/0009244}} (\bibinfo {year} {2000})}\BibitemShut {NoStop}%
\bibitem [{\citenamefont {Mukamel}(2003)}]{Mukamel2003}%
  \BibitemOpen
  \bibfield  {author} {\bibinfo {author} {\bibfnamefont {Shaul}\ \bibnamefont
  {Mukamel}},\ }\bibfield  {title} {\enquote {\bibinfo {title} {Quantum
  extension of the jarzynski relation: Analogy with stochastic dephasing},}\
  }\href@noop {} {\bibfield  {journal} {\bibinfo  {journal} {Phys. Rev. Lett.}\
  }\textbf {\bibinfo {volume} {90}},\ \bibinfo {pages} {170604} (\bibinfo
  {year} {2003})}\BibitemShut {NoStop}%
\bibitem [{\citenamefont {{De Roeck}}\ and\ \citenamefont
  {Maes}(2004)}]{DeRoeckMaes2004}%
  \BibitemOpen
  \bibfield  {author} {\bibinfo {author} {\bibfnamefont {Wojciech}\
  \bibnamefont {{De Roeck}}}\ and\ \bibinfo {author} {\bibfnamefont
  {Christian}\ \bibnamefont {Maes}},\ }\bibfield  {title} {\enquote {\bibinfo
  {title} {Quantum version of free-energy--irreversible-work relations},}\
  }\href@noop {} {\bibfield  {journal} {\bibinfo  {journal} {Phys. Rev. E}\
  }\textbf {\bibinfo {volume} {69}},\ \bibinfo {pages} {026115} (\bibinfo
  {year} {2004})}\BibitemShut {NoStop}%
\bibitem [{\citenamefont {Esposito}\ and\ \citenamefont
  {Mukamel}(2006)}]{EspositoMukamel2006}%
  \BibitemOpen
  \bibfield  {author} {\bibinfo {author} {\bibfnamefont {Massimiliano}\
  \bibnamefont {Esposito}}\ and\ \bibinfo {author} {\bibfnamefont {Shaul}\
  \bibnamefont {Mukamel}},\ }\bibfield  {title} {\enquote {\bibinfo {title}
  {Fluctuation theorems for quantum master equations},}\ }\href@noop {}
  {\bibfield  {journal} {\bibinfo  {journal} {Phys. Rev. E}\ }\textbf {\bibinfo
  {volume} {73}},\ \bibinfo {pages} {046129} (\bibinfo {year}
  {2006})}\BibitemShut {NoStop}%
\bibitem [{\citenamefont {Talkner}\ \emph {et~al.}(2007)\citenamefont
  {Talkner}, \citenamefont {Lutz},\ and\ \citenamefont
  {H\"anggi}}]{TalknerLutzHanggi2007}%
  \BibitemOpen
  \bibfield  {author} {\bibinfo {author} {\bibfnamefont {Peter}\ \bibnamefont
  {Talkner}}, \bibinfo {author} {\bibfnamefont {Eric}\ \bibnamefont {Lutz}}, \
  and\ \bibinfo {author} {\bibfnamefont {Peter}\ \bibnamefont {H\"anggi}},\
  }\bibfield  {title} {\enquote {\bibinfo {title} {Fluctuation theorems: Work
  is not an observable},}\ }\href@noop {} {\bibfield  {journal} {\bibinfo
  {journal} {Phys. Rev. E}\ }\textbf {\bibinfo {volume} {75}},\ \bibinfo
  {pages} {050102(R)} (\bibinfo {year} {2007})}\BibitemShut {NoStop}%
\bibitem [{\citenamefont {Talkner}\ and\ \citenamefont
  {H\"anggi}(2007)}]{TalknerHanggi2007}%
  \BibitemOpen
  \bibfield  {author} {\bibinfo {author} {\bibfnamefont {Peter}\ \bibnamefont
  {Talkner}}\ and\ \bibinfo {author} {\bibfnamefont {Peter}\ \bibnamefont
  {H\"anggi}},\ }\bibfield  {title} {\enquote {\bibinfo {title} {The
  tasaki-crooks quantum fluctuation theorem},}\ }\href@noop {} {\bibfield
  {journal} {\bibinfo  {journal} {J. Phys. A: Math. Theor.}\ }\textbf {\bibinfo
  {volume} {40}},\ \bibinfo {pages} {F569--F571} (\bibinfo {year}
  {2007})}\BibitemShut {NoStop}%
\bibitem [{\citenamefont {Crooks}(2008)}]{Crooks2008}%
  \BibitemOpen
  \bibfield  {author} {\bibinfo {author} {\bibfnamefont {Gavin~E.}\
  \bibnamefont {Crooks}},\ }\bibfield  {title} {\enquote {\bibinfo {title}
  {Quantum operation time reversal},}\ }\href@noop {} {\bibfield  {journal}
  {\bibinfo  {journal} {Phys. Rev. A}\ }\textbf {\bibinfo {volume} {77}},\
  \bibinfo {pages} {034101} (\bibinfo {year} {2008})}\BibitemShut {NoStop}%
\bibitem [{\citenamefont {Campisi}\ \emph {et~al.}(2009)\citenamefont
  {Campisi}, \citenamefont {Talkner},\ and\ \citenamefont
  {H\"anggi}}]{CampisiTalknerHanggi2009}%
  \BibitemOpen
  \bibfield  {author} {\bibinfo {author} {\bibfnamefont {Michele}\ \bibnamefont
  {Campisi}}, \bibinfo {author} {\bibfnamefont {Peter}\ \bibnamefont
  {Talkner}}, \ and\ \bibinfo {author} {\bibfnamefont {Peter}\ \bibnamefont
  {H\"anggi}},\ }\bibfield  {title} {\enquote {\bibinfo {title} {Fluctuation
  theorem for arbitrary open quantum systems},}\ }\href@noop {} {\bibfield
  {journal} {\bibinfo  {journal} {Phys. Rev. Lett.}\ }\textbf {\bibinfo
  {volume} {102}},\ \bibinfo {pages} {210401} (\bibinfo {year}
  {2009})}\BibitemShut {NoStop}%
\bibitem [{\citenamefont {Esposito}\ \emph {et~al.}(2009)\citenamefont
  {Esposito}, \citenamefont {Harbola},\ and\ \citenamefont
  {Mukamel}}]{EspositoHarbolaMukamel2009}%
  \BibitemOpen
  \bibfield  {author} {\bibinfo {author} {\bibfnamefont {Massimiliano}\
  \bibnamefont {Esposito}}, \bibinfo {author} {\bibfnamefont {Upendra}\
  \bibnamefont {Harbola}}, \ and\ \bibinfo {author} {\bibfnamefont {Shaul}\
  \bibnamefont {Mukamel}},\ }\bibfield  {title} {\enquote {\bibinfo {title}
  {Nonequilibrium fluctuations, fluctuation theorems, and counting statistics
  in quantum systems},}\ }\href@noop {} {\bibfield  {journal} {\bibinfo
  {journal} {Rev. Mod. Phys.}\ }\textbf {\bibinfo {volume} {81}},\ \bibinfo
  {pages} {1665} (\bibinfo {year} {2009})}\BibitemShut {NoStop}%
\bibitem [{\citenamefont {Campisi}\ \emph {et~al.}(2011)\citenamefont
  {Campisi}, \citenamefont {H\"anggi},\ and\ \citenamefont
  {Talkner}}]{CampisiHanggiTalkner2011}%
  \BibitemOpen
  \bibfield  {author} {\bibinfo {author} {\bibfnamefont {Michele}\ \bibnamefont
  {Campisi}}, \bibinfo {author} {\bibfnamefont {Peter}\ \bibnamefont
  {H\"anggi}}, \ and\ \bibinfo {author} {\bibfnamefont {Peter}\ \bibnamefont
  {Talkner}},\ }\bibfield  {title} {\enquote {\bibinfo {title}
  {\textit{Colloquium} : Quantum fluctuation relations: Foundations and
  applications},}\ }\href@noop {} {\bibfield  {journal} {\bibinfo  {journal}
  {Rev. Mod. Phys.}\ }\textbf {\bibinfo {volume} {83}},\ \bibinfo {pages}
  {1653} (\bibinfo {year} {2011})}\BibitemShut {NoStop}%
\bibitem [{\citenamefont {Jarzynski}(1997)}]{Jarzynski1997}%
  \BibitemOpen
  \bibfield  {author} {\bibinfo {author} {\bibfnamefont {C.}~\bibnamefont
  {Jarzynski}},\ }\bibfield  {title} {\enquote {\bibinfo {title}
  {Nonequilibrium equality for free energy difference},}\ }\href@noop {}
  {\bibfield  {journal} {\bibinfo  {journal} {Phys. Rev. Lett.}\ }\textbf
  {\bibinfo {volume} {78}},\ \bibinfo {pages} {2690--2693} (\bibinfo {year}
  {1997})}\BibitemShut {NoStop}%
\bibitem [{\citenamefont {Crooks}(1999)}]{Crooks1999}%
  \BibitemOpen
  \bibfield  {author} {\bibinfo {author} {\bibfnamefont {Gavin~E.}\
  \bibnamefont {Crooks}},\ }\bibfield  {title} {\enquote {\bibinfo {title}
  {Entropy production fluctuation theorem and the nonequilibrium work relation
  for free energy difference},}\ }\href@noop {} {\bibfield  {journal} {\bibinfo
   {journal} {Phys. Rev. E}\ }\textbf {\bibinfo {volume} {60}},\ \bibinfo
  {pages} {2721--2726} (\bibinfo {year} {1999})}\BibitemShut {NoStop}%
\bibitem [{\citenamefont {Seifert}(2005)}]{Seifert2005}%
  \BibitemOpen
  \bibfield  {author} {\bibinfo {author} {\bibfnamefont {Udo}\ \bibnamefont
  {Seifert}},\ }\bibfield  {title} {\enquote {\bibinfo {title} {Entropy
  production along a stochastic trajectory and an integral fluctuation
  theorem},}\ }\href@noop {} {\bibfield  {journal} {\bibinfo  {journal} {Phys.
  Rev. Lett.}\ }\textbf {\bibinfo {volume} {95}},\ \bibinfo {pages} {040602}
  (\bibinfo {year} {2005})}\BibitemShut {NoStop}%
\bibitem [{\citenamefont {Jarzynski}(2011)}]{Jarzynski2011}%
  \BibitemOpen
  \bibfield  {author} {\bibinfo {author} {\bibfnamefont {C.}~\bibnamefont
  {Jarzynski}},\ }\bibfield  {title} {\enquote {\bibinfo {title} {Equalities
  and inequalities: Irreversibility and the second law of thermodynamics at the
  nanoscale},}\ }\href@noop {} {\bibfield  {journal} {\bibinfo  {journal}
  {Annual Review of Condensed Matter Physics}\ }\textbf {\bibinfo {volume}
  {2}},\ \bibinfo {pages} {329} (\bibinfo {year} {2011})}\BibitemShut {NoStop}%
\bibitem [{\citenamefont {Seifert}(2012)}]{Seifert2012}%
  \BibitemOpen
  \bibfield  {author} {\bibinfo {author} {\bibfnamefont {Udo}\ \bibnamefont
  {Seifert}},\ }\bibfield  {title} {\enquote {\bibinfo {title} {Stochastic
  thermodynamics, fluctuation theorems and molecular machines},}\ }\href@noop
  {} {\bibfield  {journal} {\bibinfo  {journal} {Rep. Prog. Phys.}\ }\textbf
  {\bibinfo {volume} {75}},\ \bibinfo {pages} {126001} (\bibinfo {year}
  {2012})}\BibitemShut {NoStop}%
\bibitem [{\citenamefont {Van~den Broeck}\ and\ \citenamefont
  {Esposito}(2015)}]{EspositoVandenBroeck2015}%
  \BibitemOpen
  \bibfield  {author} {\bibinfo {author} {\bibfnamefont {C.}~\bibnamefont
  {Van~den Broeck}}\ and\ \bibinfo {author} {\bibfnamefont {M.}~\bibnamefont
  {Esposito}},\ }\bibfield  {title} {\enquote {\bibinfo {title} {Ensemble and
  trajectory thermodynamics: A brief introduction},}\ }\href {\doibase
  10.1016/j.physa.2014.04.035} {\bibfield  {journal} {\bibinfo  {journal}
  {Physica A}\ }\textbf {\bibinfo {volume} {418}},\ \bibinfo {pages} {6}
  (\bibinfo {year} {2015})}\BibitemShut {NoStop}%
\bibitem [{\citenamefont {Parrondo}\ \emph {et~al.}(2015)\citenamefont
  {Parrondo}, \citenamefont {Horowitz},\ and\ \citenamefont
  {Sagawa}}]{ParrondoHorowitzSagawa2015}%
  \BibitemOpen
  \bibfield  {author} {\bibinfo {author} {\bibfnamefont {Juan M.~R.}\
  \bibnamefont {Parrondo}}, \bibinfo {author} {\bibfnamefont {Jordan~M.}\
  \bibnamefont {Horowitz}}, \ and\ \bibinfo {author} {\bibfnamefont {Takahiro}\
  \bibnamefont {Sagawa}},\ }\bibfield  {title} {\enquote {\bibinfo {title}
  {Thermodynamics of information},}\ }\href@noop {} {\bibfield  {journal}
  {\bibinfo  {journal} {Nature Phys.}\ }\textbf {\bibinfo {volume} {11}},\
  \bibinfo {pages} {131} (\bibinfo {year} {2015})}\BibitemShut {NoStop}%
\bibitem [{\citenamefont {Maruyama}\ \emph {et~al.}(2009)\citenamefont
  {Maruyama}, \citenamefont {Nori},\ and\ \citenamefont
  {Vedral}}]{MaruyamaNoriVedral2009}%
  \BibitemOpen
  \bibfield  {author} {\bibinfo {author} {\bibfnamefont {Koji}\ \bibnamefont
  {Maruyama}}, \bibinfo {author} {\bibfnamefont {Franco}\ \bibnamefont {Nori}},
  \ and\ \bibinfo {author} {\bibfnamefont {Vlatko}\ \bibnamefont {Vedral}},\
  }\bibfield  {title} {\enquote {\bibinfo {title} {Colloquium: The physics of
  maxwell's demon and information},}\ }\href@noop {} {\bibfield  {journal}
  {\bibinfo  {journal} {Rev. Mod. Phys.}\ }\textbf {\bibinfo {volume} {81}},\
  \bibinfo {pages} {1} (\bibinfo {year} {2009})}\BibitemShut {NoStop}%
\bibitem [{\citenamefont {Sagawa}(2012)}]{Sagawa2012}%
  \BibitemOpen
  \bibfield  {author} {\bibinfo {author} {\bibfnamefont {Takahiro}\
  \bibnamefont {Sagawa}},\ }\href@noop {} {\emph {\bibinfo {title}
  {Thermodynamics of Information Processing in Small Systems}}}\ (\bibinfo
  {publisher} {Springer},\ \bibinfo {year} {2012})\BibitemShut {NoStop}%
\bibitem [{\citenamefont {Sagawa}\ and\ \citenamefont
  {Ueda}(2008)}]{SagawaUeda2008}%
  \BibitemOpen
  \bibfield  {author} {\bibinfo {author} {\bibfnamefont {Takahiro}\
  \bibnamefont {Sagawa}}\ and\ \bibinfo {author} {\bibfnamefont {Masahito}\
  \bibnamefont {Ueda}},\ }\bibfield  {title} {\enquote {\bibinfo {title}
  {Second law of thermodynamics with discrete quantum feedback control},}\
  }\href@noop {} {\bibfield  {journal} {\bibinfo  {journal} {Phys. Rev. Lett.}\
  }\textbf {\bibinfo {volume} {100}},\ \bibinfo {pages} {080403} (\bibinfo
  {year} {2008})}\BibitemShut {NoStop}%
\bibitem [{\citenamefont {Sagawa}\ and\ \citenamefont
  {Ueda}(2009)}]{SagawaUeda2009}%
  \BibitemOpen
  \bibfield  {author} {\bibinfo {author} {\bibfnamefont {Takahiro}\
  \bibnamefont {Sagawa}}\ and\ \bibinfo {author} {\bibfnamefont {Masahito}\
  \bibnamefont {Ueda}},\ }\bibfield  {title} {\enquote {\bibinfo {title}
  {Minimal energy cost for thermodynamics information processing: Measurement
  and information erasure},}\ }\href@noop {} {\bibfield  {journal} {\bibinfo
  {journal} {Phys. Rev. Lett.}\ }\textbf {\bibinfo {volume} {102}},\ \bibinfo
  {pages} {250602} (\bibinfo {year} {2009})}\BibitemShut {NoStop}%
\bibitem [{\citenamefont {Sagawa}\ and\ \citenamefont
  {Ueda}(2010)}]{SagawaUeda2010}%
  \BibitemOpen
  \bibfield  {author} {\bibinfo {author} {\bibfnamefont {Takahiro}\
  \bibnamefont {Sagawa}}\ and\ \bibinfo {author} {\bibfnamefont {Masahito}\
  \bibnamefont {Ueda}},\ }\bibfield  {title} {\enquote {\bibinfo {title}
  {Generalized jarzynski equality under nonequilibrium feedback control},}\
  }\href@noop {} {\bibfield  {journal} {\bibinfo  {journal} {Phys. Rev. Lett.}\
  }\textbf {\bibinfo {volume} {104}},\ \bibinfo {pages} {090602} (\bibinfo
  {year} {2010})}\BibitemShut {NoStop}%
\bibitem [{\citenamefont {Horowitz}\ and\ \citenamefont
  {Vaikuntanathan}(2010)}]{HorowitzVaikuntanathan2010}%
  \BibitemOpen
  \bibfield  {author} {\bibinfo {author} {\bibfnamefont {Jordan~M.}\
  \bibnamefont {Horowitz}}\ and\ \bibinfo {author} {\bibfnamefont
  {Suriyanarayanan}\ \bibnamefont {Vaikuntanathan}},\ }\bibfield  {title}
  {\enquote {\bibinfo {title} {Nonequilibrium detailed fluctuation theorem for
  repeated discrete feedback},}\ }\href@noop {} {\bibfield  {journal} {\bibinfo
   {journal} {Phys. Rev. E}\ }\textbf {\bibinfo {volume} {82}},\ \bibinfo
  {pages} {061120} (\bibinfo {year} {2010})}\BibitemShut {NoStop}%
\bibitem [{\citenamefont {Sagawa}\ and\ \citenamefont
  {Ueda}(2012)}]{SagawaUeda2012PRE}%
  \BibitemOpen
  \bibfield  {author} {\bibinfo {author} {\bibfnamefont {Takahiro}\
  \bibnamefont {Sagawa}}\ and\ \bibinfo {author} {\bibfnamefont {Masahito}\
  \bibnamefont {Ueda}},\ }\bibfield  {title} {\enquote {\bibinfo {title}
  {Nonequilibrium thermodynamics of feedback control},}\ }\href@noop {}
  {\bibfield  {journal} {\bibinfo  {journal} {Phys. Rev. E}\ }\textbf {\bibinfo
  {volume} {85}},\ \bibinfo {pages} {021104} (\bibinfo {year}
  {2012})}\BibitemShut {NoStop}%
\bibitem [{\citenamefont {Abreu}\ and\ \citenamefont
  {Seifert}(2012)}]{AbreuSeifert2012}%
  \BibitemOpen
  \bibfield  {author} {\bibinfo {author} {\bibfnamefont {D.}~\bibnamefont
  {Abreu}}\ and\ \bibinfo {author} {\bibfnamefont {U.}~\bibnamefont
  {Seifert}},\ }\bibfield  {title} {\enquote {\bibinfo {title} {Thermodynamics
  of genuine non-equilibrium states under feedback control},}\ }\href@noop {}
  {\bibfield  {journal} {\bibinfo  {journal} {Phys. Rev. Lett.}\ }\textbf
  {\bibinfo {volume} {108}},\ \bibinfo {pages} {030601} (\bibinfo {year}
  {2012})}\BibitemShut {NoStop}%
\bibitem [{\citenamefont {Lahiri}\ \emph {et~al.}(2012)\citenamefont {Lahiri},
  \citenamefont {Rana},\ and\ \citenamefont
  {Jayannavar}}]{LahiriRanaJayannavar2012}%
  \BibitemOpen
  \bibfield  {author} {\bibinfo {author} {\bibfnamefont {S.}~\bibnamefont
  {Lahiri}}, \bibinfo {author} {\bibfnamefont {S.}~\bibnamefont {Rana}}, \ and\
  \bibinfo {author} {\bibfnamefont {A.~M.}\ \bibnamefont {Jayannavar}},\
  }\bibfield  {title} {\enquote {\bibinfo {title} {Fluctuation theorems in the
  presence of information gain and feedback},}\ }\href@noop {} {\bibfield
  {journal} {\bibinfo  {journal} {J. Phys. A: Math. Theor.}\ }\textbf {\bibinfo
  {volume} {45}},\ \bibinfo {pages} {162001} (\bibinfo {year}
  {2012})}\BibitemShut {NoStop}%
\bibitem [{\citenamefont {Funo}\ \emph {et~al.}(2013)\citenamefont {Funo},
  \citenamefont {Watanabe},\ and\ \citenamefont {Ueda}}]{FunoWatanabeUeda2013}%
  \BibitemOpen
  \bibfield  {author} {\bibinfo {author} {\bibfnamefont {Ken}\ \bibnamefont
  {Funo}}, \bibinfo {author} {\bibfnamefont {Yu}~\bibnamefont {Watanabe}}, \
  and\ \bibinfo {author} {\bibfnamefont {Masahito}\ \bibnamefont {Ueda}},\
  }\bibfield  {title} {\enquote {\bibinfo {title} {Integral quantum fluctuation
  theorems under measurement and feedback control},}\ }\href@noop {} {\bibfield
   {journal} {\bibinfo  {journal} {Phys. Rev. E}\ }\textbf {\bibinfo {volume}
  {88}},\ \bibinfo {pages} {052121} (\bibinfo {year} {2013})}\BibitemShut
  {NoStop}%
\bibitem [{\citenamefont {Rastegin}(2013)}]{Rastegin2013}%
  \BibitemOpen
  \bibfield  {author} {\bibinfo {author} {\bibfnamefont {Alexey~E.}\
  \bibnamefont {Rastegin}},\ }\bibfield  {title} {\enquote {\bibinfo {title}
  {Non-equilibrium equalities with unital quantum channels},}\ }\href@noop {}
  {\bibfield  {journal} {\bibinfo  {journal} {J. Stat. Mech.: Theor. Exp.}\ ,\
  \bibinfo {pages} {P06016}} (\bibinfo {year} {2013})}\BibitemShut {NoStop}%
\bibitem [{\citenamefont {Goold}\ \emph {et~al.}(2015)\citenamefont {Goold},
  \citenamefont {Paternostro},\ and\ \citenamefont
  {Modi}}]{GooldPaternostroModi2015}%
  \BibitemOpen
  \bibfield  {author} {\bibinfo {author} {\bibfnamefont {John}\ \bibnamefont
  {Goold}}, \bibinfo {author} {\bibfnamefont {Mauro}\ \bibnamefont
  {Paternostro}}, \ and\ \bibinfo {author} {\bibfnamefont {Kavan}\ \bibnamefont
  {Modi}},\ }\bibfield  {title} {\enquote {\bibinfo {title} {Nonequilibrium
  quantum {L}andauer principle},}\ }\href@noop {} {\bibfield  {journal}
  {\bibinfo  {journal} {Phys. Rev. Lett.}\ }\textbf {\bibinfo {volume} {114}},\
  \bibinfo {pages} {060602} (\bibinfo {year} {2015})}\BibitemShut {NoStop}%
\bibitem [{\citenamefont {Sung}(2005)}]{Sung2005}%
  \BibitemOpen
  \bibfield  {author} {\bibinfo {author} {\bibfnamefont {Jaeyoung}\
  \bibnamefont {Sung}},\ }\bibfield  {title} {\enquote {\bibinfo {title}
  {Validity condition of the jarzynski relation for a classical mechanical
  systems},}\ }\href@noop {} {\bibfield  {journal} {\bibinfo  {journal}
  {arXiv}\ }\textbf {\bibinfo {volume} {cond-mat}},\ \bibinfo {pages} {0506214}
  (\bibinfo {year} {2005})}\BibitemShut {NoStop}%
\bibitem [{\citenamefont {Lua}\ and\ \citenamefont
  {Grosberg}(2005)}]{LuaGrosberg2005}%
  \BibitemOpen
  \bibfield  {author} {\bibinfo {author} {\bibfnamefont {Rhonald~C.}\
  \bibnamefont {Lua}}\ and\ \bibinfo {author} {\bibfnamefont {Alexander~Y.}\
  \bibnamefont {Grosberg}},\ }\bibfield  {title} {\enquote {\bibinfo {title}
  {Practical applicability of the jarzynski relation in statistical mechanics:
  A pedagogical example},}\ }\href@noop {} {\bibfield  {journal} {\bibinfo
  {journal} {J. Phys. Chem. B}\ }\textbf {\bibinfo {volume} {109}},\ \bibinfo
  {pages} {6805} (\bibinfo {year} {2005})}\BibitemShut {NoStop}%
\bibitem [{\citenamefont {Gross}(2005)}]{Gro05Aug}%
  \BibitemOpen
  \bibfield  {author} {\bibinfo {author} {\bibfnamefont {D.~H.~E.}\
  \bibnamefont {Gross}},\ }\bibfield  {title} {\enquote {\bibinfo {title} {Flaw
  of jarzynski's equality when applied to systems with several degrees of
  freedom},}\ }\href@noop {} {\bibfield  {journal} {\bibinfo  {journal}
  {arXiv}\ }\textbf {\bibinfo {volume} {cond-mat}},\ \bibinfo {pages} {0508721}
  (\bibinfo {year} {2005})}\BibitemShut {NoStop}%
\bibitem [{\citenamefont {Jarzynski}(2005)}]{Jar05}%
  \BibitemOpen
  \bibfield  {author} {\bibinfo {author} {\bibfnamefont {C.}~\bibnamefont
  {Jarzynski}},\ }\bibfield  {title} {\enquote {\bibinfo {title} {Reply to
  comments by d.h.e. gross},}\ }\href@noop {} {\bibfield  {journal} {\bibinfo
  {journal} {arXiv}\ }\textbf {\bibinfo {volume} {cond-mat}},\ \bibinfo {pages}
  {0509344} (\bibinfo {year} {2005})}\BibitemShut {NoStop}%
\bibitem [{\citenamefont {Murashita}\ \emph {et~al.}(2014)\citenamefont
  {Murashita}, \citenamefont {Funo},\ and\ \citenamefont
  {Ueda}}]{MurashitaFunoUeda2014}%
  \BibitemOpen
  \bibfield  {author} {\bibinfo {author} {\bibfnamefont {Y\^uto}\ \bibnamefont
  {Murashita}}, \bibinfo {author} {\bibfnamefont {Ken}\ \bibnamefont {Funo}}, \
  and\ \bibinfo {author} {\bibfnamefont {Masahito}\ \bibnamefont {Ueda}},\
  }\bibfield  {title} {\enquote {\bibinfo {title} {Nonequilirium equalities in
  absolutely irreversible processes},}\ }\href@noop {} {\bibfield  {journal}
  {\bibinfo  {journal} {Phys. Rev. E}\ }\textbf {\bibinfo {volume} {90}},\
  \bibinfo {pages} {042110} (\bibinfo {year} {2014})}\BibitemShut {NoStop}%
\bibitem [{\citenamefont {Ashida}\ \emph {et~al.}(2014)\citenamefont {Ashida},
  \citenamefont {Funo}, \citenamefont {Murashita},\ and\ \citenamefont
  {Ueda}}]{AshidaFunoMurashitaUeda2014}%
  \BibitemOpen
  \bibfield  {author} {\bibinfo {author} {\bibfnamefont {Yuto}\ \bibnamefont
  {Ashida}}, \bibinfo {author} {\bibfnamefont {Ken}\ \bibnamefont {Funo}},
  \bibinfo {author} {\bibfnamefont {Y\^uto}\ \bibnamefont {Murashita}}, \ and\
  \bibinfo {author} {\bibfnamefont {Masahito}\ \bibnamefont {Ueda}},\
  }\bibfield  {title} {\enquote {\bibinfo {title} {General achievable bound of
  extractable work under feedback control},}\ }\href@noop {} {\bibfield
  {journal} {\bibinfo  {journal} {Phys. Rev. E}\ }\textbf {\bibinfo {volume}
  {90}},\ \bibinfo {pages} {052125} (\bibinfo {year} {2014})}\BibitemShut
  {NoStop}%
\bibitem [{\citenamefont {Funo}\ \emph {et~al.}(2015)\citenamefont {Funo},
  \citenamefont {Murashita},\ and\ \citenamefont
  {Ueda}}]{FunoMurashitaUeda2015}%
  \BibitemOpen
  \bibfield  {author} {\bibinfo {author} {\bibfnamefont {Ken}\ \bibnamefont
  {Funo}}, \bibinfo {author} {\bibfnamefont {Y\^uto}\ \bibnamefont
  {Murashita}}, \ and\ \bibinfo {author} {\bibfnamefont {Masahito}\
  \bibnamefont {Ueda}},\ }\bibfield  {title} {\enquote {\bibinfo {title}
  {Quantum nonequilibrium equalities with absolute irreversibility},}\
  }\href@noop {} {\bibfield  {journal} {\bibinfo  {journal} {New J. Phys.}\
  }\textbf {\bibinfo {volume} {17}},\ \bibinfo {pages} {075005} (\bibinfo
  {year} {2015})}\BibitemShut {NoStop}%
\bibitem [{\citenamefont {Barreiro}\ \emph {et~al.}(2011)\citenamefont
  {Barreiro}, \citenamefont {M\"uller}, \citenamefont {Schindler},
  \citenamefont {Nigg}, \citenamefont {Monz}, \citenamefont {Chwalla},
  \citenamefont {Hennrich}, \citenamefont {Roos}, \citenamefont {Zoller},\ and\
  \citenamefont {Blatt}}]{Barreiro_Blatt2011}%
  \BibitemOpen
  \bibfield  {author} {\bibinfo {author} {\bibfnamefont {J.~T.}\ \bibnamefont
  {Barreiro}}, \bibinfo {author} {\bibfnamefont {M.}~\bibnamefont {M\"uller}},
  \bibinfo {author} {\bibfnamefont {P.}~\bibnamefont {Schindler}}, \bibinfo
  {author} {\bibfnamefont {D.}~\bibnamefont {Nigg}}, \bibinfo {author}
  {\bibfnamefont {T.}~\bibnamefont {Monz}}, \bibinfo {author} {\bibfnamefont
  {M.}~\bibnamefont {Chwalla}}, \bibinfo {author} {\bibfnamefont
  {M.}~\bibnamefont {Hennrich}}, \bibinfo {author} {\bibfnamefont {C.~F.}\
  \bibnamefont {Roos}}, \bibinfo {author} {\bibfnamefont {P.}~\bibnamefont
  {Zoller}}, \ and\ \bibinfo {author} {\bibfnamefont {R.}~\bibnamefont
  {Blatt}},\ }\bibfield  {title} {\enquote {\bibinfo {title} {An open-system
  quantum simulator with trapped ions},}\ }\href@noop {} {\bibfield  {journal}
  {\bibinfo  {journal} {Nature}\ }\textbf {\bibinfo {volume} {470}},\ \bibinfo
  {pages} {486} (\bibinfo {year} {2011})}\BibitemShut {NoStop}%
\bibitem [{\citenamefont {Petersson}\ \emph {et~al.}(2012)\citenamefont
  {Petersson}, \citenamefont {McFaul}, \citenamefont {Schroer}, \citenamefont
  {Jung}, \citenamefont {Tayler}, \citenamefont {Houck},\ and\ \citenamefont
  {Petta}}]{Petersson_Petta2012}%
  \BibitemOpen
  \bibfield  {author} {\bibinfo {author} {\bibfnamefont {K.~D.}\ \bibnamefont
  {Petersson}}, \bibinfo {author} {\bibfnamefont {L.~W.}\ \bibnamefont
  {McFaul}}, \bibinfo {author} {\bibfnamefont {M.~D.}\ \bibnamefont {Schroer}},
  \bibinfo {author} {\bibfnamefont {M.}~\bibnamefont {Jung}}, \bibinfo {author}
  {\bibfnamefont {J.~M.}\ \bibnamefont {Tayler}}, \bibinfo {author}
  {\bibfnamefont {A.~A.}\ \bibnamefont {Houck}}, \ and\ \bibinfo {author}
  {\bibfnamefont {J.~R.}\ \bibnamefont {Petta}},\ }\bibfield  {title} {\enquote
  {\bibinfo {title} {Circuit quantum electrodynamics with a spin qubit},}\
  }\href@noop {} {\bibfield  {journal} {\bibinfo  {journal} {Nature}\ }\textbf
  {\bibinfo {volume} {490}},\ \bibinfo {pages} {380} (\bibinfo {year}
  {2012})}\BibitemShut {NoStop}%
\bibitem [{\citenamefont {Devoret}\ and\ \citenamefont
  {Schoelkopf}(2013)}]{DevoretSchoelkopf2013}%
  \BibitemOpen
  \bibfield  {author} {\bibinfo {author} {\bibfnamefont {M.~H.}\ \bibnamefont
  {Devoret}}\ and\ \bibinfo {author} {\bibfnamefont {R.~J.}\ \bibnamefont
  {Schoelkopf}},\ }\bibfield  {title} {\enquote {\bibinfo {title}
  {Superconducting circuit for quantum inforamtion},}\ }\href@noop {}
  {\bibfield  {journal} {\bibinfo  {journal} {Science}\ }\textbf {\bibinfo
  {volume} {339}},\ \bibinfo {pages} {1169} (\bibinfo {year}
  {2013})}\BibitemShut {NoStop}%
\bibitem [{\citenamefont {Deffner}\ and\ \citenamefont
  {Lutz}(2011)}]{DeffnerLutz2011}%
  \BibitemOpen
  \bibfield  {author} {\bibinfo {author} {\bibfnamefont {Sebastian}\
  \bibnamefont {Deffner}}\ and\ \bibinfo {author} {\bibfnamefont {Eric}\
  \bibnamefont {Lutz}},\ }\bibfield  {title} {\enquote {\bibinfo {title}
  {Nonequilibrium entropy production for open quantum systems},}\ }\href@noop
  {} {\bibfield  {journal} {\bibinfo  {journal} {Phys. Rev. Lett.}\ }\textbf
  {\bibinfo {volume} {107}},\ \bibinfo {pages} {140404} (\bibinfo {year}
  {2011})}\BibitemShut {NoStop}%
\bibitem [{\citenamefont {Deffner}(2013)}]{Deffner2013}%
  \BibitemOpen
  \bibfield  {author} {\bibinfo {author} {\bibfnamefont {Sebastian}\
  \bibnamefont {Deffner}},\ }\bibfield  {title} {\enquote {\bibinfo {title}
  {Quantum entropy production in phase space},}\ }\href@noop {} {\bibfield
  {journal} {\bibinfo  {journal} {Europhys. Lett.}\ }\textbf {\bibinfo {volume}
  {103}},\ \bibinfo {pages} {30001} (\bibinfo {year} {2013})}\BibitemShut
  {NoStop}%
\bibitem [{\citenamefont {Lorenzo}\ \emph {et~al.}(2015)\citenamefont
  {Lorenzo}, \citenamefont {McCloskey}, \citenamefont {Ciccarello},
  \citenamefont {Paternostro},\ and\ \citenamefont
  {Palma}}]{Lorenzo_Palma2015}%
  \BibitemOpen
  \bibfield  {author} {\bibinfo {author} {\bibfnamefont {S.}~\bibnamefont
  {Lorenzo}}, \bibinfo {author} {\bibfnamefont {R.}~\bibnamefont {McCloskey}},
  \bibinfo {author} {\bibfnamefont {F.}~\bibnamefont {Ciccarello}}, \bibinfo
  {author} {\bibfnamefont {M.}~\bibnamefont {Paternostro}}, \ and\ \bibinfo
  {author} {\bibfnamefont {G.~M.}\ \bibnamefont {Palma}},\ }\bibfield  {title}
  {\enquote {\bibinfo {title} {Landauer's principle in multipartite open
  quantum system dynamics},}\ }\href@noop {} {\bibfield  {journal} {\bibinfo
  {journal} {Phys. Rev. Lett.}\ }\textbf {\bibinfo {volume} {115}},\ \bibinfo
  {pages} {120403} (\bibinfo {year} {2015})}\BibitemShut {NoStop}%
\bibitem [{\citenamefont {Binder}\ \emph {et~al.}(2015)\citenamefont {Binder},
  \citenamefont {Vinjanampathy}, \citenamefont {Modi},\ and\ \citenamefont
  {Goold}}]{Binder_Goold2015}%
  \BibitemOpen
  \bibfield  {author} {\bibinfo {author} {\bibfnamefont {Felix}\ \bibnamefont
  {Binder}}, \bibinfo {author} {\bibfnamefont {Sai}\ \bibnamefont
  {Vinjanampathy}}, \bibinfo {author} {\bibfnamefont {Kavan}\ \bibnamefont
  {Modi}}, \ and\ \bibinfo {author} {\bibfnamefont {John}\ \bibnamefont
  {Goold}},\ }\bibfield  {title} {\enquote {\bibinfo {title} {Quantum
  thermodynamics of general quantum processes},}\ }\href@noop {} {\bibfield
  {journal} {\bibinfo  {journal} {Phys. Rev. E}\ }\textbf {\bibinfo {volume}
  {91}},\ \bibinfo {pages} {032119} (\bibinfo {year} {2015})}\BibitemShut
  {NoStop}%
\bibitem [{\citenamefont {Kutvonen}\ \emph {et~al.}(2016)\citenamefont
  {Kutvonen}, \citenamefont {Sagawa},\ and\ \citenamefont
  {Ala-Nissila}}]{KutvonenSagawaAlaNissila2016}%
  \BibitemOpen
  \bibfield  {author} {\bibinfo {author} {\bibfnamefont {Aki}\ \bibnamefont
  {Kutvonen}}, \bibinfo {author} {\bibfnamefont {Takahiro}\ \bibnamefont
  {Sagawa}}, \ and\ \bibinfo {author} {\bibfnamefont {Tapio}\ \bibnamefont
  {Ala-Nissila}},\ }\bibfield  {title} {\enquote {\bibinfo {title}
  {Thermodynamics of information exchange between two coupled quantum dots},}\
  }\href@noop {} {\bibfield  {journal} {\bibinfo  {journal} {Phys. Rev. E}\
  }\textbf {\bibinfo {volume} {93}},\ \bibinfo {pages} {032147} (\bibinfo
  {year} {2016})}\BibitemShut {NoStop}%
\bibitem [{\citenamefont {Talkner}\ and\ \citenamefont
  {H\"anggi}(2016)}]{TalknerHanggi2016}%
  \BibitemOpen
  \bibfield  {author} {\bibinfo {author} {\bibfnamefont {Peter}\ \bibnamefont
  {Talkner}}\ and\ \bibinfo {author} {\bibfnamefont {Peter}\ \bibnamefont
  {H\"anggi}},\ }\bibfield  {title} {\enquote {\bibinfo {title} {Open system
  trajectories specify fluctuating work but not heat},}\ }\href@noop {}
  {\bibfield  {journal} {\bibinfo  {journal} {Phys. Rev. E}\ }\textbf {\bibinfo
  {volume} {94}},\ \bibinfo {pages} {022143} (\bibinfo {year}
  {2016})}\BibitemShut {NoStop}%
\bibitem [{\citenamefont {Pigeon}\ \emph {et~al.}(2016)\citenamefont {Pigeon},
  \citenamefont {Fusco}, \citenamefont {Xuereb}, \citenamefont {Chiara},\ and\
  \citenamefont {Paternostro}}]{Pigeon_Paternostro2016}%
  \BibitemOpen
  \bibfield  {author} {\bibinfo {author} {\bibfnamefont {Simon}\ \bibnamefont
  {Pigeon}}, \bibinfo {author} {\bibfnamefont {Lorenzo}\ \bibnamefont {Fusco}},
  \bibinfo {author} {\bibfnamefont {Andr\'e}\ \bibnamefont {Xuereb}}, \bibinfo
  {author} {\bibfnamefont {Gabriele~De}\ \bibnamefont {Chiara}}, \ and\
  \bibinfo {author} {\bibfnamefont {Mauro}\ \bibnamefont {Paternostro}},\
  }\bibfield  {title} {\enquote {\bibinfo {title} {Thermodynamics of
  trajectories and local fluctuation theorems for harmonic quantum networks},}\
  }\href@noop {} {\bibfield  {journal} {\bibinfo  {journal} {New J. Phys.}\
  }\textbf {\bibinfo {volume} {18}},\ \bibinfo {pages} {013009} (\bibinfo
  {year} {2016})}\BibitemShut {NoStop}%
\bibitem [{\citenamefont {Alonso}\ \emph {et~al.}(2016)\citenamefont {Alonso},
  \citenamefont {Lutz},\ and\ \citenamefont {Romito}}]{AlonsLutzRomito2016}%
  \BibitemOpen
  \bibfield  {author} {\bibinfo {author} {\bibfnamefont {Jose~Joaquin}\
  \bibnamefont {Alonso}}, \bibinfo {author} {\bibfnamefont {Eric}\ \bibnamefont
  {Lutz}}, \ and\ \bibinfo {author} {\bibfnamefont {Alessandro}\ \bibnamefont
  {Romito}},\ }\bibfield  {title} {\enquote {\bibinfo {title} {Thermodynamics
  of weakly measured quantum systems},}\ }\href@noop {} {\bibfield  {journal}
  {\bibinfo  {journal} {Phys. Rev. Lett.}\ }\textbf {\bibinfo {volume} {116}},\
  \bibinfo {pages} {080403} (\bibinfo {year} {2016})}\BibitemShut {NoStop}%
\bibitem [{\citenamefont {Goold}\ \emph {et~al.}(2016)\citenamefont {Goold},
  \citenamefont {Huber}, \citenamefont {Riera}, \citenamefont {{del Rio}},\
  and\ \citenamefont {Skrzypczyk}}]{Goold_Skrzypczyk2016}%
  \BibitemOpen
  \bibfield  {author} {\bibinfo {author} {\bibfnamefont {John}\ \bibnamefont
  {Goold}}, \bibinfo {author} {\bibfnamefont {Marcus}\ \bibnamefont {Huber}},
  \bibinfo {author} {\bibfnamefont {Arnau}\ \bibnamefont {Riera}}, \bibinfo
  {author} {\bibfnamefont {L\'dia}\ \bibnamefont {{del Rio}}}, \ and\ \bibinfo
  {author} {\bibfnamefont {Paul}\ \bibnamefont {Skrzypczyk}},\ }\bibfield
  {title} {\enquote {\bibinfo {title} {The role of quantum information in
  thermodynamics-a topical review},}\ }\href@noop {} {\bibfield  {journal}
  {\bibinfo  {journal} {J. Phys. A: Math. Theor.}\ }\textbf {\bibinfo {volume}
  {49}},\ \bibinfo {pages} {143001} (\bibinfo {year} {2016})}\BibitemShut
  {NoStop}%
\bibitem [{\citenamefont {Horowitz}(2012)}]{Horowitz2012}%
  \BibitemOpen
  \bibfield  {author} {\bibinfo {author} {\bibfnamefont {Jordan~M.}\
  \bibnamefont {Horowitz}},\ }\bibfield  {title} {\enquote {\bibinfo {title}
  {Quantum-trajectory approach to the stochastic thermodynamics of a forced
  harmonic oscillator},}\ }\href@noop {} {\bibfield  {journal} {\bibinfo
  {journal} {Phys. Rev. E}\ }\textbf {\bibinfo {volume} {85}},\ \bibinfo
  {pages} {031110} (\bibinfo {year} {2012})}\BibitemShut {NoStop}%
\bibitem [{\citenamefont {Leggio}\ \emph {et~al.}(2013)\citenamefont {Leggio},
  \citenamefont {Napoli}, \citenamefont {Messina},\ and\ \citenamefont
  {Breuer}}]{LeggioNapoliMessinaBreuer2013}%
  \BibitemOpen
  \bibfield  {author} {\bibinfo {author} {\bibfnamefont {B.}~\bibnamefont
  {Leggio}}, \bibinfo {author} {\bibfnamefont {A.}~\bibnamefont {Napoli}},
  \bibinfo {author} {\bibfnamefont {A.}~\bibnamefont {Messina}}, \ and\
  \bibinfo {author} {\bibfnamefont {{H.-P.}}\ \bibnamefont {Breuer}},\
  }\bibfield  {title} {\enquote {\bibinfo {title} {Entropy production and
  information fluctuations along quantum trajectories},}\ }\href@noop {}
  {\bibfield  {journal} {\bibinfo  {journal} {Phys. Rev. A}\ }\textbf {\bibinfo
  {volume} {88}},\ \bibinfo {pages} {042111} (\bibinfo {year}
  {2013})}\BibitemShut {NoStop}%
\bibitem [{\citenamefont {Hekking}\ and\ \citenamefont
  {Pekola}(2013)}]{HekkingPekola2013}%
  \BibitemOpen
  \bibfield  {author} {\bibinfo {author} {\bibfnamefont {F.~W.~J.}\
  \bibnamefont {Hekking}}\ and\ \bibinfo {author} {\bibfnamefont {J.~P.}\
  \bibnamefont {Pekola}},\ }\bibfield  {title} {\enquote {\bibinfo {title}
  {Quantum jump approach for work and dissipation in a two-level system},}\
  }\href@noop {} {\bibfield  {journal} {\bibinfo  {journal} {Phys. Rev. Lett.}\
  }\textbf {\bibinfo {volume} {111}},\ \bibinfo {pages} {093602} (\bibinfo
  {year} {2013})}\BibitemShut {NoStop}%
\bibitem [{\citenamefont {Horowitz}\ and\ \citenamefont
  {Parrondo}(2013)}]{HorowitzParrondo2013}%
  \BibitemOpen
  \bibfield  {author} {\bibinfo {author} {\bibfnamefont {Jordan~M.}\
  \bibnamefont {Horowitz}}\ and\ \bibinfo {author} {\bibfnamefont {Juan M.~R.}\
  \bibnamefont {Parrondo}},\ }\bibfield  {title} {\enquote {\bibinfo {title}
  {Entropy production along nonequilibrium quantum jump trajectories},}\
  }\href@noop {} {\bibfield  {journal} {\bibinfo  {journal} {New J. Phys.}\
  }\textbf {\bibinfo {volume} {15}},\ \bibinfo {pages} {085028} (\bibinfo
  {year} {2013})}\BibitemShut {NoStop}%
\bibitem [{\citenamefont {Gong}\ \emph {et~al.}(2016)\citenamefont {Gong},
  \citenamefont {Ashida},\ and\ \citenamefont {Ueda}}]{GongAshidaUeda2016}%
  \BibitemOpen
  \bibfield  {author} {\bibinfo {author} {\bibfnamefont {Zongping}\
  \bibnamefont {Gong}}, \bibinfo {author} {\bibfnamefont {Yuto}\ \bibnamefont
  {Ashida}}, \ and\ \bibinfo {author} {\bibfnamefont {Masahito}\ \bibnamefont
  {Ueda}},\ }\bibfield  {title} {\enquote {\bibinfo {title} {Quantum trajectory
  thermodynamics with discrete feedback control},}\ }\href@noop {} {\bibfield
  {journal} {\bibinfo  {journal} {Phys. Rev. A}\ }\textbf {\bibinfo {volume}
  {94}},\ \bibinfo {pages} {012107} (\bibinfo {year} {2016})}\BibitemShut
  {NoStop}%
\bibitem [{\citenamefont {Lindblad}(1976)}]{Lindblad1976}%
  \BibitemOpen
  \bibfield  {author} {\bibinfo {author} {\bibfnamefont {G.}~\bibnamefont
  {Lindblad}},\ }\bibfield  {title} {\enquote {\bibinfo {title} {On the
  generators of quantum dynamical semigroups},}\ }\href@noop {} {\bibfield
  {journal} {\bibinfo  {journal} {Commun. math. Phys.}\ }\textbf {\bibinfo
  {volume} {48}},\ \bibinfo {pages} {119--130} (\bibinfo {year}
  {1976})}\BibitemShut {NoStop}%
\bibitem [{\citenamefont {Crooks}(2000)}]{Crooks2000}%
  \BibitemOpen
  \bibfield  {author} {\bibinfo {author} {\bibfnamefont {Gavin~E.}\
  \bibnamefont {Crooks}},\ }\bibfield  {title} {\enquote {\bibinfo {title}
  {Path-ensemble averages in systems driven far from equilibrium},}\
  }\href@noop {} {\bibfield  {journal} {\bibinfo  {journal} {Phys. Rev. E}\
  }\textbf {\bibinfo {volume} {61}},\ \bibinfo {pages} {2361--2366} (\bibinfo
  {year} {2000})}\BibitemShut {NoStop}%
\bibitem [{\citenamefont {Wehrl}(1977)}]{Wehrl1977}%
  \BibitemOpen
  \bibfield  {author} {\bibinfo {author} {\bibfnamefont {A.}~\bibnamefont
  {Wehrl}},\ }\bibfield  {title} {\enquote {\bibinfo {title} {Remarks on
  a-entropy},}\ }\href@noop {} {\bibfield  {journal} {\bibinfo  {journal} {Rep.
  Math. Phys.}\ }\textbf {\bibinfo {volume} {12}},\ \bibinfo {pages} {385}
  (\bibinfo {year} {1977})}\BibitemShut {NoStop}%
\bibitem [{\citenamefont {Balian}\ \emph {et~al.}(1986)\citenamefont {Balian},
  \citenamefont {V\'en\'eroni},\ and\ \citenamefont
  {Balazs}}]{BalianVeneroniBalazs1986}%
  \BibitemOpen
  \bibfield  {author} {\bibinfo {author} {\bibfnamefont {R.}~\bibnamefont
  {Balian}}, \bibinfo {author} {\bibfnamefont {M.}~\bibnamefont
  {V\'en\'eroni}}, \ and\ \bibinfo {author} {\bibfnamefont {N.}~\bibnamefont
  {Balazs}},\ }\bibfield  {title} {\enquote {\bibinfo {title} {Relevant entropy
  vs. measurement entropy},}\ }\href@noop {} {\bibfield  {journal} {\bibinfo
  {journal} {Europhys. Lett.}\ }\textbf {\bibinfo {volume} {1}},\ \bibinfo
  {pages} {1} (\bibinfo {year} {1986})}\BibitemShut {NoStop}%
\bibitem [{\citenamefont {S{\l}omczy\'nski}\ and\ \citenamefont
  {Szymusiak}(2016)}]{SlomczynskiSzymusiak2016}%
  \BibitemOpen
  \bibfield  {author} {\bibinfo {author} {\bibfnamefont {W.}~\bibnamefont
  {S{\l}omczy\'nski}}\ and\ \bibinfo {author} {\bibfnamefont {A.}~\bibnamefont
  {Szymusiak}},\ }\bibfield  {title} {\enquote {\bibinfo {title} {Highly
  symmetric {POVM}s and their informational power},}\ }\href@noop {} {\bibfield
   {journal} {\bibinfo  {journal} {Quantum Inf. Process}\ }\textbf {\bibinfo
  {volume} {15}},\ \bibinfo {pages} {565} (\bibinfo {year} {2016})}\BibitemShut
  {NoStop}%
\bibitem [{\citenamefont {Halmos}(1974)}]{Hal74}%
  \BibitemOpen
  \bibfield  {author} {\bibinfo {author} {\bibfnamefont {Paul~R.}\ \bibnamefont
  {Halmos}},\ }\href@noop {} {\emph {\bibinfo {title} {Measure Theory}}}\
  (\bibinfo  {publisher} {Springer},\ \bibinfo {year} {1974})\ pp.\ \bibinfo
  {pages} {134, 182}\BibitemShut {NoStop}%
\bibitem [{\citenamefont {Bartle}(1995)}]{Bar95}%
  \BibitemOpen
  \bibfield  {author} {\bibinfo {author} {\bibfnamefont {Robert~G.}\
  \bibnamefont {Bartle}},\ }\href@noop {} {\emph {\bibinfo {title} {The
  Elements of Integration and Lebesgue Measure}}}\ (\bibinfo  {publisher} {John
  Wiley \& Sons Ltd.},\ \bibinfo {year} {1995})\ p.~\bibinfo {pages}
  {88}\BibitemShut {NoStop}%
\bibitem [{\citenamefont {Murashita}\ and\ \citenamefont
  {Ueda}(2017)}]{MurashitaUeda2017}%
  \BibitemOpen
  \bibfield  {author} {\bibinfo {author} {\bibfnamefont {Y\^uto}\ \bibnamefont
  {Murashita}}\ and\ \bibinfo {author} {\bibfnamefont {Masahito}\ \bibnamefont
  {Ueda}},\ }\bibfield  {title} {\enquote {\bibinfo {title} {Gibbs paradox
  revisited from the fluctuation theorem with absolute irreversibility},}\
  }\href@noop {} {\bibfield  {journal} {\bibinfo  {journal} {Phys. Rev. Lett.}\
  }\textbf {\bibinfo {volume} {118}},\ \bibinfo {pages} {060601} (\bibinfo
  {year} {2017})}\BibitemShut {NoStop}%
\bibitem [{\citenamefont {Hoang}\ \emph {et~al.}(2016)\citenamefont {Hoang},
  \citenamefont {Venkatesh}, \citenamefont {Han}, \citenamefont {Jo},
  \citenamefont {Watanabe},\ and\ \citenamefont {Choi}}]{Hoang_2016}%
  \BibitemOpen
  \bibfield  {author} {\bibinfo {author} {\bibfnamefont {Danh-Tai}\
  \bibnamefont {Hoang}}, \bibinfo {author} {\bibfnamefont {B.~Prasanna}\
  \bibnamefont {Venkatesh}}, \bibinfo {author} {\bibfnamefont {Seungju}\
  \bibnamefont {Han}}, \bibinfo {author} {\bibfnamefont {Junghyo}\ \bibnamefont
  {Jo}}, \bibinfo {author} {\bibfnamefont {Gentaro}\ \bibnamefont {Watanabe}},
  \ and\ \bibinfo {author} {\bibfnamefont {Mahn-Soo}\ \bibnamefont {Choi}},\
  }\bibfield  {title} {\enquote {\bibinfo {title} {Scaling law for irreversible
  entropy production in critical systems},}\ }\href@noop {} {\bibfield
  {journal} {\bibinfo  {journal} {Sci. Rep.}\ }\textbf {\bibinfo {volume}
  {6}},\ \bibinfo {pages} {27603} (\bibinfo {year} {2016})}\BibitemShut
  {NoStop}%
\bibitem [{\citenamefont {Ueda}\ \emph {et~al.}(1992)\citenamefont {Ueda},
  \citenamefont {Imoto}, \citenamefont {Nagaoka},\ and\ \citenamefont
  {Ogawa}}]{UedaImotoNagaokaOgawa1992}%
  \BibitemOpen
  \bibfield  {author} {\bibinfo {author} {\bibfnamefont {Masahito}\
  \bibnamefont {Ueda}}, \bibinfo {author} {\bibfnamefont {Nobuyuki}\
  \bibnamefont {Imoto}}, \bibinfo {author} {\bibfnamefont {Hiroshi}\
  \bibnamefont {Nagaoka}}, \ and\ \bibinfo {author} {\bibfnamefont {Tetsuo}\
  \bibnamefont {Ogawa}},\ }\bibfield  {title} {\enquote {\bibinfo {title}
  {Continuous quantum-nondemolition measurement of photon number},}\
  }\href@noop {} {\bibfield  {journal} {\bibinfo  {journal} {Phys. Rev. A}\
  }\textbf {\bibinfo {volume} {46}},\ \bibinfo {pages} {2859} (\bibinfo {year}
  {1992})}\BibitemShut {NoStop}%
\bibitem [{\citenamefont {Kist}\ \emph {et~al.}(1999)\citenamefont {Kist},
  \citenamefont {Orszag}, \citenamefont {Brun},\ and\ \citenamefont
  {Davidovich}}]{KistOrszagBrunDavidovich1999}%
  \BibitemOpen
  \bibfield  {author} {\bibinfo {author} {\bibfnamefont {Tarso B.~L.}\
  \bibnamefont {Kist}}, \bibinfo {author} {\bibfnamefont {M.}~\bibnamefont
  {Orszag}}, \bibinfo {author} {\bibfnamefont {T.~A.}\ \bibnamefont {Brun}}, \
  and\ \bibinfo {author} {\bibfnamefont {L.}~\bibnamefont {Davidovich}},\
  }\bibfield  {title} {\enquote {\bibinfo {title} {Stochastic schr\"odinger
  equations in cavity {QED}: physical interpretation and localization},}\
  }\href@noop {} {\bibfield  {journal} {\bibinfo  {journal} {J. Opt. B: Quantum
  Semiclass. Opt.}\ }\textbf {\bibinfo {volume} {1}} (\bibinfo {year}
  {1999})}\BibitemShut {NoStop}%
\bibitem [{\citenamefont {Guerlin}\ \emph {et~al.}(2007)\citenamefont
  {Guerlin}, \citenamefont {Bernu}, \citenamefont {Del\'eglise}, \citenamefont
  {Sayrin}, \citenamefont {Gleyzes}, \citenamefont {Kuhr}, \citenamefont
  {Brune}, \citenamefont {Raimond},\ and\ \citenamefont
  {Haroche}}]{Guerlin...Haroche2007}%
  \BibitemOpen
  \bibfield  {author} {\bibinfo {author} {\bibfnamefont {Christine}\
  \bibnamefont {Guerlin}}, \bibinfo {author} {\bibfnamefont {Julien}\
  \bibnamefont {Bernu}}, \bibinfo {author} {\bibfnamefont {Samuel}\
  \bibnamefont {Del\'eglise}}, \bibinfo {author} {\bibfnamefont {Cl\'ement}\
  \bibnamefont {Sayrin}}, \bibinfo {author} {\bibfnamefont {S\'ebastien}\
  \bibnamefont {Gleyzes}}, \bibinfo {author} {\bibfnamefont {Stefan}\
  \bibnamefont {Kuhr}}, \bibinfo {author} {\bibfnamefont {Michel}\ \bibnamefont
  {Brune}}, \bibinfo {author} {\bibfnamefont {Jean-Michel}\ \bibnamefont
  {Raimond}}, \ and\ \bibinfo {author} {\bibfnamefont {Serge}\ \bibnamefont
  {Haroche}},\ }\bibfield  {title} {\enquote {\bibinfo {title} {Progressive
  field-state collapse and quantum non-demolition photon counting},}\
  }\href@noop {} {\bibfield  {journal} {\bibinfo  {journal} {Nature}\ }\textbf
  {\bibinfo {volume} {448}},\ \bibinfo {pages} {889} (\bibinfo {year}
  {2007})}\BibitemShut {NoStop}%
\bibitem [{\citenamefont {Del\'eglise}\ \emph {et~al.}(2008)\citenamefont
  {Del\'eglise}, \citenamefont {Dotsenko}, \citenamefont {Sayrin},
  \citenamefont {Bernu}, \citenamefont {Brune}, \citenamefont {Raimond},\ and\
  \citenamefont {Haroche}}]{Deleglise...Haroche2008}%
  \BibitemOpen
  \bibfield  {author} {\bibinfo {author} {\bibfnamefont {Samuel}\ \bibnamefont
  {Del\'eglise}}, \bibinfo {author} {\bibfnamefont {Igor}\ \bibnamefont
  {Dotsenko}}, \bibinfo {author} {\bibfnamefont {Cl\'ement}\ \bibnamefont
  {Sayrin}}, \bibinfo {author} {\bibfnamefont {Julien}\ \bibnamefont {Bernu}},
  \bibinfo {author} {\bibfnamefont {Michel}\ \bibnamefont {Brune}}, \bibinfo
  {author} {\bibfnamefont {Jean-Michel}\ \bibnamefont {Raimond}}, \ and\
  \bibinfo {author} {\bibfnamefont {Serge}\ \bibnamefont {Haroche}},\
  }\bibfield  {title} {\enquote {\bibinfo {title} {Reconstruction of
  non-classical cavity field states with snapshots of their decoherence},}\
  }\href@noop {} {\bibfield  {journal} {\bibinfo  {journal} {Nature}\ }\textbf
  {\bibinfo {volume} {455}},\ \bibinfo {pages} {510} (\bibinfo {year}
  {2008})}\BibitemShut {NoStop}%
\bibitem [{\citenamefont {Haroche}\ and\ \citenamefont
  {Raimond}(2013)}]{HarocheRaimond}%
  \BibitemOpen
  \bibfield  {author} {\bibinfo {author} {\bibfnamefont {S.}~\bibnamefont
  {Haroche}}\ and\ \bibinfo {author} {\bibfnamefont {J.-M.}\ \bibnamefont
  {Raimond}},\ }\href@noop {} {\emph {\bibinfo {title} {Exploring the Quantum:
  Atoms, Cavities, and Photons}}}\ (\bibinfo  {publisher} {Oxford University
  Press},\ \bibinfo {year} {2013})\BibitemShut {NoStop}%
\bibitem [{\citenamefont {Schuster}\ \emph {et~al.}(2007)\citenamefont
  {Schuster}, \citenamefont {Houck}, \citenamefont {Schreier}, \citenamefont
  {Wallraff}, \citenamefont {Gambetta}, \citenamefont {Blais}, \citenamefont
  {Frunzio}, \citenamefont {Majer}, \citenamefont {Johnson}, \citenamefont
  {Devoret}, \citenamefont {Girvin},\ and\ \citenamefont
  {Schoelkopf}}]{Schuster_2007}%
  \BibitemOpen
  \bibfield  {author} {\bibinfo {author} {\bibfnamefont {D.~I.}\ \bibnamefont
  {Schuster}}, \bibinfo {author} {\bibfnamefont {A.~A.}\ \bibnamefont {Houck}},
  \bibinfo {author} {\bibfnamefont {J.~A.}\ \bibnamefont {Schreier}}, \bibinfo
  {author} {\bibfnamefont {A.}~\bibnamefont {Wallraff}}, \bibinfo {author}
  {\bibfnamefont {J.~M.}\ \bibnamefont {Gambetta}}, \bibinfo {author}
  {\bibfnamefont {A.}~\bibnamefont {Blais}}, \bibinfo {author} {\bibfnamefont
  {L.}~\bibnamefont {Frunzio}}, \bibinfo {author} {\bibfnamefont
  {J.}~\bibnamefont {Majer}}, \bibinfo {author} {\bibfnamefont
  {B.}~\bibnamefont {Johnson}}, \bibinfo {author} {\bibfnamefont {M.~H.}\
  \bibnamefont {Devoret}}, \bibinfo {author} {\bibfnamefont {S.~M.}\
  \bibnamefont {Girvin}}, \ and\ \bibinfo {author} {\bibfnamefont {R.~J.}\
  \bibnamefont {Schoelkopf}},\ }\bibfield  {title} {\enquote {\bibinfo {title}
  {Resolving photon number states in a superconducting circuit},}\ }\href@noop
  {} {\bibfield  {journal} {\bibinfo  {journal} {Nature (London)}\ }\textbf
  {\bibinfo {volume} {445}},\ \bibinfo {pages} {515} (\bibinfo {year}
  {2007})}\BibitemShut {NoStop}%
\bibitem [{\citenamefont {Wallraff}\ \emph {et~al.}(2005)\citenamefont
  {Wallraff}, \citenamefont {Schuster}, \citenamefont {Blais}, \citenamefont
  {Frunzio}, \citenamefont {Majer}, \citenamefont {Devoret}, \citenamefont
  {Girvin},\ and\ \citenamefont {Schoelkopf}}]{Wallraff_2005}%
  \BibitemOpen
  \bibfield  {author} {\bibinfo {author} {\bibfnamefont {A.}~\bibnamefont
  {Wallraff}}, \bibinfo {author} {\bibfnamefont {D.~I.}\ \bibnamefont
  {Schuster}}, \bibinfo {author} {\bibfnamefont {A.}~\bibnamefont {Blais}},
  \bibinfo {author} {\bibfnamefont {L.}~\bibnamefont {Frunzio}}, \bibinfo
  {author} {\bibfnamefont {J.}~\bibnamefont {Majer}}, \bibinfo {author}
  {\bibfnamefont {M.~H.}\ \bibnamefont {Devoret}}, \bibinfo {author}
  {\bibfnamefont {S.~M.}\ \bibnamefont {Girvin}}, \ and\ \bibinfo {author}
  {\bibfnamefont {R.~J.}\ \bibnamefont {Schoelkopf}},\ }\bibfield  {title}
  {\enquote {\bibinfo {title} {Approaching unit visibility for control of a
  superconducting qubit with dispersive readout},}\ }\href@noop {} {\bibfield
  {journal} {\bibinfo  {journal} {Phys. Rev. Lett.}\ }\textbf {\bibinfo
  {volume} {95}},\ \bibinfo {pages} {060501} (\bibinfo {year}
  {2005})}\BibitemShut {NoStop}%
\bibitem [{\citenamefont {Paauw}\ \emph {et~al.}(2009)\citenamefont {Paauw},
  \citenamefont {Fedorov}, \citenamefont {Harmans},\ and\ \citenamefont
  {Mooij}}]{Paauw_2009}%
  \BibitemOpen
  \bibfield  {author} {\bibinfo {author} {\bibfnamefont {F.~G.}\ \bibnamefont
  {Paauw}}, \bibinfo {author} {\bibfnamefont {A.}~\bibnamefont {Fedorov}},
  \bibinfo {author} {\bibfnamefont {C.~J. P.~M.}\ \bibnamefont {Harmans}}, \
  and\ \bibinfo {author} {\bibfnamefont {J.~E.}\ \bibnamefont {Mooij}},\
  }\bibfield  {title} {\enquote {\bibinfo {title} {Tuning the gap of a
  superconducting flux qubit},}\ }\href@noop {} {\bibfield  {journal} {\bibinfo
   {journal} {Phys. Rev. Lett.}\ }\textbf {\bibinfo {volume} {102}},\ \bibinfo
  {pages} {090501} (\bibinfo {year} {2009})}\BibitemShut {NoStop}%
\bibitem [{\citenamefont {Bretheau}\ \emph {et~al.}(2016)\citenamefont
  {Bretheau}, \citenamefont {Campagne-Ibarcq}, \citenamefont {Flurin},
  \citenamefont {Mallet},\ and\ \citenamefont {Huard}}]{Bretheau_2016}%
  \BibitemOpen
  \bibfield  {author} {\bibinfo {author} {\bibfnamefont {L.}~\bibnamefont
  {Bretheau}}, \bibinfo {author} {\bibfnamefont {P.}~\bibnamefont
  {Campagne-Ibarcq}}, \bibinfo {author} {\bibfnamefont {E.}~\bibnamefont
  {Flurin}}, \bibinfo {author} {\bibfnamefont {F.}~\bibnamefont {Mallet}}, \
  and\ \bibinfo {author} {\bibfnamefont {B.}~\bibnamefont {Huard}},\ }\bibfield
   {title} {\enquote {\bibinfo {title} {Quantum dynamics of an electromagnetic
  mode that cannot contain $n$ photons},}\ }\href@noop {} {\bibfield  {journal}
  {\bibinfo  {journal} {Sicience}\ }\textbf {\bibinfo {volume} {348}},\
  \bibinfo {pages} {776} (\bibinfo {year} {2016})}\BibitemShut {NoStop}%
\bibitem [{\citenamefont {Santos}\ \emph {et~al.}(2017)\citenamefont {Santos},
  \citenamefont {C\'eleri}, \citenamefont {Landi},\ and\ \citenamefont
  {Paternostro}}]{Santos_2017}%
  \BibitemOpen
  \bibfield  {author} {\bibinfo {author} {\bibfnamefont {Jader~P.}\
  \bibnamefont {Santos}}, \bibinfo {author} {\bibfnamefont {Lucas~C.}\
  \bibnamefont {C\'eleri}}, \bibinfo {author} {\bibfnamefont {Gabriel~T.}\
  \bibnamefont {Landi}}, \ and\ \bibinfo {author} {\bibfnamefont {Mauro}\
  \bibnamefont {Paternostro}},\ }\bibfield  {title} {\enquote {\bibinfo {title}
  {The role of quantum coherence in non-equilibrium entropy production},}\
  }\href@noop {} {\bibfield  {journal} {\bibinfo  {journal} {arXiv}\ ,\
  \bibinfo {pages} {1707.08946}} (\bibinfo {year} {2017})}\BibitemShut
  {NoStop}%
\bibitem [{\citenamefont {Francica}\ \emph {et~al.}(2017)\citenamefont
  {Francica}, \citenamefont {Goold},\ and\ \citenamefont
  {Plastina}}]{FrancicaGooldPlastina2017}%
  \BibitemOpen
  \bibfield  {author} {\bibinfo {author} {\bibfnamefont {G.}~\bibnamefont
  {Francica}}, \bibinfo {author} {\bibfnamefont {J.}~\bibnamefont {Goold}}, \
  and\ \bibinfo {author} {\bibfnamefont {F.}~\bibnamefont {Plastina}},\
  }\bibfield  {title} {\enquote {\bibinfo {title} {The role of coherence in the
  non-equilibrium thermodynamics of quantum systems},}\ }\href@noop {}
  {\bibfield  {journal} {\bibinfo  {journal} {arXiv}\ ,\ \bibinfo {pages}
  {1707.06950}} (\bibinfo {year} {2017})}\BibitemShut {NoStop}%
\end{thebibliography}%

\end{document}